\documentclass[compress]{elsarticle}

\usepackage[margin=1in]{geometry}

\usepackage{amsmath,amsfonts}
\usepackage{algorithmic}
\usepackage{algorithm}
\usepackage{array}
\usepackage[caption=false,font=normalsize,labelfont=sf,textfont=sf]{subfig}
\usepackage{textcomp}
\usepackage{stfloats}
\usepackage{xurl}
\usepackage{verbatim}
\usepackage{graphicx}
\usepackage{booktabs}
\usepackage{bm}
\usepackage{tcolorbox}
\usepackage{enumitem}
\usepackage{hyperref}
\usepackage{adjustbox}
\usepackage{multirow}

\usepackage{placeins}

\usepackage{amsmath,amsfonts,amssymb,amsthm}
\newtheorem{theorem}{Theorem}[section]

\theoremstyle{definition}
\newtheorem{example}[theorem]{Example}

\newcommand{\fittt}[1]{\resizebox{\linewidth}{!}{\ttfamily\detokenize{#1}}}
\newcommand{\RQone}{{What is the impact of sequence length on model performance?}}
\newcommand{\RQtwo}{{How do different tokenization methods affect performance?}}
\newcommand{\RQthree}{{How does sample efficiency vary across model architectures?}}
\newcommand{\RQfour}{{Which model architecture performs best?}}

\begin{document}

\begin{frontmatter}

\title{On Sequence-to-Sequence Models for Automated Log Parsing}

\author[label1]{Adam Sorrenti}
\ead{adam.sorrenti@torontomu.ca}

\author[label1]{Andriy Miranskyy}
\ead{avm@torontomu.ca}

\affiliation[label1]{organization={Department of Computer Science, Toronto Metropolitan University},
            addressline={350 Victoria Street}, 
            city={Toronto},
            postcode={M5B~2K3}, 
            state={Ontario},
            country={Canada}}

\begin{abstract}

\textbf{Context:}
Log parsing is a critical standard operating procedure in software systems, enabling monitoring, anomaly detection, and failure diagnosis. However, automated log parsing remains challenging due to heterogeneous log formats, distribution shifts between training and deployment data, and the brittleness of rule-based approaches.

\textbf{Objectives:}
This study aims to systematically evaluate how sequence modelling architecture, representation choice, sequence length, and training data availability influence automated log parsing performance and computational cost.

\textbf{Methods:}
We conduct a controlled empirical study comparing four sequence modelling architectures: Transformer, Mamba state-space, monodirectional LSTM, and bidirectional LSTM models. In total, 396 models are trained across multiple dataset configurations and evaluated using relative Levenshtein edit distance with statistical significance testing. 

\textbf{Results:}
Transformer achieves the lowest mean relative edit distance (0.111), followed by Mamba (0.145), mono-LSTM (0.186), and bi-LSTM (0.265), where lower values are better.
Mamba provides competitive accuracy with substantially lower computational cost. Character-level tokenization generally improves performance, sequence length has negligible practical impact on Transformer accuracy, and both Mamba and Transformer demonstrate stronger sample efficiency than recurrent models.

\textbf{Conclusion:}
Overall, Transformers reduce parsing error by 23.4\%, while Mamba is a strong alternative under data or compute constraints.
These results also clarify the roles of representation choice, sequence length, and sample efficiency, providing practical guidance for researchers and practitioners.

\end{abstract}

\begin{keyword}
Automated log parsing \sep Recurrent neural networks \sep Sequence-to-sequence models \sep State space models

\end{keyword}

\end{frontmatter}

\section{Introduction}
Software systems produce vast amounts of log data, a rich source of information crucial for system diagnostics~\cite{DBLP:conf/asplos/YuanMXTZP10, DBLP:conf/icws/LuRWTWW17, DBLP:conf/hpdc/DasMSV18}, data analytics~\cite{DBLP:journals/pvldb/LeeLLLR12, DBLP:conf/hpdc/DasMSV18}, and anomaly detection~\cite{DBLP:conf/ccs/Du0ZS17, DBLP:conf/sigsoft/ZhangXLQZDXYCLC19, DBLP:conf/srds/Zhang0MZ20, DBLP:conf/icse/IslamP0SEM21, DBLP:conf/icse-seip/IslamRPSSM25}. Software logs are often the only available record of software runtime state. These logs, generated at an unprecedented scale \cite{DBLP:journals/tpds/MiWZLC13}, encapsulate a detailed history of system activities, errors, and transactions. However, the volume and complexity of these data present significant challenges in effective log parsing and analysis. Traditional, rule-based methods struggle to keep pace with the evolving log formats found in modern large-scale software systems. Tools for log analysis prefer a unified format \cite{DBLP:journals/software/MiranskyyHCL16}, yet many log sources are inherently diverse in structure, varying significantly across systems and applications. This diversity of log formats necessitates more insight into adaptable methods for effective log parsing.

Previous empirical studies show that cloud monitoring logs exhibit inherent structural heterogeneity due to the scale, velocity, and diversity of monitored components~\cite{pourmajidi2017challenges}, that operational log volumes render manual analysis infeasible even within a single cloud provider~\cite{pourmajidi2019dogfooding}, and that elasticity and continuous service evolution invalidate static monitoring assumptions over time~\cite{pourmajidi2021challenging}.

These deployment conditions also make the economics of automated parsing a first-order concern. We do not assume that a production system would invoke an LLM on every repeated instance of a known log template. In a stable, homogeneous system, an LLM may identify a template once, after which a lightweight rule-based parser or regular expression can be reused. The concern addressed here is different: modern cloud and enterprise infrastructures often consist of many services, frameworks, versions, and independently evolving components that generate heterogeneous and drifting log streams~\cite{DBLP:conf/icse/IslamP0SEM21,pourmajidi2021challenging}. In such settings, parser construction is not a one-time activity: new templates, previously unseen message structures, and service-specific variations arise across many contexts, requiring repeated template discovery, parser repair, and validation. 

This scaling concern is not merely hypothetical. Prior empirical studies of large-scale cloud systems report substantial structural heterogeneity in monitoring data, operational log volumes that make manual analysis infeasible, and continuous service evolution that invalidates static monitoring assumptions over time~\cite{DBLP:journals/software/MiranskyyHCL16,pourmajidi2017challenges,DBLP:conf/icse/IslamP0SEM21,pourmajidi2021challenging}. Under such conditions, even if LLM inference is reserved only for the fraction of messages associated with new or changing templates, recurring hosted-model calls can quickly dominate the operational cost of the parsing pipeline.
At production scale, where monitoring systems may process millions of log records over short operational windows, the relevant question is therefore not only whether an LLM can parse an individual log message, but how often it must be invoked before its cost can be amortized by a cheaper parser. Recent industry reporting similarly describes organizations questioning whether AI investments justify their costs~\cite{weatherbed2026uberai}, and highlights cases where unmanaged model usage produced unexpectedly large bills~\cite{warwick2026claudecost}.

To illustrate why smaller, task-specific models such as Transformer and Mamba remain practically relevant, Example~\ref{ex:llm_log_parsing_cost} shows how hosted LLM inference can become prohibitively expensive in recurring log parsing scenarios.

\example[Cost scaling of hosted LLM parsing]{\label{ex:llm_log_parsing_cost}
To make this trade-off concrete, Tables~\ref{tab:llm_common400_comparison} and~\ref{tab:llm_common400_token_usage_cost} provide a motivating comparison on a common subset of 400 log examples shared across all LLM runs. For each example, we compare the LLM predictions with predictions from two sequence models considered in this study: the word-tokenized Transformer ($M_T$) and the character-tokenized Mamba model ($M_M$), both evaluated under the high-difficulty training condition with full training data. Table~\ref{tab:llm_common400_comparison} reports the mean relative edit distance ($D_R$) of each LLM and the percentage of cases in which the LLM outperforms or ties the corresponding sequence model. Table~\ref{tab:llm_common400_token_usage_cost} reports the token usage and estimated inference cost for the same 400 examples. The LLM prompt construction is documented in~\ref{appendix:llm-prompt}. 

As a simple scaling exercise, the costs in Table~\ref{tab:llm_common400_token_usage_cost} can be extrapolated from 400 examples to one million examples by multiplying by 2,500. This yields estimated costs of approximately \$107,925 for GPT-5.5, \$15,900 for GPT-5 mini, \$2,700 for GPT-4o, \$2,375 for GPT-4.1, and \$600 for GPT-OSS-120B per million log records under the same prompting protocol. At this scale, the cost is no longer merely a technical implementation detail: expenditures ranging from hundreds to tens of thousands of dollars per million records would likely require explicit justification to operational managers and financial officers, particularly when lower-cost local models provide competitive parsing accuracy in many configurations.

These values are point-in-time estimates. They should not be read as a recommendation to invoke an LLM for every log message, but as a quantification of the marginal-cost risk when parser discovery, repair, or validation must be repeated across large heterogeneous streams. Table~\ref{tab:llm_common400_comparison} also shows that the strongest LLM result is the most expensive, while several lower-cost LLMs underperform the local Transformer and Mamba baselines on the same examples. The contrast with the local baselines is important: as reported in Table~\ref{tab:model_params}, the profiled per-batch inference time is 158.88~ms for the shortest Transformer configuration ($M_T$, $L=4188$) and only 0.46~ms for Mamba ($M_M$). These models, therefore, incur an upfront training or adaptation cost primarily, after which inference can be run locally at a predictable marginal cost. Thus, LLMs may be attractive for one-off analysis, bootstrapping, or low-volume template discovery, whereas production environments with frequent drift require attention to recurring marginal cost, latency, rate limits, data-governance constraints, and dependence on provider-side model and pricing changes.

\begin{table}[tb]
\centering
\caption{Motivating comparison of LLM parsing results against the Transformer ($M_T$) and Mamba ($M_M$) baselines. Lower $D_R$ is better. 
For each baseline, LLM win (\%) denotes the percentage of examples for which the LLM obtains a lower relative edit distance than the corresponding sequence model, and Tie (\%) denotes equal relative edit distance.
}
\label{tab:llm_common400_comparison}
\begin{tabular}{lrrrrr}
\toprule
 & & \multicolumn{2}{c}{$M_T$} & \multicolumn{2}{c}{$M_M$} \\
\cmidrule(lr){3-4}\cmidrule(lr){5-6}
LLM & $D_R$ & LLM Win (\%) & Tie (\%) & LLM Win (\%) & Tie (\%) \\
\midrule
GPT-5.5 & 0.002 & 71.2 & 25.5 & 100.0 & 0.0 \\
GPT-5 mini & 0.074 & 51.2 & 2.2 & 71.8 & 2.0 \\
GPT-4o & 0.264 & 4.0 & 0.2 & 25.2 & 1.2 \\
GPT-4.1 & 0.327 & 2.0 & 0.2 & 13.5 & 0.8 \\
GPT-OSS-120B & 0.446 & 5.5 & 0.2 & 23.5 & 0.8 \\
\bottomrule
\end{tabular}

\end{table}

\begin{table}[tb]
\centering
\caption{Token usage and estimated inference cost. Costs use direct OpenAI prices where available~\cite{openai_pricing_2026}, and the OpenRouter paid route for GPT-OSS-120B~\cite{openrouter_gptoss120b_pricing_2026}. Input-token totals can vary across models and providers because tokenizers and API-level templates differ, including system-preamble formatting for reasoning models.}
\label{tab:llm_common400_token_usage_cost}
\begin{tabular}{lrrrr}
\toprule
LLM & Input & Output & Est. cost (USD) \\
\midrule
GPT-5.5 & 282,166 & 1,392,064 & 43.17 \\
GPT-5 mini & 282,166 & 3,144,828 & 6.36 \\
GPT-4o & 282,566 & 37,534 & 1.08 \\
GPT-4.1 & 282,566 & 48,483 & 0.95 \\
GPT-OSS-120B & 306,566 & 1,239,202 & 0.24 \\
\bottomrule
\end{tabular}

\end{table}

\medskip
}

Many traditional log parsing methods rely on domain experts to craft and maintain structured templates for extraction. Related rule-based approaches \cite{8029742, 4601543} include tree parsing and heuristics. There also exists a large body of research on data-driven log parsing techniques, such as frequent pattern mining \cite{DBLP:journals/tse/DaiLCSC22}, clustering \cite{DBLP:conf/cnsm/VaarandiP15, 10.1145/2063576.2063690}, and longest common subsequence analysis \cite{8489912} that extract log templates. Neural network-based approaches to automated log parsing, such as masked language modelling \cite{9534113, 10.1007/978-3-030-67667-4_8} and large language models (LLMs) \cite{10.1145/3597503.3639150}, have also shown efficacy. More recently, a hybrid approach has been investigated using an encoder-only Transformer model \cite{NIPS2017_3f5ee243} to extract vector embeddings from log strings, followed by clustering and template extraction \cite{10301239}.

Log template extraction approaches often rely on unified log formats to distinguish between static and variable components of a log by comparing multiple, similar log strings. The advancements in neural network architectures offer a promising alternative approach that does not rely on log template extraction. Long Short-Term Memory (LSTM), a variant of the recurrent neural network (RNN), has demonstrated promising log parsing capabilities using a sequence-to-sequence approach \cite{Rand_2021} to map each input log token to its corresponding log field.  

While emerging state space and linear-time RNN models, such as S4~\cite{DBLP:journals/corr/abs-2111-00396}, xLSTM \cite{DBLP:conf/nips/BeckPSAP0KBH24}, and ParaRNN~\cite{danieli2025pararnnunlockingparalleltraining}, introduce promising efficiency characteristics, their application in log parsing remains unexplored. In particular, architectures like the Mamba state space model~\cite{DBLP:journals/corr/abs-2312-00752} have not been systematically evaluated for this application, representing a novel direction of investigation in this work.

This work is structured around several key research questions (RQs) as follows:
\begin{enumerate}[align=left,label=\textbf{RQ\arabic*.}]
    \item \RQone

    \item \RQtwo

    \item \RQthree

    \item \RQfour

\end{enumerate}
By investigating these research questions, we have the potential to inform researchers and practitioners on the best practices for applying deep sequence-to-sequence models to log parsing.

In this work, we use the term \textit{generalization} to refer specifically to robustness under log format distribution shift, including changes in field order, field presence, and formatting conventions between training and validation data. We do not evaluate the semantic understanding of log content nor cross-domain generalization across fundamentally different logging paradigms (e.g., structured JSON or event-based logs).

The \textbf{contributions} of this work are summarized as follows.

\begin{itemize}

   \item \textbf{Large-scale empirical analysis of key modelling factors.}
   We systematically quantify the impact of sequence length, tokenization strategy, and training data availability on log parsing performance, providing actionable guidance on when specific architectures are likely to succeed or fail in practice.
  
  \item \textbf{Novel application of state space models (Mamba) for log parsing.} 
  We demonstrate that Mamba-style architectures can achieve competitive accuracy while offering substantially lower computational cost, highlighting their suitability for production environments with resource constraints.

  \item \textbf{Cross-paradigm comparison of modern sequence modelling architectures.}
  We provide a controlled comparison of Transformers, bidirectional and monodirectional LSTMs, and state-space models, enabling architecture-level insights grounded in four research questions spanning accuracy, efficiency, and data sensitivity.

  \item \textbf{Quantitative characterization of dataset structure and format diversity.}
  Beyond model benchmarking, we analyze log format variability and field complexity, clarifying how real-world log heterogeneity influences parsing difficulty and model behaviour.
\end{itemize}

The remainder of this work is structured as follows. Section~\ref{sec:background} highlights background material relevant to the work. Section~\ref{sec:methodology} discusses the proposed methodology and its implementation. Evaluation and results of the aforementioned research question are given in Section~\ref{sec:evaluate}. Finally, Section~\ref{sec:conclusion} concludes the paper.

\section{Background} \label{sec:background}

Logging mechanisms construct statements that contain static and variable components, often devised by domain experts with little documentation and lax standards \cite{7202961}. The following subsection provides relevant background to the task and evaluation of automated log parsing.

\subsection{Log Parsing}
Log parsing is the act of extracting and identifying the variable log components (otherwise known as `log fields'). The differentiation between static and variable log components is often given~\cite{8804456} as a log format (otherwise known as `event template') where special tokens such as `\texttt{<*>}' represent the variable substring (the log field). 

Log fields can also be parsed using a fixed vocabulary. Figure~\ref{fig:apache-example} shows an Apache web server log and its associated log format on a per-character basis. Each character of the log input is assigned a category that corresponds to the semantic meaning of that substring. For example, the field characters which map to \texttt{08/Apr/2025:16:47:52 -0600} represent a timestamp, denoted by \texttt{t}. Additionally, the underscores denote a separator between the fields. Reference Table~\ref{tab:fields} for a description of each log field type.

\begin{table}[htbp]
\caption{A list of the relevant Apache web server log fields \cite{apache_mod_log_config}.}
\label{tab:fields}
\centering
\begin{tabular}{@{}p{0.09\columnwidth}p{0.87\columnwidth}@{}}
\toprule
Field & Field           \\                                                         
acronym & description   \\ \midrule
\texttt{h}             & IP address of the client host. Can be IPv4 or IPv6. \\
\texttt{l}             & The remote logname. We were unable to find a good example of what   kinds of values are returned by Apache servers, thus for this  paper we only supplied the commonly-given value ‘-’. This field could not be omitted as it is present in both the ELF and CLF formats.  \\
\texttt{u}             & The remote username. Can be empty (‘-’) \\
\texttt{t}             & The datetime of the request, presented in the default {[}day/month/year:hour:minute:second zone{]} format.  \\
\texttt{r}             & The request line from the client. Made up of the method, path and   querystring, and protocol. \\
\texttt{s}             & The status of the request. \\
\texttt{b}             & The number of bytes sent.   \\
\texttt{m}             & The request method.   \\
\texttt{U}             & The requested URI path.  \\
\texttt{H}             & The request protocol.  \\
\texttt{q}             & The request querystring.   \\
\texttt{v}             & The canonical servername of the server servicing the request.  \\
\texttt{V}             & The servername according to UseCanonical. In our generator, this field is identical to the \texttt{v}   field.   \\
\texttt{i}            & The user agent of the request$^{\dagger}$. \\
\texttt{R}             & The referrer of the request$^{\dagger}$. \\   
\texttt{\_}             & Represents a separator between log fields. \\
\bottomrule
\end{tabular}
\vspace{2pt}
\parbox{\textwidth}{\footnotesize
$^{\dagger}$ In a real Apache HTTP server deployment, this field is extracted from the \texttt{\%i}  log parameter, see~\cite{apache_mod_log_config} for details.
}
\end{table}

\begin{figure}[htbp]
  \centering
  \begin{tcolorbox}[
    width=\textwidth,     %
    colback=white, colframe=black,
    boxrule=2pt, arc=0pt,   %
    left=3pt, right=3pt, top=3pt, bottom=3pt %
  ]
    \fittt{197.34.164.236 - - 08/Apr/2025:16:47:52 -0600 "POST HTTP/1.1" 500 79060} \fittt{hhhhhhhhhhhhhh_l_u_tttttttttttttttttttttttttt_"rrrrrrrrrrrrr"_sss_bbbbb}
  \end{tcolorbox}
  \caption{A sample sequence-to-sequence mapping of an Apache web server log. The first line contains the raw log message. The second line contains the field type to which a given input character is mapped.}
  \label{fig:apache-example}
\end{figure}

\subsection{Benchmarks} \label{sec:benchmarks}
The systematic evaluation of log parsing techniques enables the research community to take steps in the direction of more robust and accurate methods. LogHub \cite{10301257} is a well-known benchmark dataset collection in the field of automatic log parsing and log-based anomaly detection. LogHub offers a comprehensive collection of real-world log datasets gathered from diverse software systems, including cloud platforms, distributed systems, and high-performance computing environments.

Table~\ref{tab:loghub_variable_instance_summary} summarizes the Loghub-2k datasets~\cite{10301257} (a preprocessed subset of the LogHub dataset containing 2000 randomly sampled examples) by the number of unique log templates and the statistics that describe the number of variables per log record.  Notice that the largest mean number of fields in a log is 4.5 (refer to the `Proxifier' dataset), and the largest number of unique log formats is 341 (refer to the `Mac' dataset). While the diversity of production logging mechanisms is a strength of Loghub-2k, the range and magnitude of unique log formats and fields are limited compared to industry-standard log formats, such as Apache web server logs.

\begin{table}[htbp]
\caption{The number of log formats (left) and unique log field statistics (right) per Loghub-2k dataset.}
\label{tab:loghub_variable_instance_summary}
\centering
\begin{tabular}{lr|rrrrr}
\toprule
 &  & \multicolumn{5}{c}{Unique Log Field} \\
\cmidrule(lr){3-7}
Dataset & Count & Mean & Median & Min. & Max. & Std. Dev. \\
\midrule
Android & 166 & 2.14 & 1.00 & 0 & 20 & 3.10 \\
Apache & 6 & 1.50 & 1.50 & 1 & 2 & 0.55 \\
BGL & 120 & 2.18 & 1.00 & 0 & 25 & 2.85 \\
HDFS & 14 & 3.64 & 3.50 & 1 & 6 & 1.34 \\
HPC & 46 & 1.24 & 1.00 & 0 & 9 & 1.80 \\
Hadoop & 114 & 1.75 & 1.00 & 0 & 12 & 2.24 \\
HealthApp & 75 & 1.32 & 1.00 & 0 & 8 & 1.72 \\
Linux & 118 & 1.19 & 1.00 & 0 & 13 & 1.65 \\
Mac & \textbf{341} & 3.06 & 1.00 & 0 & 36 & 4.69 \\
OpenSSH & 27 & 2.07 & 2.00 & 0 & 4 & 1.27 \\
OpenStack & 43 & 2.63 & 1.00 & 0 & 14 & 2.55 \\
Proxifier & 8 & \textbf{4.50} & 4.50 & 2 & 6 & 1.31 \\
Spark & 36 & 1.08 & 1.00 & 0 & 4 & 1.13 \\
Thunderbird & 149 & 1.58 & 1.00 & 0 & 10 & 1.99 \\
Windows & 50 & 1.40 & 1.00 & 0 & 7 & 1.62 \\
Zookeeper & 50 & 1.82 & 1.00 & 0 & 7 & 1.73 \\
\bottomrule
\end{tabular}
\end{table}

Rand~\textit{et~al.}~\cite{Rand_2021} introduced a synthetically generated Apache web server log parsing dataset (we denote this dataset as HTTPd-parse) and validated its efficacy with publicly available logs from three production-grade systems. HTTPd-parse consists of four synthetically generated training datasets of varying difficulty. $T_T$, $T_E$, $T_M$, and $T_H$ are training datasets labelled to describe their similarity to the validation datasets ($V_A$, $V_B$, and $V_C$) and their presumed modelling difficulty (trivial, easy, medium, and hard, respectively). 

The heuristic used to modulate training dataset difficulty (as described in \cite{Rand_2021}) operates by varying the percentage of standard and random Apache log formats used in the training dataset versus the validation dataset's composition. A complete description of the training and validation dataset composition is provided in Section~\ref{sec:dataset}. Table~\ref{tab:formats} describes the order in which web server log fields (whose description is given in Table~\ref{tab:fields}) appear in the source log record. A key differentiation is shown in Table~\ref{tab:log_statistics_httpd_parse}, which indicates that HTTPd-parse contains a larger range and magnitude of log formats and unique log fields compared to Loghub-2k. Furthermore, Table~\ref{tab:elf_train_dataset_summary} reports the percentage of ELF log records in each training dataset, illustrating the degree of format mismatch between training and validation data, particularly for the medium- and high-difficulty settings.

\begin{table}[htbp]
\caption{ Standard Apache log formats.}
\label{tab:formats}
\centering
\begin{tabular}{ll}
\toprule
Name                             & Fields (acronyms)       \\ 
\midrule
Common Log Format (CLF)          & \texttt{h l u t "r" s b}         \\
Combined Log Format (ELF) & \texttt{h l u t "r" s b "R" "i"} \\ 
\bottomrule
\end{tabular}
\end{table}

\begin{table}[htbp]
\caption{The number of log formats (left) and unique log field statistics (right) per \texttt{HTTPd-parse} training (top) and validation (bottom) dataset.}
\label{tab:log_statistics_httpd_parse}
\centering
\begin{tabular}{lr|rrrrr}
\toprule
 &  & \multicolumn{5}{c}{Unique Log Field} \\
\cmidrule(lr){3-7}
Dataset & Count & Mean & Median & Min. & Max. & Std. Dev. \\
\midrule
$T_T$ & 1 & \textbf{9.00} & 9.00 & 9 & 9 & 0.00 \\
$T_E$ & 7078 & 8.31 & 9.00 & 2 & 15 & 2.27 \\
$T_M$ & 47089 & 7.76 & 7.00 & 2 & 15 & 2.63 \\
$T_H$ & \textbf{92103} & 8.52 & 9.00 & 2 & 15 & 3.56 \\
\midrule
$V_A$ & 1 & 9.00 & 9.00 & 9 & 9 & 0.00 \\
$V_B$ & 1 & 9.00 & 9.00 & 9 & 9 & 0.00 \\
$V_C$ & 1 & 9.00 & 9.00 & 9 & 9 & 0.00 \\
\bottomrule
\end{tabular}
\end{table}

\begin{table}[htbp]
\centering
\caption{Summary of \texttt{HTTPd-parse} training datasets and their percentage of ELF logs.}
\label{tab:elf_train_dataset_summary}
\begin{tabular}{lrrrr}
\toprule
{Dataset} & {\# of logs} & {\# of ELF} & {\% of ELF} \\
\midrule
$T_T$ & 100000 & 100000 & 100.0 \\
$T_E$ & 20000 & 8011 & 40.1 \\
$T_M$ & 100000 & 0 & 0.0 \\
$T_H$ & 100000 & 0 & 0.0 \\
\bottomrule
\end{tabular}
\end{table}

\subsubsection{Evaluation Metrics}
\label{sec:eval-metrics}

To motivate the discussion of log parsing evaluation, we will first consider Parsing Accuracy (PA). PA is defined as the ratio of correctly parsed log records to the total number of log records parsed. For example, if a dataset under study has a set of three possible log formats $\{e_1, e_2, e_3\}$ and a total of four log records have been parsed and assigned formats $[e_2, e_2, e_3, e_1]$. Then, if the ground truth log formats for the four parsed log records are actually $[e_1, e_2, e_2, e_3]$, we would have a $\text{PA}=\frac{1}{4}$ since the second parsed log record is the only one correctly labelled.

When evaluating parsed log formats in terms of a sequence mapping (as shown in Figure~\ref{fig:apache-example}), rather than simply assigning each log record a format based on the order in which fields appear, Levenshtein edit distance can be used to quantify string (dis)similarity \cite{Rand_2021}. An absolute edit distance denoted $D_A$, measures the total number of single-character edits (insertion, substitution, or deletion) required to transform one string into another. The relative edit distance, denoted $D_R$, normalizes the $D_A$ by the expected string length.  When comparing models, it is essential to consider the $D_R$, as differences in sequence length across validation datasets may introduce scaling bias in absolute performance. Note that for all results, we truncate the predicted character-based target fields to match the input logs before calculating $D_A$ and $D_R$ between the prediction and ground truth. Truncation was performed equivalently for all experiments and is a common log parsing heuristic in production systems.

We will report Levenshtein edit distance as our primary metric, in accordance with the authors of the HTTPd-parse~\cite{Rand_2021}, because it offers an interpretable evaluation of partially correct predictions, distinguishing near misses from severe errors by penalizing the number of edits required.

\subsection{Solution Space}
Automated log parsing has been extensively studied, resulting in a range of approaches. For completeness, the following items will explore the space of log parsing solutions by briefly describing the common techniques already in the literature:

\begin{enumerate}[label=\textit{\arabic*)}, leftmargin=0pt, itemindent=3em]
    \item \textit{Frequent Pattern Mining}: 
    Frequent pattern mining-based log parsing methods leverage the idea that groups of similar log records share repeated (frequent) tokens. These approaches often rely on strategies for identifying frequently co-occurring token sets and treating the remaining substrings as variable parameters. For example, MoLFI~\cite{messaoudi2018search} employs a two-phase clustering approach to discover frequent token patterns and subsequently refines these patterns to produce event templates. In general, frequent pattern mining approaches can handle large log datasets but may require careful parameter tuning (e.g., support thresholds) to avoid over-segmentation or under-segmentation of log statements.
    
    \item \textit{Clustering}: 
    Clustering-based approaches assume that log records belonging to the same event template form natural clusters in a chosen feature space (e.g., token embeddings or string similarity measures). Techniques such as LogSig~\cite{10.1145/2063576.2063690} and Logcluster~\cite{DBLP:conf/cnsm/VaarandiP15} apply token-level similarity or n-gram-based measures to group logs into distinct clusters, and each cluster then corresponds to a log template. 
    Clustering techniques can be effective in capturing subtle differences between similar logs, although they often require determining the optimal number of clusters or similarity thresholds.
        
    \item \textit{Longest Common Subsequence}: 
    Approaches based on the longest common subsequence view two log records as sequences of tokens and identify their shared (static) components by computing the LCS. Any tokens not captured within the LCS are assumed to be variable fields. An example is the LCS-based log parsing framework proposed in~\cite{8489912}, which uses a multi-step process to identify the static subsequence, refine it to remove noise, and then construct log templates. LCS-based methods typically perform well in scenarios where logs exhibit relatively small modifications (e.g., minor differences in parameter values) around a common template. However, performance may degrade in highly diverse or noisy logs due to an explosion in pairwise comparisons.

    \item \textit{Heuristics}: 
    Heuristic-based approaches rely on systematic rules to identify variable tokens by traversing and comparing log strings. For instance, the tree-based parsing models proposed in \cite{8029742, 4601543} first hierarchically split log records by delimiters (e.g., whitespace or punctuation) and then employ domain-specific heuristics (e.g., numeric detection, IP-address pattern recognition) to recognize and replace variable fields. Another example is the Drain parser \cite{8029742}, which incrementally partitions logs based on prefix and length heuristics to group messages into templates. Heuristic-based approaches often achieve fast parsing speeds, but their success relies heavily on well-chosen rules or domain knowledge to handle edge cases.

\end{enumerate}

\subsection{Deep Sequence Models}

A deep sequence model is defined as a parameterized function $y=f_\theta(x)$ that transforms an input sequence $x\in \mathbb{R}^{L \times D_{\text{in}}}$ into an output sequence $y \in \mathbb{R}^{L' \times D_{\text{out}}}$. Here \(L\) and \(L'\) denote the input and output lengths, while \(D_{\text{in}}\) and \(D_{\text{out}}\) are the sequence element feature dimensions.
The learned parameters $\theta$ of the function $f$ are computed by a stochastic gradient descent approach via back-propagation to minimize a task-specific loss.

Deep sequence models often follow an \textit{encoder--decoder} structure in order to map sequences of varying length and dimensionality \cite{DBLP:conf/emnlp/ChoMGBBSB14, sutskever2014sequencesequencelearningneural}. The following items review the major classes of deep sequence models relevant to our study of automatic log parsing, highlighting their computational trade-offs and motivating the exploration of state space models.

\subsubsection{Recurrent Neural Network} \label{sec:rnn}

Recurrent neural networks (RNNs) process sequences by maintaining a hidden state that is updated sequentially over time. Gated variants, most notably the LSTM \cite{DBLP:journals/neco/HochreiterS97} and gated recurrent unit (GRU) \cite{DBLP:conf/emnlp/ChoMGBBSB14}, mitigate vanishing and exploding gradients through explicit gating mechanisms and have demonstrated strong performance in sequence-to-sequence log parsing formulations \cite{Rand_2021}.

Additionally, RNN-based models suffer from two fundamental limitations: (i) training requires backpropagation-through-time (BPTT), which is memory-intensive for long sequences, and (ii) computation is inherently sequential, limiting parallelism.

\subsubsection{Transformer} \label{sec:transformer}

Many sequence modelling tasks require maintaining long-term dependencies, an operation that has been difficult for RNNs \cite{gradient-flow}. The recurrent state(s) of a RNN is required to maintain relevant information across the sequence in order to realize long-term dependencies upon prediction.

Attention mechanisms allow for the effective modelling of long-term dependencies \cite{bahdanau2016neuralmachinetranslationjointly, DBLP:conf/iclr/KimDHR17}. While attention mechanisms were successfully introduced in conjunction with RNNs, the Transformer architecture \cite{NIPS2017_3f5ee243} subsequently found that removing recurrence and structuring its computation wholly around a highly optimizable self-attention mechanism both improved sequence modelling performance and parallelism. 

While this approach forgoes issues related to BPTT and non-linear recurrence, requiring only a constant number of sequential operations compared to \(O(L)\) operations in an RNN, its computational complexity grows quadratically with sequence length. The trade-off between parallelization and unfavourable computational complexity growth with sequence length has prompted innovation in self-attention, such as sparse attention, low-rank methods, and down-sampling \cite{DBLP:journals/csur/TayDBM23}.

\subsubsection{State Space Models} \label{sec:lin-seq-models}

Structured state space models (SSMs) \cite{DBLP:journals/corr/abs-2111-00396, DBLP:journals/corr/abs-2110-13985} have recently emerged as a promising class of deep sequence models aiming to build on RNN principles with analytic linear operators to reconcile efficiency and expressivity. SSMs enable highly parallel training via global convolution, while retaining recurrent inference with modest memory requirements.

\section{Methodology}
\label{sec:methodology}

This chapter outlines the proposed methodology and implementation parameters which are used to explore our research questions. Five main experimental branches were constructed to evaluate the performance of deep sequence modelling in automatic log parsing. These branches focus on training dataset composition, tokenization methods, sequence lengths, training dataset sizes, and model architectures.

\subsection{Model Architectures}
The model architectures, LSTM and Transformer, are two prominent neural network architectures used extensively in sequence modelling. LSTMs are specifically designed for sequential data, featuring recurrent computational cells and gating mechanisms that handle long-term dependencies (as described in Section~\ref{sec:rnn}). In contrast, Transformers rely on a self-attention mechanism to capture long-range dependencies (as described in Section~\ref{sec:transformer}). Recent advancements in the area of deep sequence modelling have shown promise for a new class of recurrent models that are parallelizable and efficiently scalable (as described in Section~\ref{sec:lin-seq-models}). We study one such RNN architecture: Mamba \cite{DBLP:journals/corr/abs-2312-00752} (an implementation of SSMs). We will now denote and detail specific experimental design and implementation choices for each architecture.

\paragraph{LSTM} We use a mono-directional LSTM, denoted as $M_L$, reproduced to reflect prior work \cite{Rand_2021}. Additionally, we examine a bi-directional LSTM model, $M_B$, to comprehensively evaluate and compare performance across different sequence modelling modalities for log parsing. Bi-directional LSTM considers the input sequence in both the forward and backward directions, aiming to utilize sequential dependencies between input tokens to better predict the current output (as described in Section~\ref{sec:rnn}). $M_L$ and $M_B$ utilize identical underlying recurrent units with 512 cells, featuring a dropout rate of 0.2 in the encoder and decoder LSTM layers. These hyperparameters were chosen based on the best-performing models in prior work \cite{Rand_2021} and serve as a baseline against which to compare other model architectures.

\paragraph{Transformer} Our Transformer model implementation, denoted $M_T$, aligns with the original implementation using $256$ embedding units in both the encoder and decoder, a hidden state size of $2048$ in each feed-forward block, and eight attention heads \cite{NIPS2017_3f5ee243}. Additionally, we include a dropout rate of 0.2 in the decoder layer.

\paragraph{Mamba} We explore a Mamba model, denoted $M_M$, using an embedding dimension of $128$. Additionally, we use the default parameters provided by the reference PyTorch implementation \cite{mamba-ssm-1-2-0-post1}, notably a state expansion factor of $16$, a local convolution width of $4$, and a block expansion factor of~$2$.

\subsection{Data}
\label{sec:dataset}
Training data is separated into four distinct datasets of varying composition (as introduced in Section~\ref{sec:benchmarks}). These training datasets, denoted $T_T$, $T_E$, $T_M$, and $T_H$, represent their presumed modelling difficulty: trivial ($T_T$), easy ($T_E$), medium ($T_M$), and hard ($T_H$). Table~\ref{tab:datasets} details the specific log format description for each training and validation dataset in terms of the percentage of standardized Apache log formats (ELF and CLF) and random log formats. Random log formats (which are present in $T_E$, $T_M$, and $T_H$ at varying degrees), contain logs with 2 to 15 unique Apache log fields (as described in Table~\ref{tab:fields}) arranged in a random order. To reiterate for clarity, it should be noted that validation datasets $V_A$, $V_B$, and $V_C$ contain log formats entirely constructed using the ELF. Thus, training datasets that contain more ELF examples best reflect the validation datasets.

\subsubsection{Data Preprocessing}
Architectural limitations of the Transformer model, $M_T$, with respect to sequence length (as described in Section~\ref{sec:transformer}) led to the exploration of different tokenization methods to evaluate the architecture's log parsing ability. For $M_L$, $M_B$, and $M_M$, character-based and word-based tokenization methods are performed, followed by one-hot encoding with a maximum frequency-based vocabulary size of $15,000$. Word-tokenization is performed by delimiting log strings by whitespace and punctuation (as defined by the C locale). 

$M_T$ is trained with word-based tokenization only. Additionally, $M_T$ training sequences are truncated at varying lengths, low, medium, and high, which represent sequence lengths of 256, 765, and 4188, respectively. These sequence lengths map to approximately the 70th, 99th, and 100th percentiles of log sequence lengths in the validation datasets.

All training datasets are divided into three bins denoting the percentage of observations used: 10\%, 50\%, and 100\%. Partial training datasets (i.e., 10\% and 50\%) are randomly sampled five times with different seeds to minimize sampling bias and enhance reproducibility. 

\begin{table*}[htbp]
\caption{ Datasets' descriptions adapted from \cite{Rand_2021}. $T_T$, $T_E$, $T_M$, and $T_H$, represent the presumed modelling difficulty of trivial, easy, medium, and hard, respectively. Log record length is denominated in characters of the respective (min, median, or max) log strings. The Common Log Format (CLF) and the Extended Log Format (ELF) are standardized Apache log formats.}
\centering
\resizebox{\textwidth}{!}{
\begin{tabular}{lrrrrp{0.45\textwidth}}
\toprule
Dataset & Log records&  \multicolumn{3}{c}{Log records length}   & Log records' format description \\
  \cmidrule{3-5}
 & count & min & median & max  &   \\
\midrule
$T_T$  & 100,000 & 136 & 272 & 1173              & $100\%$ ELF  \\
$T_E$  & 20,000 & 4 & 250 & 1173              & $\approx 40\%$ of the ELF format, $\approx 24\%$ of the CLF format, and $\approx 36\%$ of randomly drawn and reshuffled fields shown in Table~\ref{tab:fields}.  The random strings have $2$ to $14$ records in them.  \\
$T_M$  & 100,000 & 4 & 161 & 1294              & $\approx 50\%$ of the CLF format and $\approx 50\%$  of the randomly generated records using the same approach as in the $T_E$ case. The random strings have $2$ to $14$ fields in them.    \\
$T_H$  & 100,000 & 4 & 291 & 1528              & $100\%$ of the randomly generated records using the same approach as in the $T_E$ case. The random strings have $2$ to $15$ fields in them.    \\
\midrule
$V_A$  & 7,314 & 194 & 238 & 602 & 100\% ELF                                \\
$V_B$  & 6,539 & 79 & 238 & 4398 & 100\% ELF                                \\
$V_C$  & 10,000 & 81 & 231 & 1363  & 100\% ELF                                \\ 
\bottomrule
\end{tabular}
}
\label{tab:datasets}
\end{table*}

\subsection{Experiment Design}

Accounting for the specific permutations of training datasets ($T_T$, $T_E$, $T_M$, and $T_H$), tokenization methods (character- and word-based), sequence lengths (256, 756, and 4188, for $M_T$ only), training dataset sizes (10\%, 50\%, and 100\%, where partial datasets are resampled five times), and model architectures ($M_L$, $M_B$, $M_T$, and $M_M$) -- a total of 396 models are trained.

Training is performed using 300 epochs and a mini-batch size of 64 log records. The Adam optimizer \cite{DBLP:journals/corr/KingmaB14} is used with an initial learning rate of $10^{-3}$, $\beta_1 = 0.9$, $\beta_2 = 0.999$, and $\varepsilon = 10^{-7}$. 

\subsection{Specifications}
\label{sec:specification}

This section provides a detailed description of the software environment, computational resources, and model complexity used in the experiments, supporting reproducibility and transparency.

\paragraph{Software Environment}
All models were trained on the Digital Research Alliance of Canada's (the Alliance) compute clusters using Python~3.10 as the base runtime environment. 
The primary deep learning framework consisted of TensorFlow~2.15.1~\cite{abadi2016tensorflow} and Keras~2.15.0~\cite{chollet2015keras}. 
In addition, experiments that explored state space models utilized the \texttt{mamba\_ssm} package (version~1.2.0.post1)~\cite{DBLP:journals/corr/abs-2312-00752}, which relies on PyTorch~2.2~\cite{paszke2019pytorch} as its backend. 
Experiment orchestration and job management were handled using Hydra~1.1.1~\cite{Yadan2019Hydra} and SLURM~\cite{yoo2003slurm} scheduling on the Alliance's high-performance computing infrastructure.

\paragraph{Compute Resources}
Training runs all utilize NVIDIA A100-40GB GPUs. 
In total, the experiments consumed approximately 2.92~GPU-years.

\paragraph{Model Costs}
A range of model architectures and hyperparameter configurations were explored, including LSTM, Transformer, and Mamba models with varying tokenization strategies and sequence lengths. Table~\ref{tab:model_params} details the number of trainable parameters for each relevant model configuration used. 
As expected, increasing the input sequence length results in an increase in parameter count for Transformers, whereas the parameter count for recurrent models remains constant. 

Word-level tokenization generally resulted in higher parameter counts than character-level tokenization, primarily due to the larger vocabulary embedding matrices. 
Similarly, bi-directional LSTM architectures consistently exhibited higher parameter counts than their mono-directional counterparts, reflecting the doubled recurrence. 
Notably, the Mamba models utilized significantly fewer parameters compared to LSTM and Transformer architectures, indicating their potential for lower inference cost. 

We also report the wall-clock time for training and inference on an NVIDIA A100-40GB GPU, measured using single-sample batches after a warmup period and averaged over multiple runs to ensure stable estimates. Transformer models exhibit rapidly increasing training and inference time with increasing sequence length, incurring costs orders of magnitude higher. Recurrent models ($M_L$, $M_B$) exhibit moderate, relatively stable computational costs, consistent with their linear-time sequential processing. In contrast, Mamba ($M_M$) achieves the lowest training and inference latency across all configurations, with near-constant cost across tokenization schemes, reflecting the architectural differences of state-space sequence modelling.

\begin{table}[htbp]
\centering
\caption{Number of trainable parameters and profiled per-batch cost for different model configurations.}
\label{tab:model_params}
\begin{tabular}{lrrr}
\toprule
\textbf{Model configuration} & \textbf{\# of Params.} & \textbf{Train (ms)} & \textbf{Infer (ms)} \\
\midrule
$M_T$ ($L{=}4188$, Word) & 56,752,280 & 630.91 & 158.58 \\
$M_T$ ($L{=}765$, Word)  & 53,247,128 & 72.59  & 29.06 \\
$M_T$ ($L{=}256$, Word)  & 52,725,912 & 25.29  & 26.59 \\
$M_B$ (Word)             & 18,207,264 & 52.58  & 48.27 \\
$M_L$ (Word)             & 11,895,328 & 28.56  & 22.18 \\
$M_B$ (Char)             & 10,556,436 & 51.96  & 41.87 \\
$M_L$ (Char)             & 4,250,644  & 27.87  & 23.13 \\
$M_M$ (Word)             & 2,040,736  & 1.82   & 0.46 \\
$M_M$ (Char)             & 129,556    & 1.80   & 0.46 \\
\bottomrule
\end{tabular}
\end{table}

\section{Evaluation}
\label{sec:evaluate}

To assess the performance of our models, we use absolute and relative Levenshtein distance (also known as edit distance). A lower Levenshtein distance indicates greater similarity, as fewer edits are needed to make the two strings identical. Refer to Section~\ref{sec:eval-metrics} for further details on log parsing evaluation metrics, and \ref{appendix:full-results} for a verbose record of all validation results.

In this chapter, we systematically address each research question using experimental evidence and statistical analyses.

\subsection{RQ1: What is the impact of sequence length on model performance?}

To evaluate the impact of sequence length on the performance of $M_T$, we will analyze $D_R$ across three sequence lengths: 256, 756, and 4188. 
Statistical analyses and descriptive summaries are computed across training datasets $T_{(\cdot)}$, validation datasets $V_{(\cdot)}$, and different training data sampling levels (10\%, 50\%, and 100\%) to assess both the magnitude and significance of performance differences.

\subsubsection{Descriptive Statistics}

Across all experimental conditions, the mean relative edit distance increased with sequence length. However, Table~\ref{tab:seq_length_descriptive} shows the median relative edit distance remained stable around 0.10, suggesting that improvements primarily affect the upper distribution tail rather than the central tendency. Notably, very long sequences constitute a small fraction of the data (upper percentiles only), which may limit sensitivity to the effect of higher sequence lengths in training.

The standard deviation and the number of outliers exhibit a similar increase with sequence length, suggesting that longer sequences are associated with greater variability in performance.

These descriptive trends were consistent across most training and validation sets as well as across different training data percentages, confirming that the effect is robust across data regimes (see \ref{appendix:rq1-extended-comparison}). An exception exists for higher sequence lengths with $T_T$ training data or 100\% of training data, indicating sensitivity to outliers (as discussed in Section~\ref{sec:rq2-trivial}).

\begin{table*}[htbp]
\centering
\caption{Descriptive statistics measured in relative edit distance $D_R$ and the number of outliers across sequence lengths.} 
\label{tab:seq_length_descriptive}
\begin{tabular}{rrrrrr|rr}
  \toprule
Sequence Length & Mean & Median & Q1 & Q3 & Std. Dev. & \multicolumn{1}{r}{\# Outliers} & \% \\ 
  \midrule
256 & 0.15 & 0.10 & 0.06 & 0.18 & 0.17 & 87,432 & 9.35 \\ 
  756 & 0.18 & 0.10 & 0.04 & 0.22 & 0.22 & 106,507 & 11.39 \\ 
  4188 & 0.19 & 0.10 & 0.04 & 0.22 & 0.23 & 112,199 & 12.00 \\ 
   \bottomrule
\end{tabular}
\end{table*}

\subsubsection{Statistical Significance Testing}
\label{sec:wilcoxon-apporach}

\paragraph{Approach} To assess the statistical significance of these observed differences in relative edit distance with sequence length, pairwise comparisons were conducted using the Wilcoxon signed-rank test at a significance level of $\alpha = 0.05$. Multiple test correction was also performed using the Benjamini–Hochberg method \cite{benjamini1995controlling} to control for the false discovery rate. The effect size ($r$) was also calculated to quantify the practical magnitude of the observed differences, independent of sample size. Values of $|r| < 0.1$ indicate a \textit{negligible} effect, $0.1 \leq |r| < 0.3$ a \textit{small} effect, $0.3 \leq |r| < 0.5$ a \textit{medium} effect, and $|r| \geq 0.5$ a \textit{large} effect.

All pairwise comparisons of sequence length on relative edit distance (i.e., 256 vs 756, 256 vs 4188, and 756 vs 4188) were statistically significant at $\alpha = 0.05$ after applying the Benjamini--Hochberg correction. Despite this, the corresponding effect sizes were negligible in magnitude (i.e., $|r| < 0.1$), indicating that the observed differences, while statistically detectable, are of limited practical significance. These results suggest that the practical impact of sequence length on performance may be negligible for $M_T$ in our setting.

\subsubsection{Validation Dataset Trends}
\label{sec:rq1-val-set}

\begin{figure}[htbp]
    \centering
    \includegraphics[width=\linewidth]{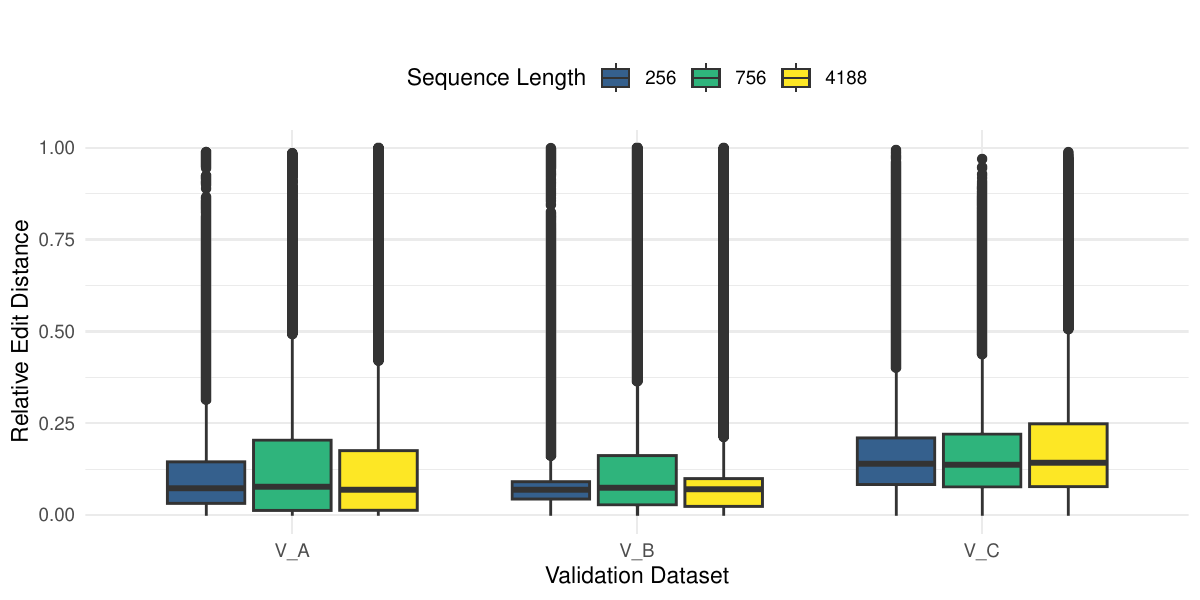}
    \caption{Box-plots of the relative edit distance of $M_T$ by sequence length and validation dataset.}
    \label{fig:rq1-boxplot-val-sets}
\end{figure}

The analysis was extended to each validation dataset ($V_A$, $V_B$, and $V_C$), revealing a largely consistent trend. Specifically, the median $D_R$ tends to increase slightly with longer sequence lengths; however, this is explained not by systematic performance changes, but rather by increased variance. Figure~\ref{fig:rq1-boxplot-val-sets} shows these differences in $D_R$ variance, and \ref{appendix:rq1-val-datasets} confirms that all statistically significant differences exhibit a negligible effect size.

\subsubsection{Discussion}

While our findings are consistent across splits, they may not generalize to settings with a different distribution of sequence length. In our setting, we observe that $M_T$ struggles to effectively utilize the additional information longer sequence lengths provide for the task of automated log parsing with the given data. The negligible effect sizes imply that longer sequences do not meaningfully improve performance on this task. Given that self-attention scales quadratically in memory and time with sequence length, the computational cost of longer sequences may outweigh any benefits. That said, environments with a higher proportion of long sequences may yield different results.

\subsection{RQ2: How do different tokenization methods affect performance?}
\label{sec:rq2}

Sequence representation is a key consideration in all sequence modelling tasks. To evaluate the impact of log string tokenization on $D_R$ for log parsing, we will analyze both character- and word-based tokenization. 
Statistical analyses and descriptive summaries are computed across training datasets $T_{(\cdot)}$, validation datasets $V_{(\cdot)}$, and different training data sampling levels (10\%, 50\%, and 100\%) to assess both the magnitude and significance of performance differences. Since $M_T$ is only trained with word tokenization, $M_T$ results will be omitted from this analysis for the sake of more direct comparisons with $M_B$, $M_L$, and $M_M$.

\paragraph{Approach} Since character- and word-based tokenization produce outputs at different levels of granularity, direct comparison would not be appropriate. To address this, predicted word sequences are post-processed through an \emph{inverse tokenization} procedure that removes artificial separators, normalizes tokens, and reconstructs a continuous character string structurally aligned with the original log format. This ensures that the edit distance reflects true parsing performance rather than tokenization effects.

\subsubsection{Descriptive Statistics}

For all model architectures considered, across all experimental configurations, character-level tokenization achieves superior performance. Table~\ref{tab:rq2_tokenization_descriptive} shows a consistent increase in mean and median edit distance when using word-level tokenization, while maintaining near-identical standard deviations. These descriptive trends remain consistent when stratifying model configurations according to training datasets $T_{(\cdot)}$, validation datasets $V_{(\cdot)}$, and varying training data sampling levels (see \ref{appendix:rq2-extended-comparison}).    

\begin{table*}[htbp]
\centering
\caption{Descriptive statistics measured in relative edit distance $D_R$ and the number of outliers across tokenization methods and model architectures.} 
\label{tab:rq2_tokenization_descriptive}
\begin{tabular}{llrrrrr|rr}
  \toprule
Model & Tokenization & Mean & Median & Q1 & Q3 & Std. Dev. & \# Outliers & \% \\ 
  \midrule
$M_B$ & char & 0.21 & 0.17 & 0.09 & 0.28 & 0.16 & 31,769 & 3.40 \\ 
  $M_B$ & word & 0.37 & 0.37 & 0.26 & 0.46 & 0.15 & 10,262 & 1.10 \\ 
  $M_L$ & char & 0.23 & 0.21 & 0.11 & 0.33 & 0.15 & 9,205 & 0.98 \\ 
  $M_L$ & word & 0.37 & 0.37 & 0.27 & 0.48 & 0.16 & 6,918 & 0.74 \\ 
  $M_M$ & char & 0.08 & 0.04 & 0.01 & 0.13 & 0.09 & 26,173 & 2.80 \\ 
  $M_M$ & word & 0.15 & 0.14 & 0.06 & 0.22 & 0.11 & 12,427 & 1.33 \\ 
   \bottomrule
\end{tabular}
\end{table*}

\subsubsection{Statistical Significance Testing}

To validate the statistical significance and magnitude of median performance difference between tokenization methods across all model configurations, we again employ a Wilcoxon signed-rank test protocol (as described in Section~\ref{sec:wilcoxon-apporach}). Averaged across all experimental configurations, statistical testing reaffirms both the significance ($\alpha = 0.05$) and ``large'' effect ($|r| \geq 0.5$) of the tokenization method on median relative edit distance.

With few exceptions, stratified model configurations also show consistent statistical significance and large effect (see \ref{appendix:rq2-extended-comparison}). The exceptions include the $M_L$ models with only 10\% of training data used, which exhibit a ``medium'' effect size ($0.3 \leq |r| < 0.5$), and certain models trained with $T_M$ and $T_H$. Next, we will further explore the exceptions in the training dataset configurations.

\subsubsection{Training Dataset Trends}

Figure~\ref{fig:rq1-boxplot-train-sets} depicts the relative edit distance by tokenization method and training dataset for each model architecture. The outcome of our Wilcoxon test suggests a qualitative difference in the effect size (ranging from ``small'' to ``medium'') for $M_L$ and $M_B$ models trained using $T_M$ and $T_H$ datasets (see \ref{appendix:rq2-extended-comparison}). 

The smaller effect size observed for $T_M$ and $T_H$ suggests that the advantage of character-level tokenization diminishes as format diversity and randomness increase. To further explore tokenization method effects, Section~\ref{sec:rq2-oov} examines out-of-vocabulary (OOV) rates and vocabulary coverage across datasets and tokenization schemes. Next, we will examine the outsized impact that word tokenization has on $T_T$ training dataset results.

\begin{figure}[htbp]
    \centering
    \includegraphics[width=\linewidth]{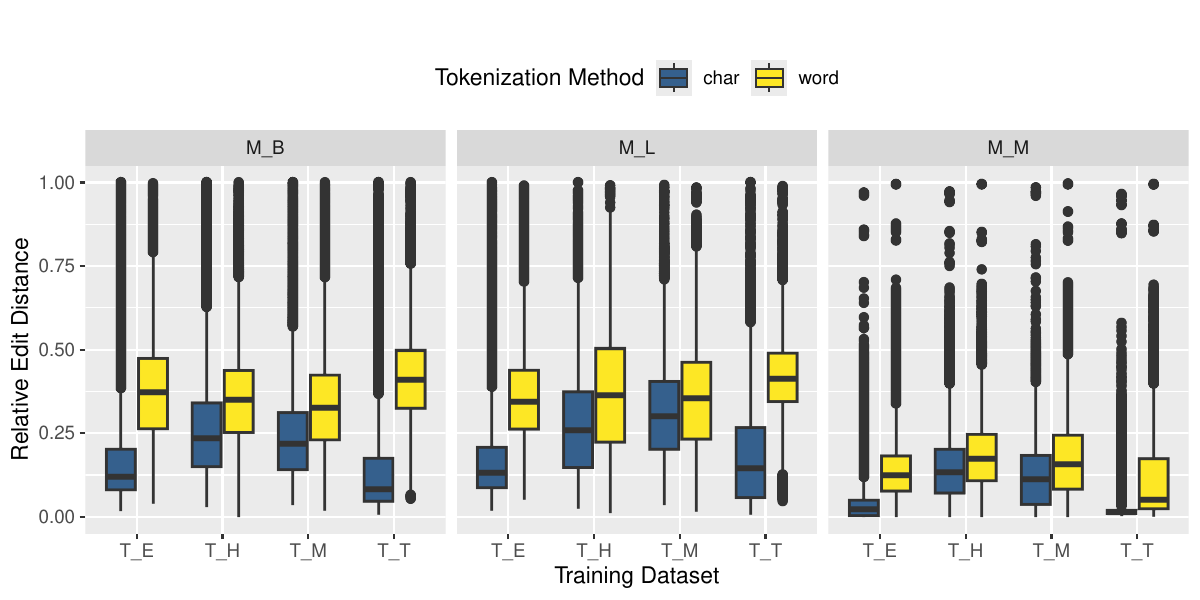}
    \caption{Box-plots of the relative edit distance by tokenization method and training dataset for each model architecture.}
    \label{fig:rq2-boxplot-train-sets}
\end{figure}

\paragraph{Trivial Dataset Analysis}  \label{sec:rq2-trivial}

In Figure~\ref{fig:rq2-boxplot-train-sets}, we observe that differences in median $D_R$ between word and character tokenization are $2\text{-}5\times$ larger for $T_T$ and $T_E$ than $T_M$ and $T_H$ (see \ref{appendix:rq2-trivial} for more details). One exception to this is $M_M$ with $T_T$, whose median performance degrades minimally between tokenization methods.

Upon further inspection of training and test examples in $T_T$ and $V_{(\cdot)}$, various syntactical differences were found. The following differences are noted: missing square brackets around timestamp fields and missing double quotes wrapping the log strings (see \ref{appendix:rq2-trivial} for a log string example). While syntactical differences were found in $T_T$, the log record format still followed the Extended Log Format (ELF) as described in Table~\ref{tab:datasets}.

To test the impact of these syntactical differences on model performance, we reran model inference on all validation datasets after standardizing the timestamp and log string syntax (denoted as $V*$). This does not require retraining the underlying model, allowing us to compare the impact of word tokenization with and without syntactical differences between the training and test sets.

Table~\ref{tab:rq2_T_T_descriptive} shows that $D_R$ decreases consistently across all models when evaluating on $V*$. Additionally, we can see that $M_T$ performed best even with non-representative training examples when compared to the recurrent models. 

\begin{table*}[htbp]
\centering
\caption{Descriptive statistics measured in relative edit distance $D_R$ for models evaluated on the V and V* datasets using word-level tokenization and 100\% of $T_T$. $M_T$ uses a sequence length of 256.} 
\label{tab:rq2_T_T_descriptive}
\begin{tabular}{llrrrrr|rr}
  \toprule
Model & Dataset & Mean & Median & Q1 & Q3 & Std. Dev. & \# Outliers & \% \\ 
  \midrule
$M_B$ & V & 0.40 & 0.41 & 0.31 & 0.49 & 0.14 & 175 & 0.82 \\ 
  $M_B$ & V* & 0.15 & 0.14 & 0.08 & 0.21 & 0.10 & 220 & 1.04 \\ 
  $M_L$ & V & 0.43 & 0.43 & 0.39 & 0.51 & 0.12 & 1429 & 6.72 \\ 
  $M_L$ & V* & 0.31 & 0.38 & 0.12 & 0.43 & 0.18 &  10 & 0.05 \\ 
  $M_T$ & V & 0.11 & 0.08 & 0.06 & 0.11 & 0.08 & 1898 & 8.93 \\ 
  $M_T$ & V* & 0.03 & 0.00 & 0.00 & 0.01 & 0.07 & 3233 & 15.21 \\ 
   \bottomrule
\end{tabular}
\end{table*}

\subsubsection{Out of Vocabulary} \label{sec:rq2-oov}

An analysis of the OOV token rate can help further contextualize the differences in tokenization methods across datasets. Table~\ref{tab:oov-summary} shows the number of OOV tokens in relation to the total number of tokens, as well as the number of unique tokens, for each dataset and tokenization method. Notice that character-based tokenization does not result in any OOV tokens since the number of unique tokens is less than our maximum vocabulary size of 15,000. While our word tokenization method makes a trade-off between the length of input logs and the size of the vocabulary, it also introduces OOV tokens.

The difficulty of each training set (as detailed in Table~\ref{tab:datasets}) does not necessarily take into account the differences in OOV token occurrences with word tokenization. A large number of OOV tokens may lead to degraded performance for certain datasets.

\begin{table*}[htbp]
\caption{OOV Token Summary by Dataset and Tokenization Method}
\label{tab:oov-summary}
\centering
\begin{tabular}{lrrrrrr}
\toprule
Dataset & OOV Count & Token Count & OOV \% & Unique OOV & Unique Tokens\\
\midrule
$T_T$-char & 0 & 117,300,000 & 0.0 & 0 & 79\\
$T_T$-word & 1,444,752 & 34,000,000 & 4.2 & 1,302,427 & 1,317,427\\
$T_E$-char & 0 & 29,260,000 & 0.0 & 0 & 81\\
$T_E$-word & 265,589 & 7,980,000 & 3.3 & 263,880 & 278,880\\
$T_M$-char & 0 & 152,700,000 & 0.0 & 0 & 81\\
$T_M$-word & 1,249,397 & 40,000,000 & 3.1 & 1147343 & 1,162,343\\
$T_H$-char & 0 & 152,800,000 & 0.0 & 0 & 81\\
$T_H$-word & 1,537,716 & 40,700,000 & 3.9 & 1,427,386 & 1,442,386\\
\bottomrule
\end{tabular}
\end{table*}

\subsubsection{Discussion}

Overall, character-level tokenization performs better than word-level tokenization on $D_R$ across most settings, and the aggregate tests indicate that this difference is statistically reliable. The advantage appears smaller for $T_M$ and $T_H$, where format diversity increases. We also observe non-zero OOV rates (about 3-4\%) for word tokenization in these datasets (as shown in Table~\ref{tab:oov-summary}). 

It is important to note that edit distance operates at the character level, which introduces an inherent interaction between the evaluation metric and the representation choice. As a result, character-level tokenization is structurally advantaged under this metric, since small boundary or delimiter mismatches at the word level may incur multiple character edits even when the semantic field assignment is largely correct. Consequently, the observed superiority of character-based tokenization should be interpreted in the context of this metric choice, rather than as a universal property of tokenization strategies for log parsing. Complementary evaluation metrics that operate at the field or span level (e.g., field-level accuracy or segment overlap) could further illuminate cases in which word-level tokenization preserves semantic correctness despite a higher character-level edit distance.

Taken together, these results suggest that character tokenization is a reasonable choice when architectural constraints permit its use, while word tokenization remains viable for diverse datasets. However, the observed performance differences may not generalize to other domains, datasets, or evaluation protocols. Furthermore, subword tokenization could offer a balanced compromise between the granularity of character-level representations and the efficiency of word-level approaches. Exploring subword-based methods, therefore, represents a sensible direction for future work.

\subsection{RQ3: How does sample efficiency vary across model architectures?}
We will now examine how performance changes as the available training data increases (10\%, 50\%, 100\%), aggregating over seeds for partial datasets and evaluating $D_R$ across all validation sets. Guided by RQ1 and RQ2, we restrict the following analysis to sequence length 256 for $M_T$ and adopt character tokenization for $M_B$, $M_L$, and $M_M$.

\subsubsection{Descriptive Statistics}

Table~\ref{tab:rq3_max_obs_pct_descriptive} shows that for $M_M$ and $M_T$, the mean $D_R$ remain stable with deviations of only $\pm0.02$. Conversely, $M_B$ and $M_L$ exhibit a monotonic decrease in median $D_R$ as the percentage of training data used increases. To validate that the differences in median $D_R$ are not attributable to distributional variance, we will again employ statistical testing.

\begin{table*}[htbp]
\centering
\caption{Descriptive statistics measured in relative edit distance $D_R$ and the number of outliers across maximum observation percentages and model architectures.}
\label{tab:rq3_max_obs_pct_descriptive}
\begin{tabular}{lrrrrrr|rr}
  \toprule
Model & Training \% & Mean & Median & Q1 & Q3 & Std. Dev. & \# Outliers & \% $^{\ddagger}$ \\ 
  \midrule
$M_B$ & 10 & 0.22 & 0.18 & 0.11 & 0.29 & 0.15 & 14,101 & 3.32 \\ 
  $M_B$ & 50 & 0.20 & 0.16 & 0.09 & 0.27 & 0.16 & 15,295 & 3.60 \\ 
  $M_B$ & 100 & 0.22 & 0.10 & 0.07 & 0.33 & 0.21 & 2,809 & 3.30 \\ 
  $M_L$ & 10 & 0.26 & 0.24 & 0.14 & 0.36 & 0.15 & 5,183 & 1.22 \\ 
  $M_L$ & 50 & 0.22 & 0.18 & 0.09 & 0.32 & 0.15 & 4,571 & 1.08 \\ 
  $M_L$ & 100 & 0.19 & 0.16 & 0.09 & 0.23 & 0.14 & 4,948 & 5.82 \\ 
  $M_M$ & 10 & 0.08 & 0.04 & 0.01 & 0.13 & 0.09 & 11,603 & 2.73 \\ 
  $M_M$ & 50 & 0.09 & 0.04 & 0.01 & 0.13 & 0.09 & 13,059 & 3.07 \\ 
  $M_M$ & 100 & 0.09 & 0.04 & 0.01 & 0.15 & 0.09 & 1,205 & 1.42 \\ 
  $M_T$ & 10 & 0.15 & 0.09 & 0.05 & 0.18 & 0.18 & 42,732 & 10.05 \\ 
  $M_T$ & 50 & 0.14 & 0.09 & 0.05 & 0.18 & 0.16 & 37,996 & 8.94 \\ 
  $M_T$ & 100 & 0.16 & 0.13 & 0.07 & 0.20 & 0.15 & 6,491 & 7.63 \\ 
   \bottomrule
\end{tabular}
\vspace{2pt}
\parbox{\textwidth}{\footnotesize
$^{\ddagger}$ Outlier percentages for 10\% and 50\% configurations are based on 5 resampled runs each, while 100\% configurations are based on a single run. Percentages in Table~\ref{tab:rq3_max_obs_pct_descriptive} are therefore not all directly comparable across training levels.
}
\end{table*}

\subsubsection{Statistical Significance Testing}
\paragraph{Approach} In addition to following the testing procedure outlined in Section~\ref{sec:wilcoxon-apporach}, we will apply additional preprocessing to enable statistical testing between models trained with 100\% of $T_{(\cdot)}$ and those with only 10\% or 50\%. Since models trained with partial training datasets are sampled five times (as described in Section~\ref{sec:dataset}), the number of test observations is not constant between 10\%, 50\%, and 100\% variants. To reconcile the number of observations before statistical testing, we calculate the median edit distance for each evaluation log parsing example in the 10\% and 50\% model variants.

All tests, with the exception of $M_B$, 50\% versus 100\%, show statistically significant results with varying effect sizes (see \ref{appendix:rq3-extended-comparison} for more details).
Both $M_M$ and $M_T$ show a negligible effect size when going from 10\% to 50\%, with only a small effect for $M_M$ and a medium effect for $M_T$ in the remaining tests (i.e., 10\% versus 100\% and 50\% versus 100\%).
$M_B$ exhibits a small effect size for both 10\% versus 50\% and 10\% versus 100\% tests. Finally, $M_L$ show a medium effect for 10\% versus 50\%, a large effect for 10\% versus 100\%, and a small effect for 50\% versus 100\%. 
Overall, these results suggest that recurrent models benefit most from additional training data, while state-space and transformer architectures achieve stable performance even under limited data conditions.

\subsubsection{Discussion}

The statistical and descriptive analyses together indicate that model architectures differ substantially in their data efficiency. The stability of $M_M$ and $M_T$ across all training proportions suggests that both architectures can generalize effectively with limited data availability.
In contrast, $M_B$ and $M_L$ exhibit stronger sensitivity to the quantity of data, showing clear improvements as the proportion of training data increases.

\subsection{RQ4: Which model architecture performs best?}

Having examined the effects of sequence length, tokenization, and training data availability in the preceding research questions, we now turn to the central question of comparative model performance.

To ensure a fair comparison, we control for the factors previously analyzed. That means, $M_L$, $M_B$, and $M_M$ are evaluated using character-level tokenization, while $M_T$ is evaluated using word-level tokenization with a sequence length of 256. All models are trained with 100\% of their respective training datasets.

This section examines per-dataset generalization: how each architecture trained on a given dataset ($T_T$, $T_E$, $T_M$, or $T_H$) performs across all validation sets. This analysis provides insight into architectural robustness, highlighting which models maintain strong performance when exposed to log formats that differ from their training distribution.

\subsubsection{Descriptive Statistics}

Table~\ref{tab:rq4_model_architecture_descriptive} summarizes the $D_R$ distributions for each model architecture across training datasets $T_T$, $T_E$, $T_M$, and $T_H$. For the simpler datasets ($T_T$ and $T_E$), $M_M$ achieves the lowest mean and median $D_R$ by a wide margin, with low variability and median values near zero. 

As the training datasets increase in complexity ($T_M$ and $T_H$), the relative performance rankings shift. For $T_M$, $M_M$ continues to outperform, while $M_T$ narrows the gap. On the most challenging dataset ($T_H$), $M_T$ attains the best overall median $D_R$ (0.09), suggesting robustness under irregular log formats.

\begin{table*}[htbp]
\centering
\caption{Descriptive statistics measured in relative edit distance $D_R$ and the number of outliers across model architectures and training datasets. \textbf{Bold} values denote the lowest mean and median $D_R$ for each training dataset group.} 
\label{tab:rq4_model_architecture_descriptive}
\begin{tabular}{llrrrrr|rr}
  \toprule
Dataset & Model & Mean & Median & Q1 & Q3 & Std. Dev. & \# Outliers & \% \\ 
  \midrule
$T_T$ & $M_B$ & 0.15 & 0.07 & 0.06 & 0.10 & 0.20 & 4190 & 19.71 \\ 
  $T_T$ & $M_L$ & 0.15 & 0.12 & 0.08 & 0.16 & 0.14 & 1843 & 8.67 \\ 
  \textbf{$T_T$} & \textbf{$M_M$} & \textbf{0.02} & \textbf{0.01} & {0.01} & {0.02} & {0.03} & {2408} & {11.33} \\ 
  $T_T$ & $M_T$ & 0.11 & 0.08 & 0.06 & 0.11 & 0.08 & 1898 & 8.93 \\ 
  \midrule
$T_E$ & $M_B$ & 0.15 & 0.07 & 0.05 & 0.09 & 0.22 & 4148 & 19.51 \\ 
  $T_E$ & $M_L$ & 0.14 & 0.12 & 0.05 & 0.18 & 0.12 & 1129 & 5.31 \\ 
  \textbf{$T_E$} & \textbf{$M_M$} & \textbf{0.05} & \textbf{0.02} & {0.01} & {0.07} & {0.07} & {905} & {4.26} \\ 
  $T_E$ & $M_T$ & 0.24 & 0.19 & 0.15 & 0.27 & 0.16 & 2124 & 9.99 \\ 
  \midrule
$T_M$ & $M_B$ & 0.24 & 0.19 & 0.11 & 0.28 & 0.17 & 1276 & 6.00 \\ 
  $T_M$ & $M_L$ & 0.26 & 0.22 & 0.17 & 0.37 & 0.14 & 210 & 0.99 \\ 
  \textbf{$T_M$} & \textbf{$M_M$} & \textbf{0.12} & \textbf{0.12} & {0.02} & {0.18} & {0.09} & {28} & {0.13} \\ 
  $T_M$ & $M_T$ & 0.18 & 0.14 & 0.08 & 0.22 & 0.17 & 2569 & 12.09 \\ 
  \midrule
$T_H$ & $M_B$ & 0.34 & 0.39 & 0.13 & 0.44 & 0.18 & 32 & 0.15 \\ 
  $T_H$ & $M_L$ & 0.19 & 0.16 & 0.08 & 0.24 & 0.12 & 457 & 2.15 \\ 
  $T_H$ & $M_M$ & 0.15 & 0.17 & 0.07 & 0.21 & 0.09 & 150 & 0.71 \\ 
  \textbf{$T_H$} & \textbf{$M_T$} & \textbf{0.12} & \textbf{0.09} & {0.04} & {0.18} & {0.12} & {718} & {3.38} \\ 
   \bottomrule
\end{tabular}
\end{table*}
 %

\subsubsection{Statistical Significance Testing}

To assess whether the observed performance differences between architectures are statistically significant, pairwise Wilcoxon signed-rank tests were conducted for each training dataset ($T_T$, $T_E$, $T_M$, and $T_H$). \ref{appendix:rq4-extended-comparison} summarizes the test outcomes, indicating whether differences were statistically significant and the corresponding effect size magnitude.

For the simpler datasets ($T_T$ and $T_E$), all tests are statistically significant and often exhibit a large or medium effect. The effect sizes are consistently large when comparing $M_M$ with other models, reaffirming its superiority on datasets with homogeneous log structures. In contrast, differences between $M_B$ and $M_T$ are negligible for $T_T$, suggesting comparable performance between these two architectures in this setting.

For $T_M$, performance differences remain statistically significant with a large or medium effect. However, the effect size between $M_M$ and $M_T$ is negligible. This indicates that both architectures achieve practically similar performance when trained with a moderately complex dataset composition.

On the most challenging dataset ($T_H$), all tests are again statistically significant. All comparisons show a large effect, with the exception of the comparison between $M_M$ and $M_T$, which only yields a medium effect. This suggests that while Mamba maintains strong performance, the Transformer gains a relative advantage under log format heterogeneity.

\subsubsection{Discussion}

Although this study does not directly evaluate semantic understanding, the sustained log parsing ability, with a median $D_R$ of $0.09$ to $0.12$ for medium and hard training datasets, suggests that models generalize field relationships rather than relying on pure memorization. To further support this suggestion, Section~\ref{sec:benchmarks} Table~\ref{tab:elf_train_dataset_summary} confirms that $T_M$ and $T_H$ do not contain any ELF log records.

Additionally, $M_M$ achieves the lowest $D_R$ with structured training datasets, such as $T_T$ and $T_E$, indicating that it models consistent log formats efficiently and with low variability.
As the diversity of log formats increases, $M_T$ achieves or outperforms the performance of all other models, suggesting a greater ability to generalize from random log formats.
Both recurrent architectures exhibit declining performance as log formats become more varied.

From an economic and operational perspective, the compute-accuracy trade-off is also significant. As shown in Table~\ref{tab:model_params}, Mamba consistently falls in the low-cost range for both training and inference, while Transformer costs are two to three orders of magnitude higher. This difference can be critical in production environments with tight computation or budget constraints.

Overall, these findings indicate that Mamba and Transformer models perform best across the evaluated settings, each offering distinct advantages.
Notably, Mamba’s comparatively low computational cost may still enable its use for heterogeneous logs where efficiency is a key constraint.

\subsection{Threats to Validity}
Following established guidance in empirical research \cite{DBLP:books/daglib/0029933,yin2009case}, we discuss threats to \emph{conclusion}, \emph{internal}, \emph{construct}, and \emph{external} validity.

\subsubsection{Conclusion Validity}
Conclusion validity concerns whether our statistical claims are supported by the data, which may be threatened by low statistical power and metric sensitivity.  

To reduce statistical risks, we complement $p$-values with effect sizes, apply non-parametric Wilcoxon tests for pairwise comparisons, and use $p$-value corrections to guard against inflated false discovery rates. We repeat all stochastic procedures with ten different random seeds and consistently apply standardized preprocessing pipelines to improve reliability. To address metric sensitivity, we normalize inputs before comparison and report both absolute and relative edit distances, thereby reducing artifacts from whitespace or tokenization differences and mitigating scale sensitivity.

\subsubsection{Internal Validity}
Internal validity addresses whether observed differences can be causally attributed to our treatments rather than to implementation errors or environmental variation.  

To reduce the risk of implementation errors, all experimental code was subjected to peer review and supported by unit tests for core components. To stabilize instrumentation, we version-controlled both software and hardware configurations via Git~\cite{git-scm} and scheduled experiments with SLURM~\cite{yoo2003slurm}, ensuring reproducibility across runs and minimizing the influence of hidden confounders.

\subsubsection{Construct Validity}
Construct validity asks whether our measures accurately reflect parsing ability, which may be challenged by our reliance on edit distance rather than binary parsing accuracy.  

We chose edit distance because it captures both exactness and partial correctness, providing a more nuanced view of parsing quality in diverse log formats. While binary parsing accuracy is common in the literature, prior work has also employed edit distance \cite{Rand_2021}, which supports our approach. Furthermore, to mitigate potential scaling biases, we report edit distance in both absolute and relative terms.

\subsubsection{External Validity}
External validity concerns the extent to which findings generalize beyond our study (including changes to datasets, architectures, and tokenization strategies).  

We conducted an extensive empirical analysis using an experimental sweep of four training dataset difficulties, three model architectures, two tokenization methods, three training data percentages, and evaluation on three real-world validation datasets, resulting in a total of 396 individually trained models. This experimental setup provides evidence for generalization to Apache-style web server logs, as well as format mismatches between training and deployment. Generalization to logs with substantially different field cardinalities and length distributions should be treated with caution (as is the case for the majority of empirical findings in software engineering due to the variability of real-world environments~\cite{wieringa2015six}). Although we characterize automated log parsing under a large variety of conditions, we do not claim universality across all logging ecosystems or training recipes. However, the experimental framework presented is reproducible and adaptable, providing a template for well-designed experiments to evaluate new log sources, representations, and model architectures.

The LLM cost analysis in Example~\ref{ex:llm_log_parsing_cost} should also be interpreted with caution. Model prices, provider offerings, context-window sizes, and inference optimizations change rapidly, so the absolute dollar estimates reported in Table~\ref{tab:llm_common400_token_usage_cost} represent a point-in-time snapshot rather than a durable benchmark. In addition, our extrapolation assumes the same prompting protocol and token usage pattern as the 400-example motivating comparison. Production systems may reduce LLM usage through caching, template reuse, batching, prompt compression, fine-tuning, or hybrid pipelines that invoke an LLM only for uncertain or previously unseen messages. Therefore, the cost extrapolation is intended to characterize an economic boundary condition rather than a universal deployment cost. The more stable conclusion is comparative: hosted LLM inference introduces a recurring marginal cost that scales with the number of LLM invocations, whereas local sequence models incur primarily upfront training or adaptation costs followed by more predictable local inference costs.

\section{Conclusion}
\label{sec:conclusion}
This study presented a systematic empirical evaluation of sequence-to-sequence deep learning architectures for automated log parsing, addressing four research questions spanning sequence length, tokenization strategy, sample efficiency, and model architecture. Relative edit distance was used as the primary evaluation metric, and statistical conclusions were supported through Wilcoxon signed-rank testing with multiple-comparison correction and effect size analysis.

First, we showed that sequence length has negligible practical impact on Transformer parsing performance in our experimental setting. Although differences were statistically detectable, effect sizes were consistently negligible, indicating that longer input sequences do not translate into meaningful parsing improvements for the evaluated workloads. Given the quadratic computational scaling of self-attention, these results suggest that increasing sequence length is unlikely to be cost-effective for log parsing under similar data distributions.

Second, character-level tokenization generally produced superior performance compared to word-level tokenization across architectures. However, this advantage decreases as log format diversity increases, highlighting an interaction between representation granularity and structural variability in logs. Importantly, the results should be interpreted in light of the character-level evaluation metric, suggesting that future work should explore field-level or span-level metrics to better capture semantic correctness.

Third, we observed substantial differences in sample efficiency across architectures. Mamba and Transformer models maintained relatively stable performance even with limited training data, whereas recurrent models benefited more strongly from increased data availability. This finding suggests that modern sequence architectures may reduce data requirements for practical deployment scenarios where labelled logs are scarce.

Finally, comparative architectural analysis revealed that Mamba and Transformer models represent the strongest overall choices, but under different operating conditions. Mamba consistently performed best on structured or moderately variable log datasets and offered significantly lower computational cost. In contrast, Transformers demonstrated stronger robustness under highly heterogeneous or distribution-shifted log formats. From an operational perspective, this highlights a practical compute-accuracy trade-off: Mamba provides strong performance in cost-constrained environments, while Transformers may be preferred when maximum robustness is required.

Taken together, these results provide a structured empirical foundation for selecting sequence modelling approaches for log parsing. In this work, generalization is defined in terms of syntactic and structural variation in log formats, including field ordering, presence, and formatting changes, rather than semantic or cross-domain transfer. The experimental framework also provides a reproducible template for evaluating future architectures under controlled distribution-shift conditions.

Several research directions remain open. Hybrid architectures combining structured sequence modelling with pretrained foundation models may further improve robustness. Subword tokenization approaches (e.g., Byte Pair Encoding~\cite{DBLP:journals/corr/SennrichHB15}, WordPiece~\cite{DBLP:journals/corr/WuSCLNMKCGMKSJL16}, and SentencePiece~\cite{DBLP:journals/corr/abs-1808-06226})  offer a promising middle ground between character and word representations. Additionally, modelling temporal dependencies across log streams rather than parsing records independently may better reflect real-world production monitoring pipelines. Finally, systematic evaluation of inference latency, energy cost, and deployment-scale efficiency will be critical for translating model advances into production-ready log analytics systems.

\section*{Acknowledgments}
This work was partially supported by the Natural Sciences and Engineering Research Council of Canada (grant \# RGPIN-2022-03886). The authors thank the Digital Research Alliance of Canada for providing computational resources.

The authors used large language models for language editing, proofreading, and assistance in generating and refining portions of the code used in this study. After using these models, the authors reviewed and edited the content as needed and take full responsibility for the content of the published article.

\bibliographystyle{elsarticle-num} 
\bibliography{refs}

\pagebreak
\appendix

\section{LLM Benchmark Prompt} \label{appendix:llm-prompt}

The LLM benchmark used the same character-labelling task as the sequence-to-sequence models. For each request, the benchmark code constructed a single user message from the prompt template below:

\begin{verbatim}
You are parsing Apache HTTP server logs.

Return only the target label string. Do not explain. Do not include markdown,
code fences, or extra text.

For every character in the input log, emit exactly one output character:
- replace every character in the client host/IP span with h
- replace every character in the remote logname span with l
- replace every character in the remote username span with u
- replace every character in the timestamp span with t
- replace every character in the full request-line span with r
- replace every character in the status-code span with s
- replace every character in the byte-count span with b
- replace every character in the request method span with m
- replace every character in the URI path span with U
- replace every character in the protocol span with H
- replace every character in the query string span with q
- replace every character in the canonical server-name span with v
- replace every character in the UseCanonical server-name span with V
- replace every character in the referrer span with R
- replace every character in the user-agent span with i
- replace field-separator spaces with _
- preserve only structural double quotes, [ brackets, and ] brackets that
  delimit fields

The output must be one single line.
The output must contain exactly the expected number of characters.
Allowed output characters are: h l u t r s b m U H q v V i R _ [ ] and
double quote.
Do not output field names. Do not output copied input substrings. Do not
summarize. Do not compress repeated labels or use counts such as h*12.
If a span has 13 characters, output the same label 13 times.

Example 1 expected output length: {few_shot_target_length}
Example 1 input:
{few_shot_source}
Example 1 output:
{few_shot_target}

Expected output length: {target_length}
Input:
{source}
Output:
\end{verbatim}

Each prediction was normalized by stripping surrounding whitespace, removing a single enclosing Markdown code fence if present, and taking the first non-empty output line. The reported edit-distance metrics compare this normalized prediction against the target character-label string.

\section{Extended Evaluation Results} \label{appendix:exp}

\subsection{Sequence Length Extended Analysis} \label{appendix:rq1-extended-comparison}
\begin{figure}[htbp]
    \centering
    \includegraphics[width=\linewidth]{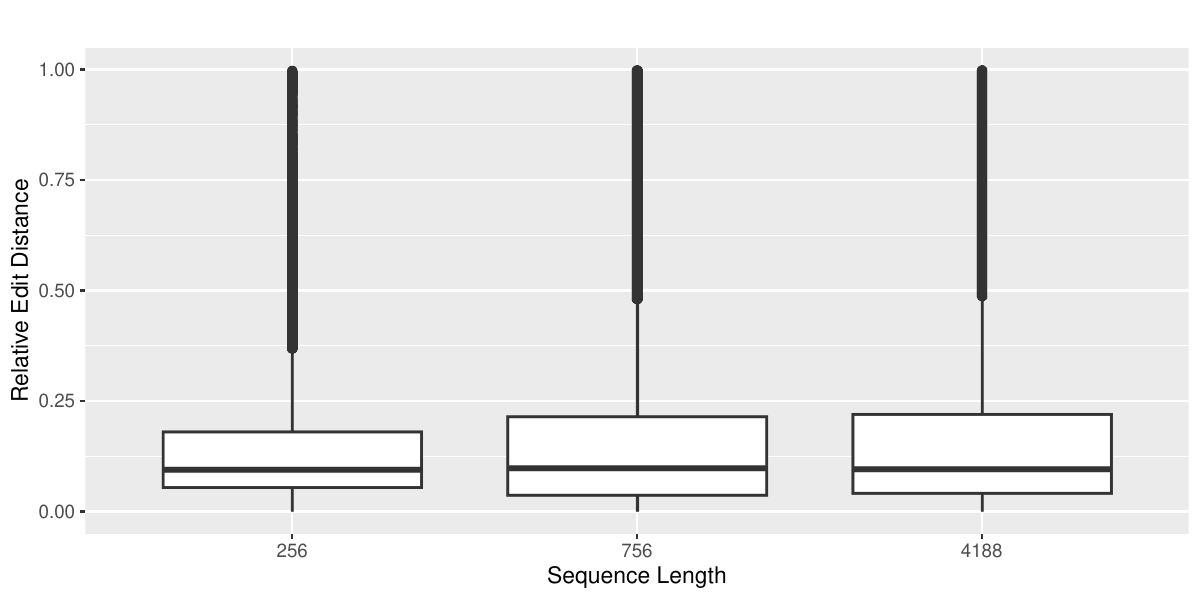}
    \caption{Box-plots of the relative edit distance of $M_T$ by sequence length.}
    \label{fig:rq1-boxplot-overall}
\end{figure}

The significance analysis in Table \ref{tab:seq_len_significance} confirms that overall differences are statistically significant but of negligible magnitude. Figure~\ref{fig:rq1-boxplot-overall} depicts the minimal performance differences between sequence lengths over all experimental configurations.
\begin{table}[htbp]
\centering
\caption{Wilcoxon test results for sequence length. Only significance and effect size magnitude are reported.} 
\label{tab:seq_len_significance}
\begin{tabular}{rrcc}
  \toprule
Group 1 & Group 2 & Significant & Magnitude \\ 
  \midrule
256 & 756 & TRUE & negligible \\ 
  256 & 4188 & TRUE & negligible \\ 
  756 & 4188 & TRUE & negligible \\ 
   \bottomrule
\end{tabular}
\end{table}

\FloatBarrier
\subsection{Validation Dataset Trends} \label{appendix:rq1-val-datasets}

Table~\ref{tab:val_sets_descriptive} shows the full descriptive statistics for changes in sequence length across each validation dataset. As discussed in Section~\ref{sec:rq1-val-set}, these trends are consistent with increased parsing error for higher sequence length due to more varied model performance.

\begin{table*}[htbp]
\centering
\caption{Descriptive statistics measured in relative edit distance $D_R$ and the number of outliers across validation datasets and sequence lengths.} 
\label{tab:val_sets_descriptive}
\begin{tabular}{rlrrrrr|rr}
  \toprule
Seq. Length & Dataset & Mean & Median & Q1 & Q3 & Std. Dev. & \# Outliers & \% \\ 
  \midrule
256 & $V_A$ & 0.14 & 0.07 & 0.03 & 0.15 & 0.19 & 28780 & 11.68 \\ 
  756 & $V_A$ & 0.19 & 0.08 & 0.01 & 0.21 & 0.26 & 34922 & 14.17 \\ 
  4188 & $V_A$ & 0.18 & 0.07 & 0.01 & 0.18 & 0.27 & 40182 & 16.30 \\ 
  256 & $V_B$ & 0.13 & 0.07 & 0.05 & 0.09 & 0.19 & 39595 & 15.86 \\ 
  756 & $V_B$ & 0.19 & 0.08 & 0.03 & 0.16 & 0.28 & 48513 & 19.43 \\ 
  4188 & $V_B$ & 0.17 & 0.07 & 0.03 & 0.10 & 0.27 & 52569 & 21.06 \\ 
  256 & $V_C$ & 0.17 & 0.14 & 0.08 & 0.21 & 0.13 & 31795 & 7.24 \\ 
  756 & $V_C$ & 0.17 & 0.14 & 0.08 & 0.22 & 0.15 & 32810 & 7.47 \\ 
  4188 & $V_C$ & 0.20 & 0.14 & 0.08 & 0.25 & 0.19 & 39769 & 9.06 \\ 
   \bottomrule
\end{tabular}
\end{table*}

\begin{table}[htbp]
\centering
\caption{Wilcoxon test results by validation dataset.} 
\label{tab:val_dataset_significance}
\begin{tabular}{lrrcc}
  \toprule
Dataset & Group 1 & Group 2 & Significant & Magnitude \\ 
  \midrule
$V_A$ & 256 & 4188 & TRUE & negligible \\ 
  $V_A$ & 256 & 756 & TRUE & negligible \\ 
  $V_A$ & 756 & 4188 & FALSE & negligible \\ 
  $V_B$ & 256 & 4188 & FALSE & negligible \\ 
  $V_B$ & 256 & 756 & TRUE & negligible \\ 
  $V_B$ & 756 & 4188 & TRUE & negligible \\ 
  $V_C$ & 256 & 4188 & TRUE & negligible \\ 
  $V_C$ & 256 & 756 & TRUE & negligible \\ 
  $V_C$ & 756 & 4188 & TRUE & negligible \\ 
   \bottomrule
\end{tabular}
\end{table}

\FloatBarrier
\subsubsection{Training Dataset Trends}

Across most training datasets, increasing the sequence length resulted in minimal changes in variability and median performance (Table~\ref{tab:training_set_descriptive}). Further analysis of the Wilcoxon test results in Table~\ref{tab:train_dataset_significance} shows the differences in median performance are statistically significant and exhibit meaningful effect size for $T_E$, $T_M$, and $T_T$.

\begin{figure}[htbp]
    \centering
    \includegraphics[width=\linewidth]{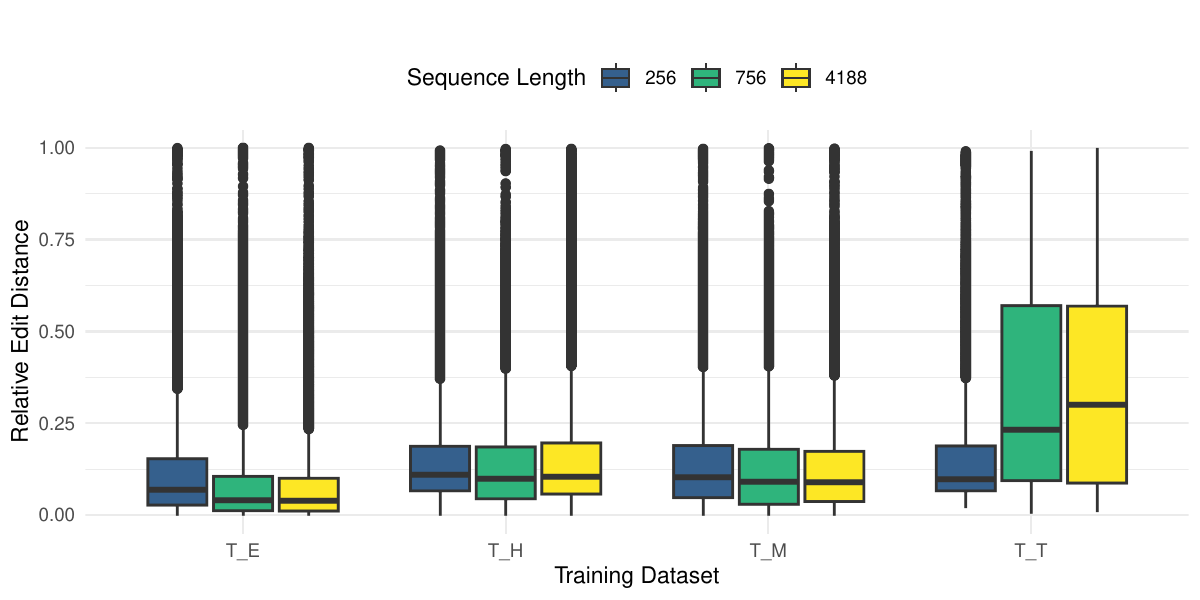}
    \caption{Box-plots of the relative edit distance of $M_T$ by sequence length and training dataset.}
    \label{fig:rq1-boxplot-train-sets}
\end{figure}

A notable observation for $T_T$ shows both the median and variance increase substantially at 756 and 4188 tokens. This deviation is clearly visible in Figure~\ref{fig:rq1-boxplot-train-sets}, indicating that $T_T$ is more sensitive to sequence length than the other training datasets.

\begin{table*}[htbp]
\centering
\caption{Descriptive statistics measured in relative edit distance $D_R$ and the number of outliers across training datasets and sequence lengths.} 
\label{tab:training_set_descriptive}
\begin{tabular}{rlrrrrr|rr}
  \toprule
Seq. Length & Dataset & Mean & Median & Q1 & Q3 & Std. Dev. & \# Outliers & \% \\ 
  \midrule
256 & $T_E$ & 0.11 & 0.07 & 0.03 & 0.15 & 0.13 & 13112 & 5.61 \\ 
  756 & $T_E$ & 0.09 & 0.04 & 0.01 & 0.11 & 0.13 & 21799 & 9.32 \\ 
  4188 & $T_E$ & 0.08 & 0.04 & 0.01 & 0.10 & 0.11 & 20553 & 8.79 \\ 
  256 & $T_H$ & 0.14 & 0.11 & 0.07 & 0.19 & 0.13 & 13743 & 5.88 \\ 
  756 & $T_H$ & 0.14 & 0.10 & 0.05 & 0.19 & 0.13 & 12425 & 5.31 \\ 
  4188 & $T_H$ & 0.18 & 0.11 & 0.06 & 0.20 & 0.22 & 27221 & 11.64 \\ 
  256 & $T_M$ & 0.15 & 0.10 & 0.05 & 0.19 & 0.16 & 22646 & 9.69 \\ 
  756 & $T_M$ & 0.14 & 0.09 & 0.03 & 0.18 & 0.15 & 18642 & 7.97 \\ 
  4188 & $T_M$ & 0.13 & 0.09 & 0.04 & 0.18 & 0.13 & 12289 & 5.26 \\ 
  256 & $T_T$ & 0.19 & 0.10 & 0.07 & 0.19 & 0.22 & 37612 & 16.09 \\ 
  756 & $T_T$ & 0.36 & 0.23 & 0.10 & 0.57 & 0.31 &   0 & 0.00 \\ 
  4188 & $T_T$ & 0.37 & 0.30 & 0.09 & 0.57 & 0.31 &   0 & 0.00 \\ 
   \bottomrule
\end{tabular}
\end{table*}

\begin{table}[htbp]
\centering
\caption{Wilcoxon test results by training dataset.} 
\label{tab:train_dataset_significance}
\begin{tabular}{lrrcc}
  \toprule
Dataset & Group 1 & Group 2 & Significant & Magnitude \\ 
  \midrule
$T_T$ & 256 & 4188 & TRUE & medium \\ 
  $T_T$ & 256 & 756 & TRUE & medium \\ 
  $T_T$ & 756 & 4188 & TRUE & negligible \\ 
  $T_E$ & 256 & 4188 & TRUE & medium \\ 
  $T_E$ & 256 & 756 & TRUE & medium \\ 
  $T_E$ & 756 & 4188 & TRUE & negligible \\ 
  $T_M$ & 256 & 4188 & TRUE & small \\ 
  $T_M$ & 256 & 756 & TRUE & small \\ 
  $T_M$ & 756 & 4188 & TRUE & negligible \\ 
  $T_H$ & 256 & 4188 & TRUE & negligible \\ 
  $T_H$ & 256 & 756 & TRUE & negligible \\ 
  $T_H$ & 756 & 4188 & TRUE & negligible \\ 
   \bottomrule
\end{tabular}
\end{table}

\FloatBarrier
\subsubsection{Data Percentage Trends}

Similarly, across different training data percentages, longer sequences are associated with greater variability but only minor shifts in median performance (Table~\ref{tab:max_obs_pct_descriptive}). This is confirmed by the statistical testing results in Table~\ref{tab:max_obs_significance}, which show statistically significant but negligible effect sizes, consistent with the global sequence length trend (Table~\ref{tab:seq_len_significance}).

\begin{figure}[htbp]
    \centering
    \includegraphics[width=\linewidth]{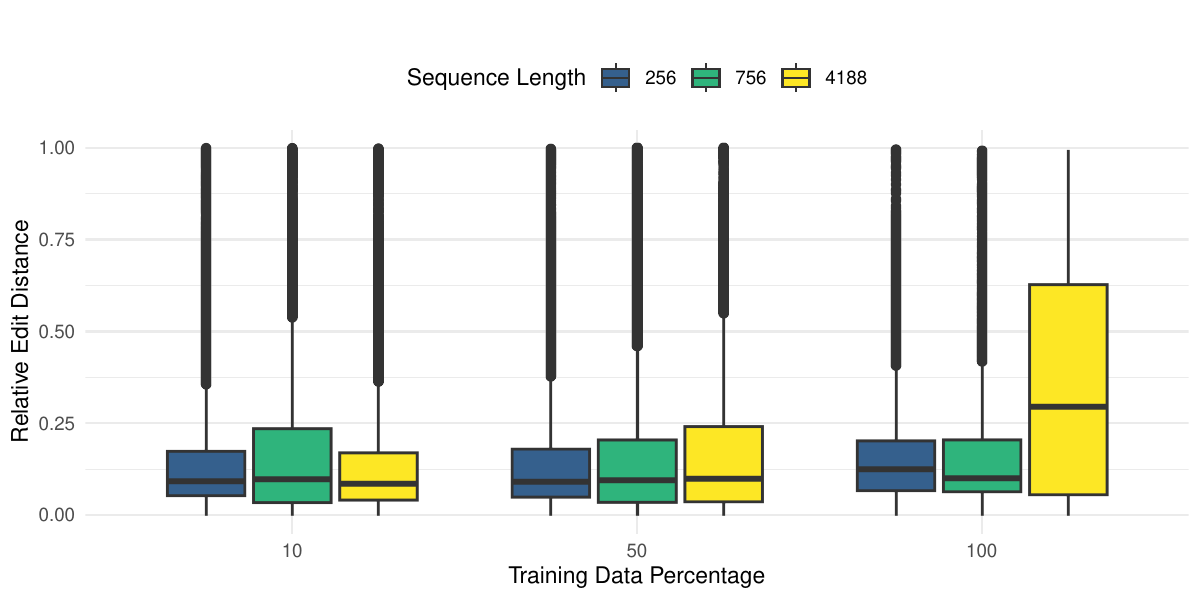}
    \caption{Box-plots of the relative edit distance of $M_T$ by sequence length.}
    \label{fig:rq1-boxplot-max_obs_pct}
\end{figure}

However, at 100\% training data, the longest sequence length (4188) shows a noticeable upward shift in both median error and variability (Figure~\ref{fig:rq1-boxplot-max_obs_pct}), representing a meaningful deviation from the otherwise stable trend.

\begin{table*}[htbp]
\centering
\caption{Descriptive statistics measured in relative edit distance $D_R$ and the number of outliers across training data percentages and sequence lengths.} 
\label{tab:max_obs_pct_descriptive}
\begin{tabular}{rrrrrrr|rr}
  \toprule
Seq. Length & \% & Mean & Median & Q1 & Q3 & Std. Dev. & \# Outliers & \% \\ 
  \midrule
256 & 10 & 0.15 & 0.09 & 0.05 & 0.18 & 0.18 & 42732 & 10.05 \\ 
  756 & 10 & 0.20 & 0.10 & 0.04 & 0.24 & 0.25 & 58013 & 13.65 \\ 
  4188 & 10 & 0.16 & 0.09 & 0.04 & 0.17 & 0.22 & 43042 & 10.12 \\ 
  256 & 50 & 0.14 & 0.09 & 0.05 & 0.18 & 0.16 & 37996 & 8.94 \\ 
  756 & 50 & 0.15 & 0.10 & 0.04 & 0.21 & 0.17 & 30309 & 7.13 \\ 
  4188 & 50 & 0.18 & 0.10 & 0.04 & 0.24 & 0.22 & 33959 & 7.99 \\ 
  256 & 100 & 0.16 & 0.13 & 0.07 & 0.20 & 0.15 & 6491 & 7.63 \\ 
  756 & 100 & 0.22 & 0.10 & 0.06 & 0.21 & 0.30 & 12602 & 14.82 \\ 
  4188 & 100 & 0.36 & 0.30 & 0.06 & 0.63 & 0.31 &   0 & 0.00 \\ 
   \bottomrule
\end{tabular}
\end{table*}

\begin{table}[htbp]
\centering
\caption{Wilcoxon test results by maximum observation percentage.} 
\label{tab:max_obs_significance}
\begin{tabular}{rrrcc}
  \toprule
Training \% & Group 1 & Group 2 & Significant & Magnitude \\ 
  \midrule
10 & 256 & 4188 & TRUE & small \\ 
  10 & 256 & 756 & TRUE & negligible \\ 
  10 & 756 & 4188 & TRUE & small \\ 
  50 & 256 & 4188 & TRUE & negligible \\ 
  50 & 256 & 756 & TRUE & negligible \\ 
  50 & 756 & 4188 & TRUE & negligible \\ 
  100 & 256 & 4188 & TRUE & medium \\ 
  100 & 256 & 756 & TRUE & small \\ 
  100 & 756 & 4188 & TRUE & medium \\ 
   \bottomrule
\end{tabular}
\end{table}

\FloatBarrier
\subsection{Tokenization Method Extended Analysis} \label{appendix:rq2-extended-comparison}

\begin{figure}[htbp]
    \centering
    \includegraphics[width=\linewidth]{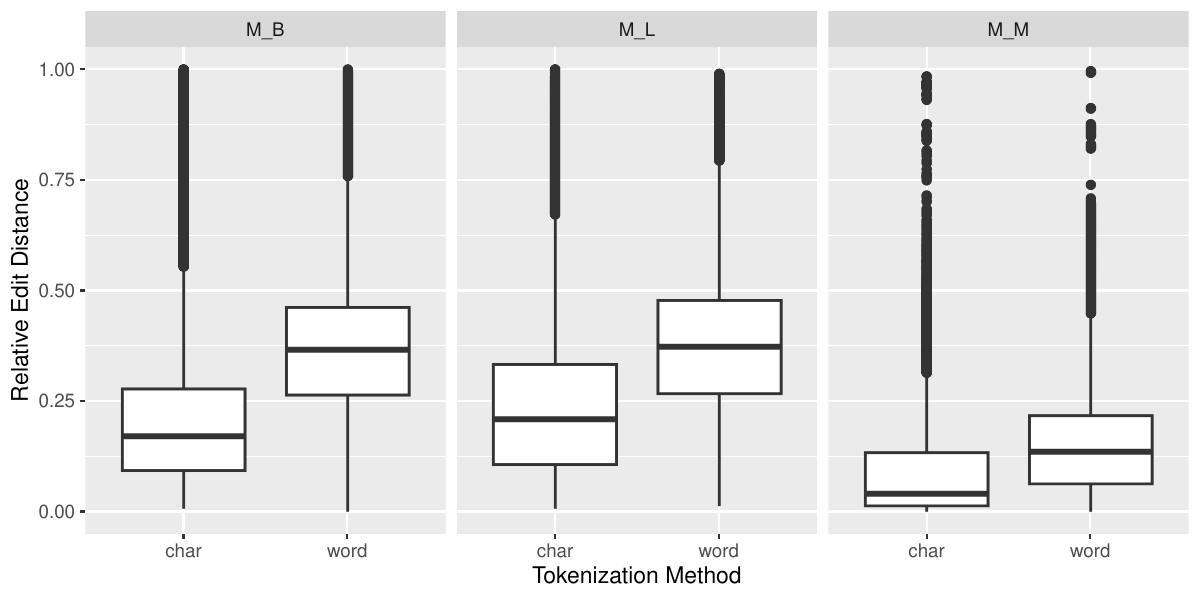}
    \caption{Box-plots of the relative edit distance by tokenization method for each model architecture.}
    \label{fig:rq2-boxplot-overall}
\end{figure}

Table~\ref{tab:rq2_tokenization_method_significance} shows that all tokenization method tests are statistically significant and exhibit a large effect. Figure~\ref{fig:rq2-boxplot-overall} shows the increase in $D_R$ when using word tokenization for all recurrent models.

\begin{table}[htbp]
\centering
\caption{Wilcoxon test results for tokenization method for each model architecture.} 
\label{tab:rq2_tokenization_method_significance}
\begin{tabular}{lllcc}
  \toprule
Model & Group 1 & Group 2 & Significant & Magnitude \\ 
  \midrule
$M_L$ & char & word & TRUE & large \\ 
  $M_B$ & char & word & TRUE & large \\ 
  $M_M$ & char & word & TRUE & large \\ 
   \bottomrule
\end{tabular}
\end{table}

\FloatBarrier
\subsubsection{Validation Dataset Trends} \label{appendix:rq2-val-datasets}

\begin{figure}[htbp]
    \centering
    \includegraphics[width=\linewidth]{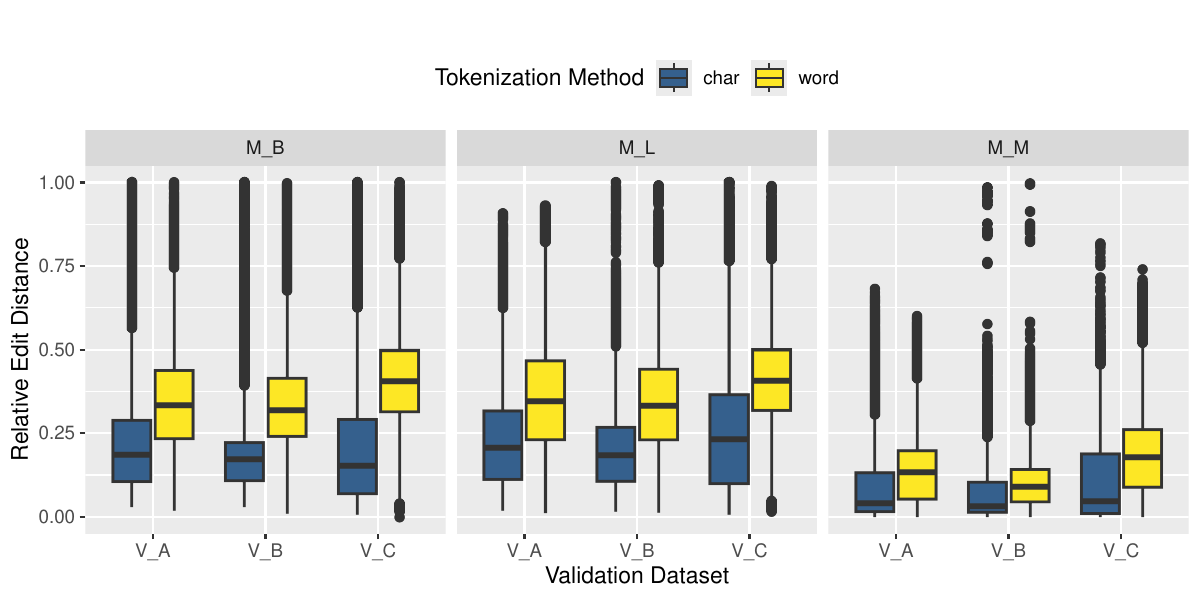}
    \caption{Box-plots of the relative edit distance by validation dataset and tokenization method for each model architecture.}
    \label{fig:rq2-boxplot-valset}
\end{figure}

Table~\ref{tab:rq2_val_set_descriptive} and Figure~\ref{fig:rq2-boxplot-valset} both show consistent out-performance with character-level autoionization across all three validation datasets. Statistical testing show in Table~\ref{tab:rq2_tokenization_method_per_validation_significance} confirms the median performance differences are significant with a large effect.

\begin{table*}[htbp]
\centering
\caption{Descriptive statistics measured in relative edit distance $D_R$ and the number of outliers across validation datasets, tokenization methods, and model architectures.} 
\label{tab:rq2_val_set_descriptive}
\begin{tabular}{lllrrrrr|rr}
  \toprule
Arch. & Rep. & Dataset & Mean & Median & Q1 & Q3 & Std. Dev. & \# Outlier & \% \\ 
  \midrule
$M_B$ & char & $V_A$ & 0.23 & 0.19 & 0.11 & 0.29 & 0.16 & 9176 & 3.72 \\ 
  $M_B$ & word & $V_A$ & 0.35 & 0.33 & 0.23 & 0.44 & 0.16 & 5794 & 2.35 \\ 
  $M_B$ & char & $V_B$ & 0.21 & 0.17 & 0.11 & 0.22 & 0.15 & 28751 & 11.52 \\ 
  $M_B$ & word & $V_B$ & 0.33 & 0.32 & 0.24 & 0.42 & 0.13 & 4285 & 1.72 \\ 
  $M_B$ & char & $V_C$ & 0.20 & 0.15 & 0.07 & 0.29 & 0.17 & 9185 & 2.09 \\ 
  $M_B$ & word & $V_C$ & 0.41 & 0.41 & 0.31 & 0.50 & 0.14 & 3762 & 0.86 \\ 
  $M_L$ & char & $V_A$ & 0.23 & 0.21 & 0.11 & 0.32 & 0.15 & 4568 & 1.85 \\ 
  $M_L$ & word & $V_A$ & 0.35 & 0.35 & 0.23 & 0.47 & 0.16 & 459 & 0.19 \\ 
  $M_L$ & char & $V_B$ & 0.20 & 0.18 & 0.11 & 0.27 & 0.12 & 3321 & 1.33 \\ 
  $M_L$ & word & $V_B$ & 0.33 & 0.33 & 0.23 & 0.44 & 0.14 & 846 & 0.34 \\ 
  $M_L$ & char & $V_C$ & 0.25 & 0.23 & 0.10 & 0.37 & 0.17 & 1810 & 0.41 \\ 
  $M_L$ & word & $V_C$ & 0.41 & 0.41 & 0.32 & 0.50 & 0.15 & 8424 & 1.92 \\ 
  $M_M$ & char & $V_A$ & 0.08 & 0.04 & 0.02 & 0.13 & 0.08 & 3399 & 1.38 \\ 
  $M_M$ & word & $V_A$ & 0.14 & 0.13 & 0.05 & 0.20 & 0.10 & 2498 & 1.01 \\ 
  $M_M$ & char & $V_B$ & 0.06 & 0.03 & 0.01 & 0.11 & 0.07 & 2542 & 1.02 \\ 
  $M_M$ & word & $V_B$ & 0.10 & 0.09 & 0.05 & 0.14 & 0.08 & 7005 & 2.81 \\ 
  $M_M$ & char & $V_C$ & 0.10 & 0.05 & 0.01 & 0.19 & 0.11 & 438 & 0.10 \\ 
  $M_M$ & word & $V_C$ & 0.18 & 0.18 & 0.09 & 0.26 & 0.12 & 3136 & 0.71 \\ 
   \bottomrule
\end{tabular}
\end{table*}

\begin{table}[htbp]
\centering
\caption{Wilcoxon test results for tokenization method by validation dataset for each model architecture.} 
\label{tab:rq2_tokenization_method_per_validation_significance}
\begin{tabular}{llllcc}
  \toprule
Model & Dataset & Group 1 & Group 2 & Significant & Magnitude \\ 
  \midrule
$M_L$ & $V_A$ & char & word & TRUE & large \\ 
  $M_L$ & $V_B$ & char & word & TRUE & large \\ 
  $M_L$ & $V_C$ & char & word & TRUE & large \\ 
  $M_B$ & $V_A$ & char & word & TRUE & large \\ 
  $M_B$ & $V_B$ & char & word & TRUE & large \\ 
  $M_B$ & $V_C$ & char & word & TRUE & large \\ 
  $M_M$ & $V_A$ & char & word & TRUE & large \\ 
  $M_M$ & $V_B$ & char & word & TRUE & large \\ 
  $M_M$ & $V_C$ & char & word & TRUE & large \\ 
   \bottomrule
\end{tabular}
\end{table}

\FloatBarrier
\subsubsection{Training Dataset Trends}

Table~\ref{tab:rq2_training_set_descriptive} shows consistently high $D_R$ metrics with word tokenization across each training dataset and model architecture. Table~\ref{tab:rq2_tokenization_method_per_training_dataset_significance} validates the significance and show most a large and medium effect size with $M_L$ trained on $T_M$ exhibiting a small effect.

\begin{table*}[htbp]
\centering
\caption{Descriptive statistics measured in relative edit distance $D_R$ and the number of outliers across training datasets, tokenization methods, and model architectures.} 
\label{tab:rq2_training_set_descriptive}
\begin{tabular}{lllrrrrr|rr}
  \toprule
Model & Rep. & Dataset & Mean & Median & Q1 & Q3 & Std. Dev. & \# Outliers & \% \\ 
  \midrule
$M_B$ & char & $T_E$ & 0.18 & 0.12 & 0.08 & 0.20 & 0.15 & 26510 & 11.34 \\ 
  $M_B$ & word & $T_E$ & 0.38 & 0.37 & 0.26 & 0.47 & 0.14 & 1713 & 0.73 \\ 
  $M_B$ & char & $T_H$ & 0.27 & 0.24 & 0.15 & 0.34 & 0.17 & 10264 & 4.39 \\ 
  $M_B$ & word & $T_H$ & 0.35 & 0.35 & 0.25 & 0.44 & 0.15 & 3928 & 1.68 \\ 
  $M_B$ & char & $T_M$ & 0.25 & 0.22 & 0.14 & 0.31 & 0.14 & 6891 & 2.95 \\ 
  $M_B$ & word & $T_M$ & 0.33 & 0.33 & 0.23 & 0.43 & 0.15 & 3070 & 1.31 \\ 
  $M_B$ & char & $T_T$ & 0.14 & 0.08 & 0.05 & 0.18 & 0.14 & 18088 & 7.74 \\ 
  $M_B$ & word & $T_T$ & 0.42 & 0.41 & 0.33 & 0.50 & 0.13 & 3231 & 1.38 \\ 
  $M_L$ & char & $T_E$ & 0.17 & 0.13 & 0.09 & 0.21 & 0.12 & 12823 & 5.48 \\ 
  $M_L$ & word & $T_E$ & 0.36 & 0.35 & 0.26 & 0.44 & 0.14 & 8191 & 3.50 \\ 
  $M_L$ & char & $T_H$ & 0.28 & 0.26 & 0.15 & 0.38 & 0.15 & 1329 & 0.57 \\ 
  $M_L$ & word & $T_H$ & 0.36 & 0.36 & 0.22 & 0.50 & 0.19 & 113 & 0.05 \\ 
  $M_L$ & char & $T_M$ & 0.31 & 0.30 & 0.20 & 0.41 & 0.14 & 2248 & 0.96 \\ 
  $M_L$ & word & $T_M$ & 0.36 & 0.36 & 0.23 & 0.46 & 0.16 & 465 & 0.20 \\ 
  $M_L$ & char & $T_T$ & 0.17 & 0.15 & 0.06 & 0.27 & 0.14 & 2024 & 0.87 \\ 
  $M_L$ & word & $T_T$ & 0.41 & 0.41 & 0.35 & 0.49 & 0.11 & 2975 & 1.27 \\ 
  $M_M$ & char & $T_E$ & 0.04 & 0.02 & 0.00 & 0.05 & 0.07 & 17480 & 7.48 \\ 
  $M_M$ & word & $T_E$ & 0.14 & 0.12 & 0.08 & 0.18 & 0.10 & 9946 & 4.25 \\ 
  $M_M$ & char & $T_H$ & 0.15 & 0.13 & 0.07 & 0.20 & 0.10 & 2375 & 1.02 \\ 
  $M_M$ & word & $T_H$ & 0.19 & 0.18 & 0.11 & 0.25 & 0.11 & 3683 & 1.58 \\ 
  $M_M$ & char & $T_M$ & 0.13 & 0.11 & 0.04 & 0.18 & 0.09 & 798 & 0.34 \\ 
  $M_M$ & word & $T_M$ & 0.17 & 0.16 & 0.08 & 0.25 & 0.11 & 1762 & 0.75 \\ 
  $M_M$ & char & $T_T$ & 0.02 & 0.02 & 0.01 & 0.02 & 0.04 & 21818 & 9.33 \\ 
  $M_M$ & word & $T_T$ & 0.11 & 0.05 & 0.03 & 0.18 & 0.11 & 5175 & 2.21 \\ 
   \bottomrule
\end{tabular}
\end{table*}

\begin{table}[htbp]
\centering
\caption{Wilcoxon test results for tokenization method by training dataset for each model architecture.} 
\label{tab:rq2_tokenization_method_per_training_dataset_significance}
\begin{tabular}{llllcc}
  \toprule
Model & Dataset & Group 1 & Group 2 & Significant & Magnitude \\ 
  \midrule
$M_L$ & $T_T$ & char & word & TRUE & large \\ 
  $M_L$ & $T_E$ & char & word & TRUE & large \\ 
  $M_L$ & $T_M$ & char & word & TRUE & small \\ 
  $M_L$ & $T_H$ & char & word & TRUE & medium \\ 
  $M_B$ & $T_T$ & char & word & TRUE & large \\ 
  $M_B$ & $T_E$ & char & word & TRUE & large \\ 
  $M_B$ & $T_M$ & char & word & TRUE & large \\ 
  $M_B$ & $T_H$ & char & word & TRUE & large \\ 
  $M_M$ & $T_T$ & char & word & TRUE & large \\ 
  $M_M$ & $T_E$ & char & word & TRUE & large \\ 
  $M_M$ & $T_M$ & char & word & TRUE & medium \\ 
  $M_M$ & $T_H$ & char & word & TRUE & medium \\ 
   \bottomrule
\end{tabular}
\end{table}

\paragraph{Trivial Dataset Trends} \label{appendix:rq2-trivial}

Section~\ref{sec:rq2-trivial} describes structural inconsistencies between $T_T$ and the validation datasets as shown in Figure~\ref{fig:apache-T_T-example}. This analysis was motivated by the disproportionate increase in $D_R$ exhibited by the word tokenization model trained on $T_T$ and $T_E$ models. Table~\ref{tab:rq2_median_comparison} shows the median $D_R$ for different architectures and training datasets, along with the absolute difference between them. In addition, we calculate the ratio of these differences to those of $T_M$ and $T_H$ within the same architecture as a baseline. For example, $M_B$ trained on $T_T$ has a median $D_R$ difference of $0.41-0.08=0.33$, a $T_M$ ratio of $0.328/0.108=3.029$, and a $T_H$ ratio of $0.328/0.115=2.856$.

\begin{table*}[ht]
\centering
\caption{Median relative edit distance ($D_R$) comparison between character-level and word-level tokenization by training dataset and model architecture. The difference is calculated as word-level minus character-level. Ratios show the relative magnitude of each difference compared to $T_M$ and $T_H$ baselines.} 
\label{tab:rq2_median_comparison}
\begin{tabular}{llrrr|rr}
  \toprule
Model & Dataset & $D_R$ Char & $D_R$ Word & Difference & Ratio to $T_M$ & Ratio to $T_H$ \\ 
  \midrule
$M_B$ & $T_E$ & 0.121 & 0.374 & 0.253 & 2.334 & 2.201 \\ 
  $M_B$ & $T_H$ & 0.237 & 0.352 & 0.115 & 1.061 & 1.000 \\ 
  $M_B$ & $T_M$ & 0.220 & 0.328 & 0.108 & 1.000 & 0.943 \\ 
  $M_B$ & $T_T$ & 0.084 & 0.412 & 0.328 & 3.029 & 2.856 \\ 
  $M_L$ & $T_E$ & 0.133 & 0.346 & 0.213 & 3.999 & 2.040 \\ 
  $M_L$ & $T_H$ & 0.260 & 0.364 & 0.104 & 1.960 & 1.000 \\ 
  $M_L$ & $T_M$ & 0.303 & 0.356 & 0.053 & 1.000 & 0.510 \\ 
  $M_L$ & $T_T$ & 0.146 & 0.414 & 0.268 & 5.044 & 2.573 \\ 
  $M_M$ & $T_E$ & 0.024 & 0.125 & 0.101 & 2.219 & 2.458 \\ 
  $M_M$ & $T_H$ & 0.134 & 0.175 & 0.041 & 0.902 & 1.000 \\ 
  $M_M$ & $T_M$ & 0.113 & 0.159 & 0.045 & 1.000 & 1.108 \\ 
  $M_M$ & $T_T$ & 0.015 & 0.052 & 0.037 & 0.820 & 0.908 \\ 
   \bottomrule
\end{tabular}
\end{table*}

\begin{figure*}[htbp]
  \centering
  \begin{tcolorbox}[
    width=\textwidth,     %
    colback=white, colframe=black,
    boxrule=2pt, arc=0pt,   %
    left=6pt, right=6pt, top=6pt, bottom=6pt %
  ]
    \fittt{T_T: 119.29.60.133 - - 26/Sep/9824:17:05:35 +0800 "OPTIONS 46rpb HTTP/1.1" 400 65690 ...} 
    \fittt{V_A: "192.168.4.164 - - [22/Dec/2016:15:19:05 +0300] "GET /DVWA/ HTTP/1.1" 200 2020 ... "}
  \end{tcolorbox}
  \caption{An example log string from both $T_T$ and $V_A$ which highlights the syntax differences present between the train and test datasets. A portion of the log strings is truncated for illustrative purposes and denoted by `$\cdots$'.}
  \label{fig:apache-T_T-example}
\end{figure*}

\FloatBarrier
\subsubsection{Data Percentage Trends}

\begin{figure}[htbp]
    \centering
    \includegraphics[width=\linewidth]{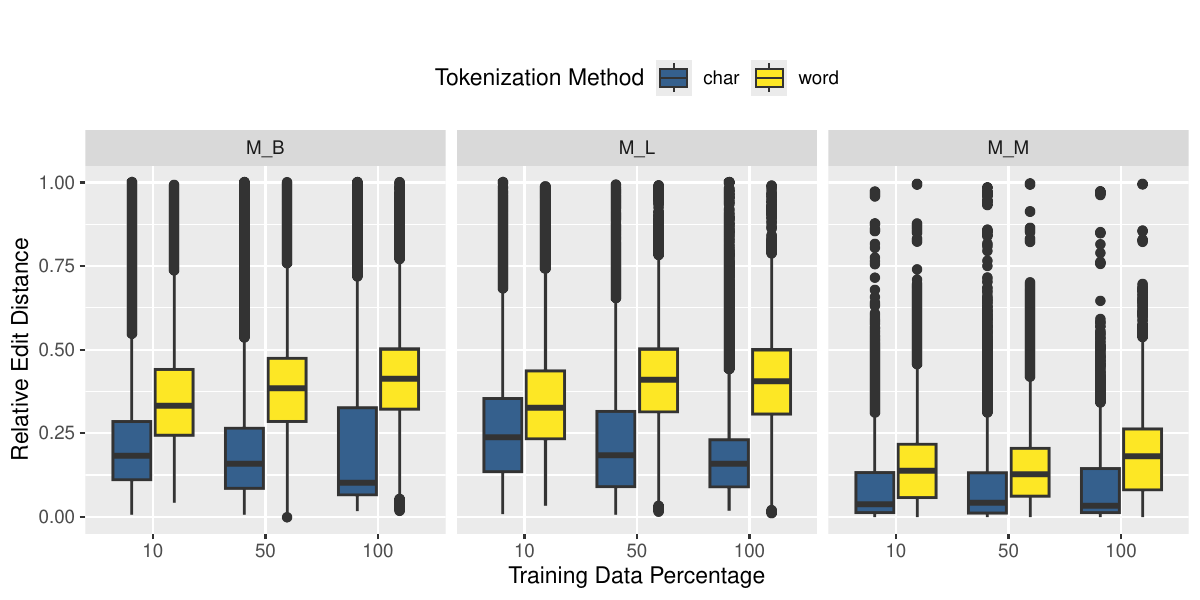}
    \caption{Box-plots of the relative edit distance by tokenization method and training data percentage for each model architecture.}
    \label{fig:rq2-boxplot-max_obs_pct}
\end{figure}

Table~\ref{tab:rq2_max_obs_pct_descriptive} and Figure~\ref{fig:rq2-boxplot-max_obs_pct} show consistent underperformance of word tokenization across all training data percentages and model architectures. Table~\ref{tab:rq2_tokenization_method_per_max_obs_pct_significance} validates the significance and effect size of the difference in median $D_R$. All tests show a large effect except for $M_L$ with 10\%, which exhibits a medium effect.

\begin{table*}[htbp]
\centering
\caption{Descriptive statistics measured in relative edit distance $D_R$ and the number of outliers across maximum observation percentages, tokenization methods, and model architectures.} 
\label{tab:rq2_max_obs_pct_descriptive}
\begin{tabular}{llrrrrrr|rr}
  \toprule
Model & Rep. & \% & Mean & Median & Q1 & Q3 & Std. Dev. & \# Outliers & \% \\ 
  \midrule
$M_B$ & char & 10 & 0.22 & 0.18 & 0.11 & 0.29 & 0.15 & 14101 & 3.32 \\ 
  $M_B$ & word & 10 & 0.35 & 0.33 & 0.25 & 0.44 & 0.14 & 4590 & 1.08 \\ 
  $M_B$ & char & 50 & 0.20 & 0.16 & 0.09 & 0.27 & 0.16 & 15295 & 3.60 \\ 
  $M_B$ & word & 50 & 0.38 & 0.39 & 0.29 & 0.48 & 0.16 & 5371 & 1.26 \\ 
  $M_B$ & char & 100 & 0.22 & 0.10 & 0.07 & 0.33 & 0.21 & 2809 & 3.30 \\ 
  $M_B$ & word & 100 & 0.41 & 0.41 & 0.32 & 0.50 & 0.15 & 2507 & 2.95 \\ 
  $M_L$ & char & 10 & 0.26 & 0.24 & 0.14 & 0.36 & 0.15 & 5183 & 1.22 \\ 
  $M_L$ & word & 10 & 0.34 & 0.33 & 0.23 & 0.44 & 0.15 & 7454 & 1.75 \\ 
  $M_L$ & char & 50 & 0.22 & 0.18 & 0.09 & 0.32 & 0.15 & 4571 & 1.08 \\ 
  $M_L$ & word & 50 & 0.40 & 0.41 & 0.31 & 0.50 & 0.15 & 4262 & 1.00 \\ 
  $M_L$ & char & 100 & 0.19 & 0.16 & 0.09 & 0.23 & 0.14 & 4948 & 5.82 \\ 
  $M_L$ & word & 100 & 0.39 & 0.41 & 0.31 & 0.50 & 0.15 & 1195 & 1.41 \\ 
  $M_M$ & char & 10 & 0.08 & 0.04 & 0.01 & 0.13 & 0.09 & 11603 & 2.73 \\ 
  $M_M$ & word & 10 & 0.15 & 0.14 & 0.06 & 0.22 & 0.11 & 4666 & 1.10 \\ 
  $M_M$ & char & 50 & 0.09 & 0.04 & 0.01 & 0.13 & 0.09 & 13059 & 3.07 \\ 
  $M_M$ & word & 50 & 0.15 & 0.13 & 0.06 & 0.21 & 0.11 & 8277 & 1.95 \\ 
  $M_M$ & char & 100 & 0.09 & 0.04 & 0.01 & 0.15 & 0.09 & 1205 & 1.42 \\ 
  $M_M$ & word & 100 & 0.19 & 0.18 & 0.08 & 0.26 & 0.12 & 488 & 0.57 \\ 
   \bottomrule
\end{tabular}
\end{table*}

\begin{table}[htbp]
\centering
\caption{Wilcoxon test results for tokenization method by training data percentage for each model architecture.} 
\label{tab:rq2_tokenization_method_per_max_obs_pct_significance}
\begin{tabular}{lrllcc}
  \toprule
Model & Training \% & Group 1 & Group 2 & Significant & Magnitude \\ 
  \midrule
$M_L$ & 10 & char & word & TRUE & medium \\ 
  $M_L$ & 50 & char & word & TRUE & large \\ 
  $M_L$ & 100 & char & word & TRUE & large \\ 
  $M_B$ & 10 & char & word & TRUE & large \\ 
  $M_B$ & 50 & char & word & TRUE & large \\ 
  $M_B$ & 100 & char & word & TRUE & large \\ 
  $M_M$ & 10 & char & word & TRUE & large \\ 
  $M_M$ & 50 & char & word & TRUE & large \\ 
  $M_M$ & 100 & char & word & TRUE & large \\ 
   \bottomrule
\end{tabular}
\end{table}

\FloatBarrier
\subsection{Sample Efficiency Extended Analysis} \label{appendix:rq3-extended-comparison}

\begin{figure}[htbp]
    \centering
    \includegraphics[width=\linewidth]{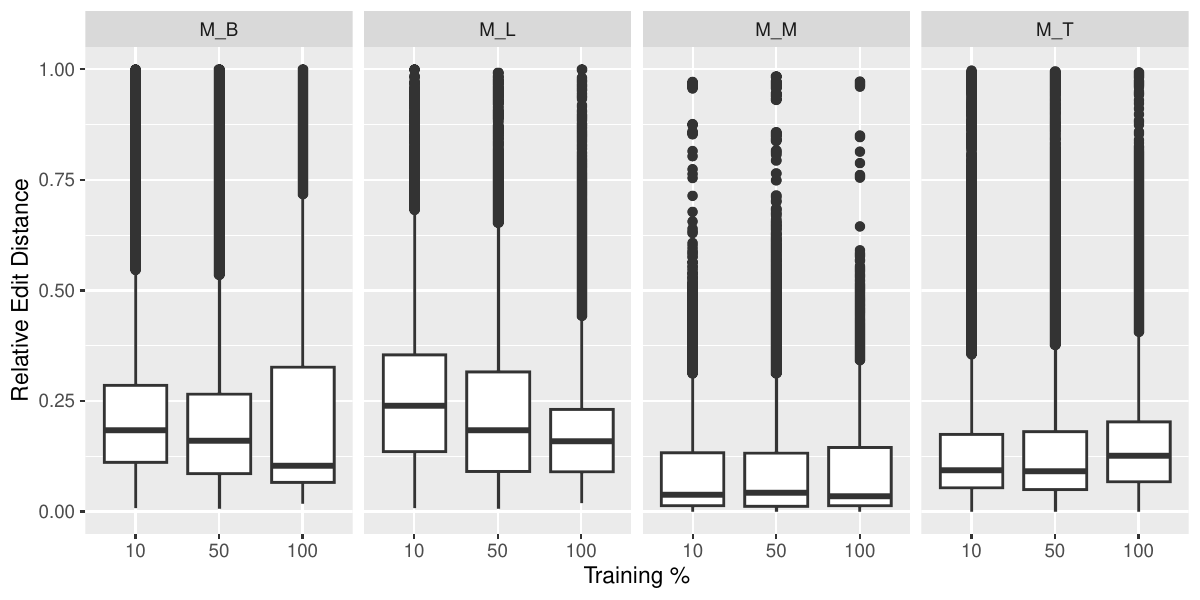}
    \caption{Box-plots of the relative edit distance by training data percentage for each model architecture.}
    \label{fig:rq3-boxplot-max_obs_pct_overall}
\end{figure}

Figure~\ref{fig:rq3-boxplot-max_obs_pct_overall} shows the $D_R$ of various training dataset percentages for each model architecture. Table~\ref{tab:rq3_max_obs_pct_significance_overall} verifies the mixed statistical significance and effect size depending on model architecture.

\begin{table}[htbp]
\centering
\caption{Wilcoxon test results for max observations percentage for each model architecture.} 
\label{tab:rq3_max_obs_pct_significance_overall}
\begin{tabular}{lrrcc}
  \toprule
Model & Group 1 & Group 2 & Significant & Magnitude \\ 
  \midrule
$M_L$ & 10 & 50 & TRUE & medium \\ 
  $M_L$ & 10 & 100 & TRUE & large \\ 
  $M_L$ & 50 & 100 & TRUE & small \\ 
  $M_B$ & 10 & 50 & TRUE & small \\ 
  $M_B$ & 10 & 100 & TRUE & small \\ 
  $M_B$ & 50 & 100 & FALSE & negligible \\ 
  $M_T$ & 10 & 50 & TRUE & negligible \\ 
  $M_T$ & 10 & 100 & TRUE & medium \\ 
  $M_T$ & 50 & 100 & TRUE & medium \\ 
  $M_M$ & 10 & 50 & TRUE & negligible \\ 
  $M_M$ & 10 & 100 & TRUE & small \\ 
  $M_M$ & 50 & 100 & TRUE & small \\ 
   \bottomrule
\end{tabular}
\end{table}

\FloatBarrier
\subsubsection{Validation Dataset Trends}

\begin{figure}[htbp]
    \centering
    \includegraphics[width=\linewidth]{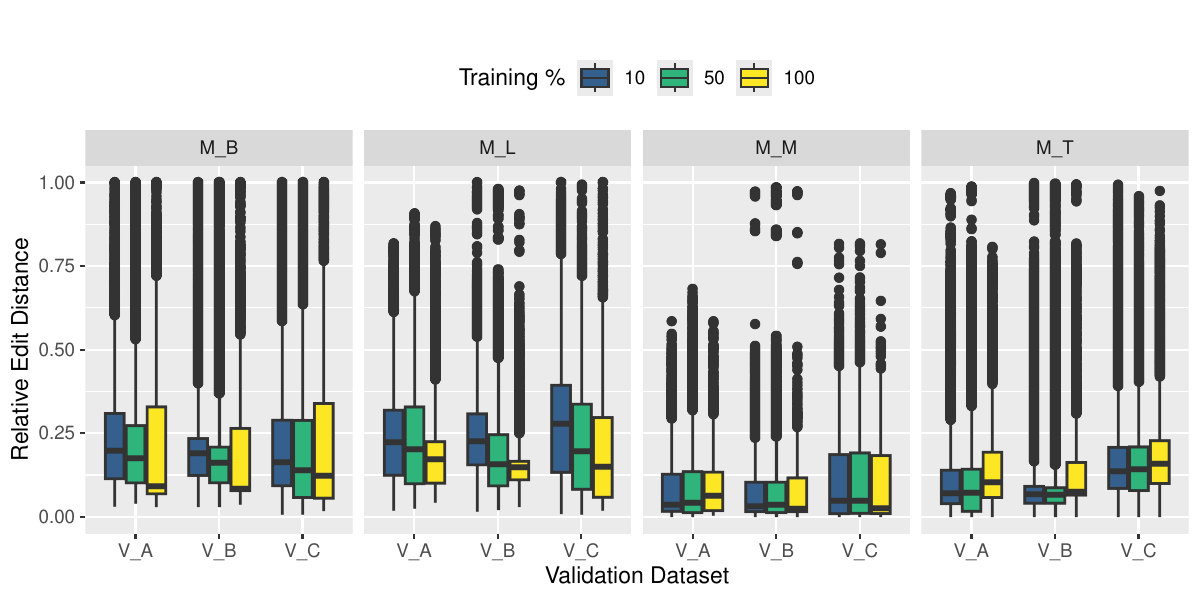}
    \caption{Box-plots of the relative edit distance by validation dataset and training data percentage for each model architecture.}
    \label{fig:rq3-boxplot-max_obs_pct_valset}
\end{figure}

Tables~\ref{tab:rq3_val_set_descriptive}~and~\ref{tab:rq3_max_obs_pct_significance_per_eval} and Figure~\ref{fig:rq3-boxplot-max_obs_pct_valset} show trends consistent with the aggregated results. Specifically, $M_M$ and $M_T$ exhibit a higher number of negligible effect sizes. This suggests that the effect of training dataset percentage is consistent across the validation datasets tested.

\begin{table*}[htbp]
\centering
\caption{Descriptive statistics measured in relative edit distance $D_R$ and the number of outliers across validation datasets, maximum observation percentages, and model architectures.} 
\label{tab:rq3_val_set_descriptive}
\resizebox{\textwidth}{!}{
\begin{tabular}{lrlrrrrr|rr}
  \toprule
Model & Training \% & Dataset & Mean & Median & Q1 & Q3 & Std. Dev. & \# Outliers & \% \\ 
  \midrule
$M_B$ & 10 & $V_A$ & 0.24 & 0.20 & 0.12 & 0.31 & 0.15 & 2114 & 1.89 \\ 
  $M_B$ & 50 & $V_A$ & 0.21 & 0.18 & 0.10 & 0.27 & 0.15 & 5578 & 4.98 \\ 
  $M_B$ & 100 & $V_A$ & 0.22 & 0.09 & 0.07 & 0.33 & 0.20 & 601 & 2.68 \\ 
  $M_B$ & 10 & $V_B$ & 0.23 & 0.19 & 0.12 & 0.24 & 0.15 & 18017 & 15.88 \\ 
  $M_B$ & 50 & $V_B$ & 0.19 & 0.16 & 0.10 & 0.21 & 0.14 & 6056 & 5.34 \\ 
  $M_B$ & 100 & $V_B$ & 0.20 & 0.09 & 0.08 & 0.27 & 0.20 & 1341 & 5.91 \\ 
  $M_B$ & 10 & $V_C$ & 0.20 & 0.17 & 0.09 & 0.29 & 0.15 & 4143 & 2.08 \\ 
  $M_B$ & 50 & $V_C$ & 0.19 & 0.14 & 0.06 & 0.29 & 0.17 & 4221 & 2.11 \\ 
  $M_B$ & 100 & $V_C$ & 0.22 & 0.12 & 0.06 & 0.34 & 0.22 & 1455 & 3.64 \\ 
  $M_L$ & 10 & $V_A$ & 0.24 & 0.22 & 0.13 & 0.32 & 0.14 & 2354 & 2.10 \\ 
  $M_L$ & 50 & $V_A$ & 0.24 & 0.20 & 0.10 & 0.33 & 0.16 & 1063 & 0.95 \\ 
  $M_L$ & 100 & $V_A$ & 0.20 & 0.17 & 0.10 & 0.23 & 0.13 & 2070 & 9.24 \\ 
  $M_L$ & 10 & $V_B$ & 0.23 & 0.23 & 0.16 & 0.31 & 0.12 & 1310 & 1.15 \\ 
  $M_L$ & 50 & $V_B$ & 0.19 & 0.16 & 0.09 & 0.25 & 0.12 & 2042 & 1.80 \\ 
  $M_L$ & 100 & $V_B$ & 0.15 & 0.15 & 0.11 & 0.17 & 0.07 & 925 & 4.08 \\ 
  $M_L$ & 10 & $V_C$ & 0.28 & 0.28 & 0.13 & 0.39 & 0.17 & 1100 & 0.55 \\ 
  $M_L$ & 50 & $V_C$ & 0.23 & 0.20 & 0.08 & 0.34 & 0.16 & 1206 & 0.60 \\ 
  $M_L$ & 100 & $V_C$ & 0.20 & 0.15 & 0.06 & 0.30 & 0.16 & 513 & 1.28 \\ 
  $M_M$ & 10 & $V_A$ & 0.07 & 0.04 & 0.02 & 0.13 & 0.07 & 1117 & 1.00 \\ 
  $M_M$ & 50 & $V_A$ & 0.08 & 0.04 & 0.01 & 0.14 & 0.08 & 1847 & 1.65 \\ 
  $M_M$ & 100 & $V_A$ & 0.09 & 0.06 & 0.02 & 0.14 & 0.08 & 373 & 1.66 \\ 
  $M_M$ & 10 & $V_B$ & 0.06 & 0.03 & 0.02 & 0.11 & 0.07 & 1236 & 1.09 \\ 
  $M_M$ & 50 & $V_B$ & 0.06 & 0.04 & 0.01 & 0.11 & 0.07 & 1134 & 1.00 \\ 
  $M_M$ & 100 & $V_B$ & 0.07 & 0.03 & 0.02 & 0.12 & 0.08 & 140 & 0.62 \\ 
  $M_M$ & 10 & $V_C$ & 0.10 & 0.05 & 0.01 & 0.19 & 0.11 & 406 & 0.20 \\ 
  $M_M$ & 50 & $V_C$ & 0.10 & 0.05 & 0.01 & 0.19 & 0.11 & 152 & 0.08 \\ 
  $M_M$ & 100 & $V_C$ & 0.09 & 0.03 & 0.01 & 0.18 & 0.10 &  30 & 0.08 \\ 
  $M_T$ & 10 & $V_A$ & 0.14 & 0.07 & 0.04 & 0.14 & 0.19 & 12884 & 11.50 \\ 
  $M_T$ & 50 & $V_A$ & 0.14 & 0.07 & 0.02 & 0.14 & 0.19 & 14029 & 12.52 \\ 
  $M_T$ & 100 & $V_A$ & 0.14 & 0.11 & 0.06 & 0.19 & 0.14 & 1589 & 7.09 \\ 
  $M_T$ & 10 & $V_B$ & 0.13 & 0.07 & 0.04 & 0.09 & 0.21 & 15461 & 13.62 \\ 
  $M_T$ & 50 & $V_B$ & 0.12 & 0.07 & 0.04 & 0.09 & 0.17 & 18245 & 16.08 \\ 
  $M_T$ & 100 & $V_B$ & 0.14 & 0.08 & 0.07 & 0.16 & 0.16 & 2273 & 10.01 \\ 
  $M_T$ & 10 & $V_C$ & 0.17 & 0.14 & 0.09 & 0.21 & 0.14 & 16476 & 8.25 \\ 
  $M_T$ & 50 & $V_C$ & 0.16 & 0.14 & 0.08 & 0.21 & 0.13 & 13076 & 6.55 \\ 
  $M_T$ & 100 & $V_C$ & 0.19 & 0.16 & 0.10 & 0.23 & 0.14 & 2574 & 6.45 \\ 
   \bottomrule
\end{tabular}
}
\end{table*}

\begin{table*}[htbp]
\centering
\caption{Wilcoxon test results for max observations percentage by validation dataset for each model architecture.} 
\label{tab:rq3_max_obs_pct_significance_per_eval}
\begin{tabular}{llrrcc}
  \toprule
Model & Dataset & Group 1 & Group 2 & Significant & Magnitude \\ 
  \midrule
$M_L$ & $V_A$ & 10 & 50 & TRUE & negligible \\ 
  $M_L$ & $V_A$ & 10 & 100 & TRUE & large \\ 
  $M_L$ & $V_A$ & 50 & 100 & TRUE & small \\ 
  $M_L$ & $V_B$ & 10 & 50 & TRUE & medium \\ 
  $M_L$ & $V_B$ & 10 & 100 & TRUE & large \\ 
  $M_L$ & $V_B$ & 50 & 100 & TRUE & small \\ 
  $M_L$ & $V_C$ & 10 & 50 & TRUE & medium \\ 
  $M_L$ & $V_C$ & 10 & 100 & TRUE & large \\ 
  $M_L$ & $V_C$ & 50 & 100 & TRUE & small \\ 
  $M_B$ & $V_A$ & 10 & 50 & TRUE & small \\ 
  $M_B$ & $V_A$ & 10 & 100 & TRUE & small \\ 
  $M_B$ & $V_A$ & 50 & 100 & TRUE & small \\ 
  $M_B$ & $V_B$ & 10 & 50 & TRUE & small \\ 
  $M_B$ & $V_B$ & 10 & 100 & TRUE & small \\ 
  $M_B$ & $V_B$ & 50 & 100 & TRUE & small \\ 
  $M_B$ & $V_C$ & 10 & 50 & TRUE & small \\ 
  $M_B$ & $V_C$ & 10 & 100 & TRUE & negligible \\ 
  $M_B$ & $V_C$ & 50 & 100 & TRUE & small \\ 
  $M_T$ & $V_A$ & 10 & 50 & TRUE & negligible \\ 
  $M_T$ & $V_A$ & 10 & 100 & TRUE & small \\ 
  $M_T$ & $V_A$ & 50 & 100 & TRUE & medium \\ 
  $M_T$ & $V_B$ & 10 & 50 & TRUE & negligible \\ 
  $M_T$ & $V_B$ & 10 & 100 & TRUE & medium \\ 
  $M_T$ & $V_B$ & 50 & 100 & TRUE & medium \\ 
  $M_T$ & $V_C$ & 10 & 50 & TRUE & negligible \\ 
  $M_T$ & $V_C$ & 10 & 100 & TRUE & medium \\ 
  $M_T$ & $V_C$ & 50 & 100 & TRUE & medium \\ 
  $M_M$ & $V_A$ & 10 & 50 & TRUE & negligible \\ 
  $M_M$ & $V_A$ & 10 & 100 & TRUE & large \\ 
  $M_M$ & $V_A$ & 50 & 100 & TRUE & medium \\ 
  $M_M$ & $V_B$ & 10 & 50 & TRUE & negligible \\ 
  $M_M$ & $V_B$ & 10 & 100 & TRUE & medium \\ 
  $M_M$ & $V_B$ & 50 & 100 & TRUE & medium \\ 
  $M_M$ & $V_C$ & 10 & 50 & TRUE & small \\ 
  $M_M$ & $V_C$ & 10 & 100 & TRUE & small \\ 
  $M_M$ & $V_C$ & 50 & 100 & TRUE & small \\ 
   \bottomrule
\end{tabular}
\end{table*}

\FloatBarrier
\subsubsection{Training Dataset Trends}

\begin{figure}[htbp]
    \centering
    \includegraphics[width=\linewidth]{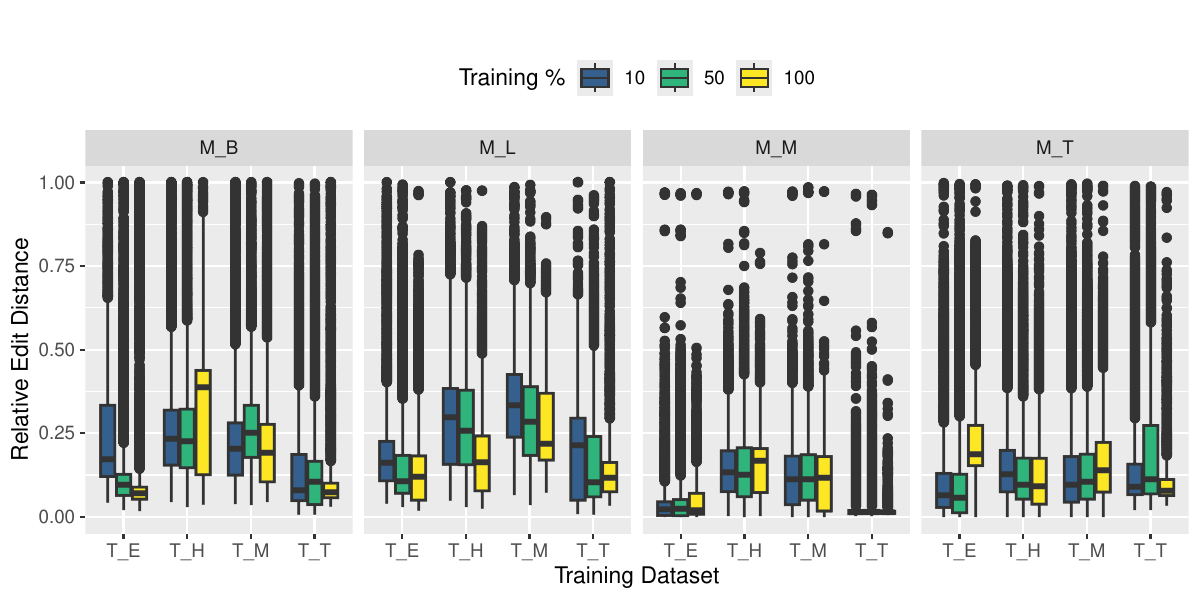}
    \caption{Box-plots of the relative edit distance by training dataset and training data percentage for each model architecture.}
    \label{fig:rq3-boxplot-max_obs_pct_trainset}
\end{figure}

Tables~\ref{tab:rq3_training_set_descriptive}~and~\ref{tab:rq3_max_obs_pct_significance_per_training} and Figure~\ref{fig:rq3-boxplot-max_obs_pct_trainset} may suggest that training datasets are disproportionately impacted by the percentage of data used. For example, both $M_B$ trained on $T_H$ and $M_T$ trained on $T_E$ exhibit an increased median $D_R$ with 100\% of data.

\begin{table*}[htbp]
\centering
\caption{Descriptive statistics measured in relative edit distance $D_R$ and the number of outliers across training datasets, maximum observation percentages, and model architectures.} 
\label{tab:rq3_training_set_descriptive}
\resizebox{0.9\textwidth}{!}{
\begin{tabular}{lrlrrrrr|rr}
  \toprule
Model & Training \% & Dataset & Mean & Median & Q1 & Q3 & Std. Dev. & \# Outliers & \% \\ 
  \midrule
$M_B$ & 10 & $T_E$ & 0.24 & 0.17 & 0.12 & 0.33 & 0.16 & 1033 & 0.97 \\ 
  $M_B$ & 50 & $T_E$ & 0.12 & 0.10 & 0.07 & 0.13 & 0.10 & 10225 & 9.62 \\ 
  $M_B$ & 100 & $T_E$ & 0.15 & 0.07 & 0.05 & 0.09 & 0.22 & 4148 & 19.51 \\ 
  $M_B$ & 10 & $T_H$ & 0.27 & 0.24 & 0.16 & 0.32 & 0.15 & 4984 & 4.69 \\ 
  $M_B$ & 50 & $T_H$ & 0.27 & 0.23 & 0.15 & 0.32 & 0.19 & 7064 & 6.65 \\ 
  $M_B$ & 100 & $T_H$ & 0.34 & 0.39 & 0.13 & 0.44 & 0.18 &  32 & 0.15 \\ 
  $M_B$ & 10 & $T_M$ & 0.23 & 0.20 & 0.13 & 0.28 & 0.13 & 3322 & 3.13 \\ 
  $M_B$ & 50 & $T_M$ & 0.27 & 0.25 & 0.18 & 0.33 & 0.13 & 3227 & 3.04 \\ 
  $M_B$ & 100 & $T_M$ & 0.24 & 0.19 & 0.11 & 0.28 & 0.17 & 1276 & 6.00 \\ 
  $M_B$ & 10 & $T_T$ & 0.14 & 0.08 & 0.05 & 0.19 & 0.14 & 12337 & 11.61 \\ 
  $M_B$ & 50 & $T_T$ & 0.13 & 0.11 & 0.04 & 0.17 & 0.12 & 3239 & 3.05 \\ 
  $M_B$ & 100 & $T_T$ & 0.15 & 0.07 & 0.06 & 0.10 & 0.20 & 4190 & 19.71 \\ 
  $M_L$ & 10 & $T_E$ & 0.20 & 0.16 & 0.11 & 0.23 & 0.12 & 5008 & 4.71 \\ 
  $M_L$ & 50 & $T_E$ & 0.15 & 0.11 & 0.07 & 0.18 & 0.12 & 6746 & 6.35 \\ 
  $M_L$ & 100 & $T_E$ & 0.14 & 0.12 & 0.05 & 0.18 & 0.12 & 1129 & 5.31 \\ 
  $M_L$ & 10 & $T_H$ & 0.29 & 0.30 & 0.16 & 0.38 & 0.15 & 696 & 0.65 \\ 
  $M_L$ & 50 & $T_H$ & 0.28 & 0.26 & 0.16 & 0.38 & 0.15 & 486 & 0.46 \\ 
  $M_L$ & 100 & $T_H$ & 0.19 & 0.16 & 0.08 & 0.24 & 0.12 & 457 & 2.15 \\ 
  $M_L$ & 10 & $T_M$ & 0.34 & 0.33 & 0.24 & 0.43 & 0.14 & 1495 & 1.41 \\ 
  $M_L$ & 50 & $T_M$ & 0.30 & 0.28 & 0.18 & 0.39 & 0.14 & 789 & 0.74 \\ 
  $M_L$ & 100 & $T_M$ & 0.26 & 0.22 & 0.17 & 0.37 & 0.14 & 210 & 0.99 \\ 
  $M_L$ & 10 & $T_T$ & 0.20 & 0.22 & 0.05 & 0.30 & 0.14 & 374 & 0.35 \\ 
  $M_L$ & 50 & $T_T$ & 0.15 & 0.11 & 0.06 & 0.24 & 0.12 & 1370 & 1.29 \\ 
  $M_L$ & 100 & $T_T$ & 0.15 & 0.12 & 0.08 & 0.16 & 0.14 & 1843 & 8.67 \\ 
  $M_M$ & 10 & $T_E$ & 0.04 & 0.02 & 0.01 & 0.05 & 0.07 & 8089 & 7.61 \\ 
  $M_M$ & 50 & $T_E$ & 0.04 & 0.02 & 0.00 & 0.05 & 0.07 & 8499 & 8.00 \\ 
  $M_M$ & 100 & $T_E$ & 0.05 & 0.02 & 0.01 & 0.07 & 0.07 & 905 & 4.26 \\ 
  $M_M$ & 10 & $T_H$ & 0.15 & 0.13 & 0.08 & 0.20 & 0.09 & 2288 & 2.15 \\ 
  $M_M$ & 50 & $T_H$ & 0.15 & 0.13 & 0.06 & 0.21 & 0.10 & 497 & 0.47 \\ 
  $M_M$ & 100 & $T_H$ & 0.15 & 0.17 & 0.07 & 0.21 & 0.09 & 150 & 0.71 \\ 
  $M_M$ & 10 & $T_M$ & 0.13 & 0.11 & 0.04 & 0.18 & 0.09 & 389 & 0.37 \\ 
  $M_M$ & 50 & $T_M$ & 0.13 & 0.11 & 0.05 & 0.19 & 0.09 & 495 & 0.47 \\ 
  $M_M$ & 100 & $T_M$ & 0.12 & 0.12 & 0.02 & 0.18 & 0.09 &  28 & 0.13 \\ 
  $M_M$ & 10 & $T_T$ & 0.02 & 0.02 & 0.01 & 0.02 & 0.03 & 9881 & 9.30 \\ 
  $M_M$ & 50 & $T_T$ & 0.02 & 0.02 & 0.01 & 0.02 & 0.04 & 10416 & 9.80 \\ 
  $M_M$ & 100 & $T_T$ & 0.02 & 0.01 & 0.01 & 0.02 & 0.03 & 2408 & 11.33 \\ 
  $M_T$ & 10 & $T_E$ & 0.11 & 0.07 & 0.03 & 0.13 & 0.12 & 8669 & 8.16 \\ 
  $M_T$ & 50 & $T_E$ & 0.09 & 0.06 & 0.01 & 0.13 & 0.12 & 5813 & 5.47 \\ 
  $M_T$ & 100 & $T_E$ & 0.24 & 0.19 & 0.15 & 0.27 & 0.16 & 2124 & 9.99 \\ 
  $M_T$ & 10 & $T_H$ & 0.16 & 0.13 & 0.08 & 0.20 & 0.14 & 7327 & 6.89 \\ 
  $M_T$ & 50 & $T_H$ & 0.13 & 0.10 & 0.05 & 0.18 & 0.12 & 5572 & 5.24 \\ 
  $M_T$ & 100 & $T_H$ & 0.12 & 0.09 & 0.04 & 0.18 & 0.12 & 718 & 3.38 \\ 
  $M_T$ & 10 & $T_M$ & 0.15 & 0.10 & 0.05 & 0.18 & 0.17 & 12420 & 11.69 \\ 
  $M_T$ & 50 & $T_M$ & 0.14 & 0.11 & 0.05 & 0.19 & 0.14 & 7818 & 7.36 \\ 
  $M_T$ & 100 & $T_M$ & 0.18 & 0.14 & 0.08 & 0.22 & 0.17 & 2569 & 12.09 \\ 
  $M_T$ & 10 & $T_T$ & 0.19 & 0.09 & 0.07 & 0.16 & 0.24 & 17272 & 16.25 \\ 
  $M_T$ & 50 & $T_T$ & 0.21 & 0.11 & 0.07 & 0.27 & 0.21 & 7745 & 7.29 \\ 
  $M_T$ & 100 & $T_T$ & 0.11 & 0.08 & 0.06 & 0.11 & 0.08 & 1898 & 8.93 \\ 
   \bottomrule
\end{tabular}
}
\end{table*}

\begin{table*}[htbp]
\centering
\caption{Wilcoxon test results for max observations percentage by training dataset for each model architecture.} 
\label{tab:rq3_max_obs_pct_significance_per_training}
\resizebox{0.65\textwidth}{!}{
\begin{tabular}{llrrcc}
  \toprule
Model & Dataset & Group 1 & Group 2 & Significant & Magnitude \\ 
  \midrule
$M_L$ & $T_T$ & 10 & 50 & TRUE & small \\ 
  $M_L$ & $T_T$ & 10 & 100 & TRUE & large \\ 
  $M_L$ & $T_T$ & 50 & 100 & TRUE & medium \\ 
  $M_L$ & $T_E$ & 10 & 50 & TRUE & medium \\ 
  $M_L$ & $T_E$ & 10 & 100 & TRUE & large \\ 
  $M_L$ & $T_E$ & 50 & 100 & FALSE & negligible \\ 
  $M_L$ & $T_M$ & 10 & 50 & TRUE & medium \\ 
  $M_L$ & $T_M$ & 10 & 100 & TRUE & large \\ 
  $M_L$ & $T_M$ & 50 & 100 & TRUE & small \\ 
  $M_L$ & $T_H$ & 10 & 50 & TRUE & small \\ 
  $M_L$ & $T_H$ & 10 & 100 & TRUE & large \\ 
  $M_L$ & $T_H$ & 50 & 100 & TRUE & large \\ 
  $M_B$ & $T_T$ & 10 & 50 & TRUE & small \\ 
  $M_B$ & $T_T$ & 10 & 100 & TRUE & small \\ 
  $M_B$ & $T_T$ & 50 & 100 & TRUE & negligible \\ 
  $M_B$ & $T_E$ & 10 & 50 & TRUE & large \\ 
  $M_B$ & $T_E$ & 10 & 100 & TRUE & large \\ 
  $M_B$ & $T_E$ & 50 & 100 & TRUE & medium \\ 
  $M_B$ & $T_M$ & 10 & 50 & TRUE & medium \\ 
  $M_B$ & $T_M$ & 10 & 100 & TRUE & medium \\ 
  $M_B$ & $T_M$ & 50 & 100 & TRUE & large \\ 
  $M_B$ & $T_H$ & 10 & 50 & TRUE & small \\ 
  $M_B$ & $T_H$ & 10 & 100 & TRUE & large \\ 
  $M_B$ & $T_H$ & 50 & 100 & TRUE & large \\ 
  $M_T$ & $T_T$ & 10 & 50 & TRUE & small \\ 
  $M_T$ & $T_T$ & 10 & 100 & TRUE & small \\ 
  $M_T$ & $T_T$ & 50 & 100 & TRUE & medium \\ 
  $M_T$ & $T_E$ & 10 & 50 & TRUE & small \\ 
  $M_T$ & $T_E$ & 10 & 100 & TRUE & large \\ 
  $M_T$ & $T_E$ & 50 & 100 & TRUE & large \\ 
  $M_T$ & $T_M$ & 10 & 50 & TRUE & negligible \\ 
  $M_T$ & $T_M$ & 10 & 100 & TRUE & medium \\ 
  $M_T$ & $T_M$ & 50 & 100 & TRUE & large \\ 
  $M_T$ & $T_H$ & 10 & 50 & TRUE & small \\ 
  $M_T$ & $T_H$ & 10 & 100 & TRUE & medium \\ 
  $M_T$ & $T_H$ & 50 & 100 & TRUE & small \\ 
  $M_M$ & $T_T$ & 10 & 50 & TRUE & small \\ 
  $M_M$ & $T_T$ & 10 & 100 & TRUE & small \\ 
  $M_M$ & $T_T$ & 50 & 100 & TRUE & small \\ 
  $M_M$ & $T_E$ & 10 & 50 & TRUE & negligible \\ 
  $M_M$ & $T_E$ & 10 & 100 & TRUE & small \\ 
  $M_M$ & $T_E$ & 50 & 100 & TRUE & small \\ 
  $M_M$ & $T_M$ & 10 & 50 & TRUE & negligible \\ 
  $M_M$ & $T_M$ & 10 & 100 & TRUE & small \\ 
  $M_M$ & $T_M$ & 50 & 100 & TRUE & negligible \\ 
  $M_M$ & $T_H$ & 10 & 50 & TRUE & negligible \\ 
  $M_M$ & $T_H$ & 10 & 100 & TRUE & small \\ 
  $M_M$ & $T_H$ & 50 & 100 & TRUE & medium \\ 
   \bottomrule
\end{tabular}
}
\end{table*}

\FloatBarrier
\subsection{Model Performance Extended Analysis} \label{appendix:rq4-extended-comparison}

\begin{figure}[htbp]
    \centering
    \includegraphics[width=\linewidth]{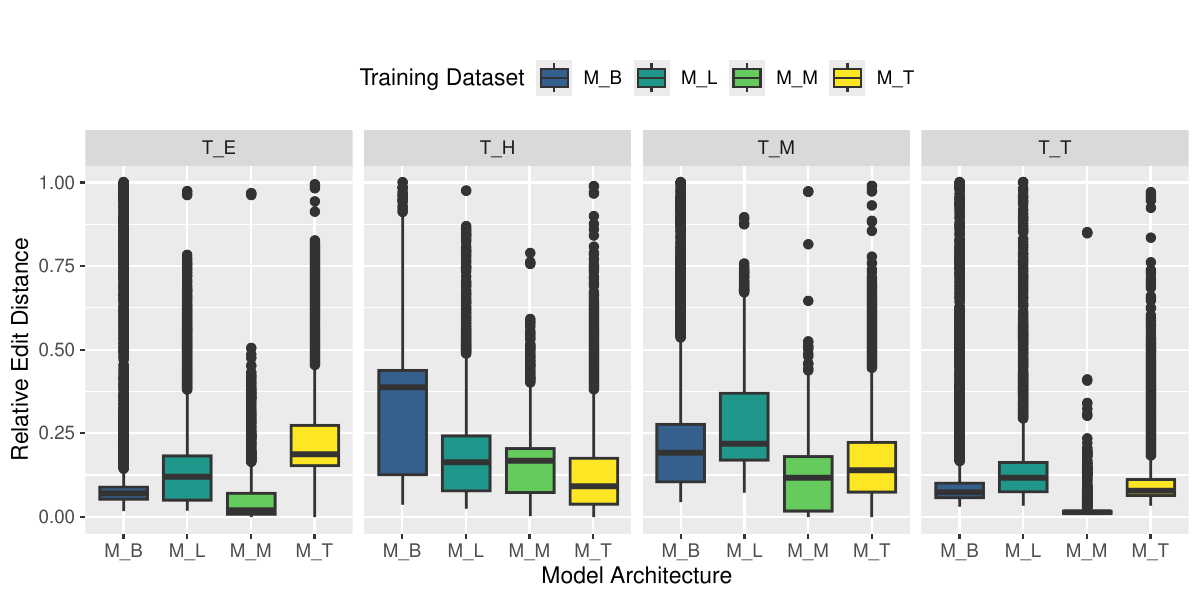}
    \caption{Box-plots of the relative edit distance by model architecture for each training dataset.}
    \label{fig:rq4-boxplot-arch-trainset}
\end{figure}

Figure~\ref{fig:rq4-boxplot-arch-trainset} shows the distribution of $D_R$ for each model architecture split by training dataset. This exhibits the differences in generalization capability of the model architectures under study. Table~\ref{tab:rq4_model_arch_significance_per_dataset} presents the significance and effect size of the difference in median $D_R$ between architectures split by the training dataset used.

\begin{table}[htbp]
\centering
\caption{Wilcoxon test results for model architecture comparisons by training dataset.} 
\label{tab:rq4_model_arch_significance_per_dataset}
\begin{tabular}{lllcc}
  \toprule
Dataset & Group 1 & Group 2 & Significant & Magnitude \\ 
  \midrule
$T_T$ & $M_L$ & $M_T$ & TRUE & medium \\ 
  $T_T$ & $M_L$ & $M_M$ & TRUE & large \\ 
  $T_T$ & $M_B$ & $M_L$ & TRUE & medium \\ 
  $T_T$ & $M_B$ & $M_T$ & TRUE & negligible \\ 
  $T_T$ & $M_B$ & $M_M$ & TRUE & large \\ 
  $T_T$ & $M_M$ & $M_T$ & TRUE & large \\ 
  $T_E$ & $M_L$ & $M_T$ & TRUE & large \\ 
  $T_E$ & $M_L$ & $M_M$ & TRUE & large \\ 
  $T_E$ & $M_B$ & $M_L$ & TRUE & medium \\ 
  $T_E$ & $M_B$ & $M_T$ & TRUE & large \\ 
  $T_E$ & $M_B$ & $M_M$ & TRUE & large \\ 
  $T_E$ & $M_M$ & $M_T$ & TRUE & large \\ 
  $T_M$ & $M_L$ & $M_T$ & TRUE & large \\ 
  $T_M$ & $M_L$ & $M_M$ & TRUE & large \\ 
  $T_M$ & $M_B$ & $M_L$ & TRUE & medium \\ 
  $T_M$ & $M_B$ & $M_T$ & TRUE & medium \\ 
  $T_M$ & $M_B$ & $M_M$ & TRUE & large \\ 
  $T_M$ & $M_M$ & $M_T$ & TRUE & negligible \\ 
  $T_H$ & $M_L$ & $M_T$ & TRUE & large \\ 
  $T_H$ & $M_L$ & $M_M$ & TRUE & large \\ 
  $T_H$ & $M_B$ & $M_L$ & TRUE & large \\ 
  $T_H$ & $M_B$ & $M_T$ & TRUE & large \\ 
  $T_H$ & $M_B$ & $M_M$ & TRUE & large \\ 
  $T_H$ & $M_M$ & $M_T$ & TRUE & medium \\ 
   \bottomrule
\end{tabular}
\end{table}

\subsection{Comprehensive Results} \label{appendix:full-results}

The following tables present comprehensive descriptive statistics of all model configurations evaluated in terms of both relative and absolute edit distance ($D_A$ and $D_R$, respectively):

\begin{table*}[htbp]
\caption{$M_L$, char Tokenization, 10\% of $T_x$}
\begin{adjustbox}{max width=\textwidth}
\begin{tabular}{@{}ll|rrrrrrrr|rrrrrrrr|rrrrrrrr@{}}
\toprule
\multicolumn{2}{c|}{Trained} & \multicolumn{8}{c|}{$V_A$} & \multicolumn{8}{c|}{$V_B$} & \multicolumn{8}{c}{$V_C$}
 \\ \multicolumn{2}{c|}{on} & min & avg & $q_{.50}$ & $q_{.75}$ & $q_{.90}$ & $q_{.95}$ & $q_{.99}$ & max & min & avg & $q_{.50}$ & $q_{.75}$ & $q_{.90}$ & $q_{.95}$ & $q_{.99}$ & max & min & avg & $q_{.50}$ & $q_{.75}$ & $q_{.90}$ & $q_{.95}$ & $q_{.99}$ & max
 \\ \midrule
\multirow{4}{*}{$D_A$}
& $T_T$ &   5 &  48 &  48 &  66 &  84 & 102 & 211 & 383 &   4 &  48 &  50 &  63 &  68 &  80 & 120 & 4166 &   2 &  55 &  58 &  79 &  94 & 119 & 262 & 1165 \\
& $T_E$ &  19 &  56 &  46 &  70 &  94 & 106 & 252 & 439 &  20 &  59 &  49 &  58 &  83 &  86 & 117 & 4200 &  14 &  39 &  29 &  43 &  74 &  99 & 152 & 1117 \\
& $T_M$ &  23 &  90 &  83 & 114 & 158 & 196 & 357 & 491 &  23 &  78 &  68 &  88 & 102 & 111 & 189 & 4303 &  19 &  92 &  93 & 115 & 133 & 149 & 234 & 1227 \\
& $T_H$ &  20 &  79 &  76 & 102 & 153 & 196 & 333 & 490 &  20 &  70 &  55 &  82 &  92 & 104 & 210 & 4373 &  16 &  77 &  81 & 106 & 119 & 129 & 191 & 1291 \\
\midrule
\multirow{4}{*}{$D_R$}
& $T_T$ & 0.02 & 0.17 & 0.20 & 0.25 & 0.30 & 0.33 & 0.47 & 0.64 & 0.02 & 0.17 & 0.21 & 0.27 & 0.31 & 0.34 & 0.41 & 0.95 & 0.01 & 0.23 & 0.27 & 0.34 & 0.43 & 0.48 & 0.65 & 1.00 \\
& $T_E$ & 0.06 & 0.20 & 0.17 & 0.27 & 0.33 & 0.38 & 0.53 & 0.73 & 0.07 & 0.22 & 0.20 & 0.23 & 0.36 & 0.37 & 0.42 & 0.96 & 0.04 & 0.18 & 0.13 & 0.20 & 0.35 & 0.52 & 0.83 & 1.00 \\
& $T_M$ & 0.10 & 0.32 & 0.30 & 0.41 & 0.52 & 0.59 & 0.71 & 0.82 & 0.10 & 0.29 & 0.29 & 0.37 & 0.42 & 0.43 & 0.59 & 0.99 & 0.07 & 0.39 & 0.39 & 0.47 & 0.55 & 0.62 & 0.75 & 0.97 \\
& $T_H$ & 0.08 & 0.27 & 0.26 & 0.37 & 0.53 & 0.60 & 0.71 & 0.82 & 0.07 & 0.26 & 0.23 & 0.34 & 0.39 & 0.40 & 0.62 & 1.00 & 0.05 & 0.33 & 0.33 & 0.41 & 0.50 & 0.59 & 0.69 & 0.95 \\
\bottomrule
\end{tabular}
\end{adjustbox}
\end{table*}

\begin{table*}[htbp]
\caption{$M_L$, char Tokenization, 50\% of $T_x$}
\begin{adjustbox}{max width=\textwidth}
\begin{tabular}{@{}ll|rrrrrrrr|rrrrrrrr|rrrrrrrr@{}}
\toprule
\multicolumn{2}{c|}{Trained} & \multicolumn{8}{c|}{$V_A$} & \multicolumn{8}{c|}{$V_B$} & \multicolumn{8}{c}{$V_C$}
 \\ \multicolumn{2}{c|}{on} & min & avg & $q_{.50}$ & $q_{.75}$ & $q_{.90}$ & $q_{.95}$ & $q_{.99}$ & max & min & avg & $q_{.50}$ & $q_{.75}$ & $q_{.90}$ & $q_{.95}$ & $q_{.99}$ & max & min & avg & $q_{.50}$ & $q_{.75}$ & $q_{.90}$ & $q_{.95}$ & $q_{.99}$ & max
 \\ \midrule
\multirow{4}{*}{$D_A$}
& $T_T$ &   5 &  43 &  31 &  57 &  82 & 105 & 283 & 404 &   4 &  40 &  25 &  49 &  60 &  77 & 140 & 4222 &   2 &  39 &  27 &  62 &  86 &  96 & 148 & 1170 \\
& $T_E$ &  11 &  56 &  46 &  63 & 104 & 145 & 309 & 491 &  11 &  42 &  26 &  46 &  58 &  63 & 135 & 4267 &   8 &  28 &  19 &  28 &  50 &  83 & 162 & 1288 \\
& $T_M$ &  17 &  87 &  81 & 113 & 159 & 198 & 315 & 491 &  15 &  68 &  52 &  81 &  96 & 115 & 204 & 4290 &  15 &  76 &  78 & 103 & 124 & 134 & 187 & 1278 \\
& $T_H$ &  16 &  87 &  72 & 118 & 169 & 196 & 347 & 545 &  15 &  62 &  42 &  88 &  93 &  99 & 181 & 4290 &  13 &  69 &  69 &  93 & 115 & 122 & 189 & 1281 \\
\midrule
\multirow{4}{*}{$D_R$}
& $T_T$ & 0.03 & 0.15 & 0.11 & 0.21 & 0.29 & 0.33 & 0.52 & 0.67 & 0.02 & 0.13 & 0.10 & 0.18 & 0.27 & 0.32 & 0.42 & 0.96 & 0.01 & 0.17 & 0.12 & 0.27 & 0.36 & 0.42 & 0.54 & 0.95 \\
& $T_E$ & 0.05 & 0.19 & 0.17 & 0.23 & 0.35 & 0.45 & 0.65 & 0.82 & 0.04 & 0.14 & 0.11 & 0.16 & 0.24 & 0.26 & 0.43 & 0.98 & 0.03 & 0.13 & 0.08 & 0.13 & 0.24 & 0.48 & 0.73 & 0.99 \\
& $T_M$ & 0.07 & 0.30 & 0.30 & 0.40 & 0.51 & 0.60 & 0.71 & 0.82 & 0.04 & 0.25 & 0.22 & 0.34 & 0.40 & 0.44 & 0.62 & 0.98 & 0.05 & 0.32 & 0.31 & 0.41 & 0.50 & 0.57 & 0.68 & 0.99 \\
& $T_H$ & 0.06 & 0.30 & 0.29 & 0.42 & 0.54 & 0.59 & 0.70 & 0.91 & 0.04 & 0.22 & 0.18 & 0.37 & 0.39 & 0.41 & 0.56 & 0.98 & 0.03 & 0.29 & 0.28 & 0.38 & 0.47 & 0.53 & 0.67 & 0.94 \\
\bottomrule
\end{tabular}
\end{adjustbox}
\end{table*}

\begin{table*}[htbp]
\caption{$M_L$, char Tokenization, 100\% of $T_x$}
\begin{adjustbox}{max width=\textwidth}
\begin{tabular}{@{}ll|rrrrrrrr|rrrrrrrr|rrrrrrrr@{}}
\toprule
\multicolumn{2}{c|}{Trained} & \multicolumn{8}{c|}{$V_A$} & \multicolumn{8}{c|}{$V_B$} & \multicolumn{8}{c}{$V_C$}
 \\ \multicolumn{2}{c|}{on} & min & avg & $q_{.50}$ & $q_{.75}$ & $q_{.90}$ & $q_{.95}$ & $q_{.99}$ & max & min & avg & $q_{.50}$ & $q_{.75}$ & $q_{.90}$ & $q_{.95}$ & $q_{.99}$ & max & min & avg & $q_{.50}$ & $q_{.75}$ & $q_{.90}$ & $q_{.95}$ & $q_{.99}$ & max
 \\ \midrule
\multirow{4}{*}{$D_A$}
& $T_T$ &  18 &  36 &  37 &  47 &  55 &  56 &  79 & 139 &  18 &  41 &  37 &  39 &  42 &  44 &  63 & 3950 &  11 &  35 &  18 &  44 &  77 & 112 & 147 & 565 \\
& $T_E$ &  24 &  65 &  54 &  66 &  91 & 143 & 333 & 460 &  15 &  41 &  29 &  32 &  44 &  75 & 137 & 4246 &   5 &  20 &  12 &  18 &  36 &  59 &  98 & 985 \\
& $T_M$ &  18 &  69 &  52 &  89 & 143 & 171 & 368 & 451 &  18 &  50 &  44 &  44 &  46 & 107 & 173 & 3938 &  14 &  83 &  84 & 112 & 134 & 151 & 207 & 1038 \\
& $T_H$ &  10 &  59 &  48 &  75 & 119 & 157 & 411 & 490 &  11 &  43 &  39 &  40 &  41 &  65 & 152 & 4288 &   9 &  49 &  54 &  73 &  80 &  89 & 121 & 1041 \\
\midrule
\multirow{4}{*}{$D_R$}
& $T_T$ & 0.06 & 0.13 & 0.13 & 0.16 & 0.17 & 0.18 & 0.25 & 0.30 & 0.06 & 0.14 & 0.15 & 0.16 & 0.17 & 0.18 & 0.19 & 0.90 & 0.03 & 0.17 & 0.07 & 0.22 & 0.43 & 0.55 & 1.00 & 1.00 \\
& $T_E$ & 0.09 & 0.23 & 0.19 & 0.23 & 0.34 & 0.50 & 0.66 & 0.77 & 0.05 & 0.14 & 0.12 & 0.13 & 0.15 & 0.25 & 0.45 & 0.97 & 0.02 & 0.09 & 0.05 & 0.08 & 0.19 & 0.41 & 0.56 & 0.78 \\
& $T_M$ & 0.07 & 0.23 & 0.20 & 0.31 & 0.45 & 0.51 & 0.69 & 0.75 & 0.08 & 0.17 & 0.18 & 0.18 & 0.20 & 0.37 & 0.52 & 0.90 & 0.08 & 0.33 & 0.34 & 0.42 & 0.47 & 0.51 & 0.67 & 0.76 \\
& $T_H$ & 0.04 & 0.20 & 0.17 & 0.24 & 0.43 & 0.50 & 0.74 & 0.87 & 0.03 & 0.15 & 0.16 & 0.16 & 0.18 & 0.22 & 0.50 & 0.97 & 0.03 & 0.20 & 0.21 & 0.28 & 0.34 & 0.41 & 0.44 & 0.76 \\
\bottomrule
\end{tabular}
\end{adjustbox}
\end{table*}

\begin{table*}[htbp]
\caption{$M_L$, word Tokenization, 10\% of $T_x$}
\begin{adjustbox}{max width=\textwidth}
\begin{tabular}{@{}ll|rrrrrrrr|rrrrrrrr|rrrrrrrr@{}}
\toprule
\multicolumn{2}{c|}{Trained} & \multicolumn{8}{c|}{$V_A$} & \multicolumn{8}{c|}{$V_B$} & \multicolumn{8}{c}{$V_C$}
 \\ \multicolumn{2}{c|}{on} & min & avg & $q_{.50}$ & $q_{.75}$ & $q_{.90}$ & $q_{.95}$ & $q_{.99}$ & max & min & avg & $q_{.50}$ & $q_{.75}$ & $q_{.90}$ & $q_{.95}$ & $q_{.99}$ & max & min & avg & $q_{.50}$ & $q_{.75}$ & $q_{.90}$ & $q_{.95}$ & $q_{.99}$ & max
 \\ \midrule
\multirow{4}{*}{$D_A$}
& $T_T$ &  24 & 111 & 108 & 125 & 140 & 157 & 288 & 465 &  23 & 102 &  95 & 116 & 124 & 128 & 164 & 4313 &  27 &  96 &  92 & 114 & 133 & 150 & 202 & 1214 \\
& $T_E$ &  28 &  86 &  71 & 101 & 158 & 194 & 329 & 502 &  32 &  68 &  57 &  63 &  73 & 105 & 215 & 4333 &  25 & 100 &  94 & 122 & 144 & 161 & 242 & 1216 \\
& $T_M$ &  12 &  74 &  61 &  88 & 128 & 168 & 391 & 532 &  17 &  65 &  49 &  77 &  83 &  94 & 178 & 4328 &  15 &  91 &  88 & 113 & 138 & 158 & 298 & 1283 \\
& $T_H$ &   9 &  74 &  66 & 105 & 144 & 180 & 368 & 498 &   7 &  69 &  56 &  78 & 115 & 135 & 206 & 4323 &   9 &  83 &  79 & 107 & 138 & 161 & 257 & 1320 \\
\midrule
\multirow{4}{*}{$D_R$}
& $T_T$ & 0.10 & 0.42 & 0.42 & 0.49 & 0.56 & 0.58 & 0.64 & 0.78 & 0.12 & 0.40 & 0.39 & 0.49 & 0.53 & 0.59 & 0.60 & 0.98 & 0.06 & 0.42 & 0.40 & 0.48 & 0.56 & 0.65 & 0.73 & 0.89 \\
& $T_E$ & 0.11 & 0.30 & 0.27 & 0.35 & 0.52 & 0.59 & 0.70 & 0.84 & 0.10 & 0.25 & 0.24 & 0.27 & 0.29 & 0.37 & 0.64 & 0.99 & 0.06 & 0.44 & 0.39 & 0.50 & 0.76 & 0.84 & 0.90 & 0.98 \\
& $T_M$ & 0.05 & 0.25 & 0.23 & 0.31 & 0.43 & 0.51 & 0.75 & 0.90 & 0.06 & 0.24 & 0.21 & 0.31 & 0.34 & 0.37 & 0.54 & 0.98 & 0.08 & 0.38 & 0.38 & 0.46 & 0.56 & 0.62 & 0.74 & 0.96 \\
& $T_H$ & 0.03 & 0.25 & 0.24 & 0.35 & 0.49 & 0.56 & 0.71 & 0.83 & 0.04 & 0.25 & 0.24 & 0.32 & 0.47 & 0.50 & 0.60 & 0.99 & 0.04 & 0.34 & 0.33 & 0.44 & 0.53 & 0.59 & 0.72 & 0.97 \\
\bottomrule
\end{tabular}
\end{adjustbox}
\end{table*}

\begin{table*}[htbp]
\caption{$M_L$, word Tokenization, 50\% of $T_x$}
\begin{adjustbox}{max width=\textwidth}
\begin{tabular}{@{}ll|rrrrrrrr|rrrrrrrr|rrrrrrrr@{}}
\toprule
\multicolumn{2}{c|}{Trained} & \multicolumn{8}{c|}{$V_A$} & \multicolumn{8}{c|}{$V_B$} & \multicolumn{8}{c}{$V_C$}
 \\ \multicolumn{2}{c|}{on} & min & avg & $q_{.50}$ & $q_{.75}$ & $q_{.90}$ & $q_{.95}$ & $q_{.99}$ & max & min & avg & $q_{.50}$ & $q_{.75}$ & $q_{.90}$ & $q_{.95}$ & $q_{.99}$ & max & min & avg & $q_{.50}$ & $q_{.75}$ & $q_{.90}$ & $q_{.95}$ & $q_{.99}$ & max
 \\ \midrule
\multirow{4}{*}{$D_A$}
& $T_T$ &  37 & 114 & 109 & 130 & 166 & 187 & 300 & 419 &  28 &  99 &  86 & 110 & 124 & 145 & 191 & 4315 &  17 & 102 &  99 & 124 & 145 & 166 & 202 & 1162 \\
& $T_E$ &  15 &  86 &  85 &  99 & 118 & 146 & 308 & 474 &  15 &  89 &  79 &  92 & 118 & 122 & 163 & 4309 &  25 &  97 &  94 & 117 & 143 & 155 & 183 & 1329 \\
& $T_M$ &   5 & 108 & 100 & 134 & 189 & 217 & 386 & 511 &   6 &  97 &  90 & 110 & 139 & 158 & 222 & 4241 &   2 & 108 & 106 & 139 & 169 & 193 & 281 & 1175 \\
& $T_H$ &   7 & 113 & 113 & 160 & 199 & 235 & 420 & 541 &   7 & 106 & 103 & 124 & 155 & 183 & 251 & 4332 &   2 & 112 & 116 & 147 & 176 & 201 & 286 & 1260 \\
\midrule
\multirow{4}{*}{$D_R$}
& $T_T$ & 0.13 & 0.42 & 0.44 & 0.50 & 0.54 & 0.57 & 0.64 & 0.74 & 0.13 & 0.38 & 0.37 & 0.47 & 0.51 & 0.52 & 0.58 & 0.98 & 0.08 & 0.43 & 0.43 & 0.50 & 0.55 & 0.58 & 0.64 & 0.99 \\
& $T_E$ & 0.06 & 0.32 & 0.31 & 0.40 & 0.45 & 0.50 & 0.65 & 0.79 & 0.06 & 0.34 & 0.32 & 0.42 & 0.49 & 0.51 & 0.51 & 0.98 & 0.05 & 0.42 & 0.41 & 0.48 & 0.55 & 0.60 & 0.71 & 0.98 \\
& $T_M$ & 0.02 & 0.38 & 0.39 & 0.50 & 0.61 & 0.66 & 0.76 & 0.85 & 0.03 & 0.36 & 0.38 & 0.46 & 0.55 & 0.58 & 0.67 & 0.97 & 0.02 & 0.44 & 0.45 & 0.54 & 0.63 & 0.69 & 0.79 & 0.88 \\
& $T_H$ & 0.03 & 0.40 & 0.43 & 0.55 & 0.68 & 0.73 & 0.82 & 0.90 & 0.03 & 0.40 & 0.44 & 0.52 & 0.63 & 0.66 & 0.76 & 0.99 & 0.02 & 0.46 & 0.49 & 0.59 & 0.68 & 0.73 & 0.78 & 0.92 \\
\bottomrule
\end{tabular}
\end{adjustbox}
\end{table*}

\begin{table*}[htbp]
\caption{$M_L$, word Tokenization, 100\% of $T_x$}
\begin{adjustbox}{max width=\textwidth}
\begin{tabular}{@{}ll|rrrrrrrr|rrrrrrrr|rrrrrrrr@{}}
\toprule
\multicolumn{2}{c|}{Trained} & \multicolumn{8}{c|}{$V_A$} & \multicolumn{8}{c|}{$V_B$} & \multicolumn{8}{c}{$V_C$}
 \\ \multicolumn{2}{c|}{on} & min & avg & $q_{.50}$ & $q_{.75}$ & $q_{.90}$ & $q_{.95}$ & $q_{.99}$ & max & min & avg & $q_{.50}$ & $q_{.75}$ & $q_{.90}$ & $q_{.95}$ & $q_{.99}$ & max & min & avg & $q_{.50}$ & $q_{.75}$ & $q_{.90}$ & $q_{.95}$ & $q_{.99}$ & max
 \\ \midrule
\multirow{4}{*}{$D_A$}
& $T_T$ &  32 & 135 & 124 & 149 & 175 & 189 & 254 & 333 &  36 & 119 & 103 & 116 & 141 & 169 & 221 & 4263 &  11 &  90 &  95 & 115 & 130 & 146 & 213 & 1054 \\
& $T_E$ &  18 &  98 &  94 & 123 & 152 & 173 & 480 & 558 &  27 & 108 & 102 & 123 & 130 & 137 & 188 & 4347 &  33 &  94 &  88 & 119 & 135 & 147 & 193 & 965 \\
& $T_M$ &   8 & 111 & 101 & 141 & 177 & 206 & 405 & 490 &   8 &  86 &  93 &  93 & 110 & 158 & 211 & 4138 &  14 &  92 &  94 & 117 & 138 & 155 & 214 & 1113 \\
& $T_H$ &   3 &  92 & 101 & 129 & 170 & 199 & 415 & 499 &   3 & 101 & 101 & 120 & 121 & 154 & 221 & 4294 &  11 & 108 & 113 & 151 & 177 & 197 & 219 & 1311 \\
\midrule
\multirow{4}{*}{$D_R$}
& $T_T$ & 0.16 & 0.51 & 0.51 & 0.54 & 0.56 & 0.58 & 0.62 & 0.68 & 0.17 & 0.47 & 0.45 & 0.52 & 0.54 & 0.56 & 0.63 & 0.97 & 0.05 & 0.36 & 0.38 & 0.43 & 0.48 & 0.52 & 0.65 & 0.82 \\
& $T_E$ & 0.07 & 0.34 & 0.35 & 0.43 & 0.49 & 0.54 & 0.90 & 0.93 & 0.10 & 0.42 & 0.42 & 0.52 & 0.55 & 0.55 & 0.59 & 0.99 & 0.09 & 0.40 & 0.38 & 0.46 & 0.58 & 0.71 & 0.80 & 0.94 \\
& $T_M$ & 0.03 & 0.40 & 0.40 & 0.52 & 0.57 & 0.62 & 0.76 & 0.82 & 0.04 & 0.32 & 0.39 & 0.39 & 0.42 & 0.53 & 0.65 & 0.94 & 0.07 & 0.37 & 0.37 & 0.46 & 0.55 & 0.64 & 0.65 & 0.81 \\
& $T_H$ & 0.01 & 0.32 & 0.37 & 0.47 & 0.56 & 0.63 & 0.77 & 0.83 & 0.01 & 0.38 & 0.43 & 0.50 & 0.50 & 0.54 & 0.67 & 0.98 & 0.05 & 0.43 & 0.46 & 0.58 & 0.63 & 0.66 & 0.69 & 0.96 \\
\bottomrule
\end{tabular}
\end{adjustbox}
\end{table*}

\begin{table*}[htbp]
\caption{$M_B$, char Tokenization, 10\% of $T_x$}
\begin{adjustbox}{max width=\textwidth}
\begin{tabular}{@{}ll|rrrrrrrr|rrrrrrrr|rrrrrrrr@{}}
\toprule
\multicolumn{2}{c|}{Trained} & \multicolumn{8}{c|}{$V_A$} & \multicolumn{8}{c|}{$V_B$} & \multicolumn{8}{c}{$V_C$}
 \\ \multicolumn{2}{c|}{on} & min & avg & $q_{.50}$ & $q_{.75}$ & $q_{.90}$ & $q_{.95}$ & $q_{.99}$ & max & min & avg & $q_{.50}$ & $q_{.75}$ & $q_{.90}$ & $q_{.95}$ & $q_{.99}$ & max & min & avg & $q_{.50}$ & $q_{.75}$ & $q_{.90}$ & $q_{.95}$ & $q_{.99}$ & max
 \\ \midrule
\multirow{4}{*}{$D_A$}
& $T_T$ &   8 &  57 &  42 &  82 & 128 & 152 & 208 & 351 &   7 &  55 &  39 &  60 & 109 & 112 & 150 & 4219 &   2 &  17 &  12 &  18 &  31 &  44 & 111 & 1066 \\
& $T_E$ &  19 &  80 &  65 & 109 & 146 & 177 & 277 & 497 &  18 &  76 &  49 & 103 & 131 & 134 & 192 & 4214 &  11 &  41 &  32 &  49 &  71 &  86 & 182 & 1245 \\
& $T_M$ &  18 &  61 &  53 &  74 &  98 & 133 & 353 & 560 &  17 &  53 &  47 &  51 &  55 &  90 & 215 & 4343 &  13 &  62 &  60 &  82 &  98 & 114 & 183 & 1192 \\
& $T_H$ &  19 &  73 &  62 &  86 & 148 & 181 & 357 & 566 &  20 &  63 &  52 &  56 &  88 & 154 & 230 & 4161 &  16 &  68 &  67 &  87 & 102 & 119 & 198 & 1143 \\
\midrule
\multirow{4}{*}{$D_R$}
& $T_T$ & 0.03 & 0.21 & 0.17 & 0.29 & 0.45 & 0.50 & 0.54 & 0.99 & 0.03 & 0.20 & 0.18 & 0.21 & 0.46 & 0.46 & 0.47 & 0.99 & 0.01 & 0.08 & 0.05 & 0.08 & 0.14 & 0.21 & 0.44 & 1.00 \\
& $T_E$ & 0.08 & 0.29 & 0.23 & 0.44 & 0.52 & 0.55 & 0.67 & 1.00 & 0.06 & 0.29 & 0.20 & 0.44 & 0.53 & 0.55 & 0.60 & 0.96 & 0.04 & 0.18 & 0.14 & 0.21 & 0.33 & 0.47 & 0.66 & 1.00 \\
& $T_M$ & 0.07 & 0.21 & 0.20 & 0.26 & 0.32 & 0.42 & 0.83 & 1.00 & 0.05 & 0.19 & 0.19 & 0.21 & 0.23 & 0.29 & 0.78 & 1.00 & 0.04 & 0.26 & 0.25 & 0.33 & 0.47 & 0.50 & 0.64 & 1.00 \\
& $T_H$ & 0.08 & 0.25 & 0.23 & 0.31 & 0.48 & 0.55 & 0.73 & 1.00 & 0.07 & 0.23 & 0.22 & 0.24 & 0.31 & 0.54 & 0.98 & 1.00 & 0.05 & 0.30 & 0.29 & 0.35 & 0.45 & 0.57 & 0.90 & 1.00 \\
\bottomrule
\end{tabular}
\end{adjustbox}
\end{table*}

\begin{table*}[htbp]
\caption{$M_B$, char Tokenization, 50\% of $T_x$}
\begin{adjustbox}{max width=\textwidth}
\begin{tabular}{@{}ll|rrrrrrrr|rrrrrrrr|rrrrrrrr@{}}
\toprule
\multicolumn{2}{c|}{Trained} & \multicolumn{8}{c|}{$V_A$} & \multicolumn{8}{c|}{$V_B$} & \multicolumn{8}{c}{$V_C$}
 \\ \multicolumn{2}{c|}{on} & min & avg & $q_{.50}$ & $q_{.75}$ & $q_{.90}$ & $q_{.95}$ & $q_{.99}$ & max & min & avg & $q_{.50}$ & $q_{.75}$ & $q_{.90}$ & $q_{.95}$ & $q_{.99}$ & max & min & avg & $q_{.50}$ & $q_{.75}$ & $q_{.90}$ & $q_{.95}$ & $q_{.99}$ & max
 \\ \midrule
\multirow{4}{*}{$D_A$}
& $T_T$ &  11 &  53 &  45 &  67 &  87 & 113 & 246 & 396 &   8 &  43 &  31 &  46 &  55 &  70 & 122 & 4219 &   2 &  17 &   8 &  16 &  37 &  70 & 170 & 1064 \\
& $T_E$ &  11 &  45 &  28 &  50 &  81 & 121 & 292 & 514 &  12 &  39 &  25 &  30 &  47 &  63 & 159 & 4279 &   7 &  22 &  16 &  24 &  40 &  54 &  79 & 1126 \\
& $T_M$ &  15 &  69 &  63 &  88 & 111 & 160 & 364 & 530 &  16 &  61 &  47 &  76 &  81 & 100 & 176 & 4291 &  13 &  73 &  72 &  92 & 119 & 135 & 176 & 1155 \\
& $T_H$ &  15 &  79 &  65 &  92 & 173 & 209 & 410 & 558 &  15 &  66 &  43 &  68 & 138 & 213 & 238 & 4395 &  14 &  67 &  66 &  84 & 117 & 130 & 209 & 1269 \\
\midrule
\multirow{4}{*}{$D_R$}
& $T_T$ & 0.04 & 0.19 & 0.16 & 0.25 & 0.29 & 0.35 & 0.50 & 0.96 & 0.03 & 0.15 & 0.13 & 0.18 & 0.23 & 0.24 & 0.38 & 0.96 & 0.01 & 0.08 & 0.04 & 0.07 & 0.15 & 0.30 & 0.90 & 1.00 \\
& $T_E$ & 0.04 & 0.15 & 0.11 & 0.18 & 0.27 & 0.36 & 0.62 & 1.00 & 0.03 & 0.13 & 0.11 & 0.13 & 0.20 & 0.22 & 0.50 & 1.00 & 0.02 & 0.10 & 0.07 & 0.11 & 0.20 & 0.27 & 0.47 & 1.00 \\
& $T_M$ & 0.07 & 0.24 & 0.22 & 0.31 & 0.37 & 0.52 & 0.76 & 1.00 & 0.04 & 0.22 & 0.20 & 0.30 & 0.34 & 0.41 & 0.57 & 1.00 & 0.04 & 0.31 & 0.30 & 0.38 & 0.48 & 0.53 & 0.72 & 1.00 \\
& $T_H$ & 0.06 & 0.27 & 0.23 & 0.33 & 0.56 & 0.63 & 0.90 & 1.00 & 0.03 & 0.24 & 0.18 & 0.25 & 0.51 & 0.92 & 1.00 & 1.00 & 0.04 & 0.29 & 0.26 & 0.35 & 0.47 & 0.59 & 1.00 & 1.00 \\
\bottomrule
\end{tabular}
\end{adjustbox}
\end{table*}

\begin{table*}[htbp]
\caption{$M_B$, char Tokenization, 100\% of $T_x$}
\begin{adjustbox}{max width=\textwidth}
\begin{tabular}{@{}ll|rrrrrrrr|rrrrrrrr|rrrrrrrr@{}}
\toprule
\multicolumn{2}{c|}{Trained} & \multicolumn{8}{c|}{$V_A$} & \multicolumn{8}{c|}{$V_B$} & \multicolumn{8}{c}{$V_C$}
 \\ \multicolumn{2}{c|}{on} & min & avg & $q_{.50}$ & $q_{.75}$ & $q_{.90}$ & $q_{.95}$ & $q_{.99}$ & max & min & avg & $q_{.50}$ & $q_{.75}$ & $q_{.90}$ & $q_{.95}$ & $q_{.99}$ & max & min & avg & $q_{.50}$ & $q_{.75}$ & $q_{.90}$ & $q_{.95}$ & $q_{.99}$ & max
 \\ \midrule
\multirow{4}{*}{$D_A$}
& $T_T$ &  12 &  41 &  18 &  52 & 102 & 118 & 299 & 365 &  11 &  31 &  19 &  20 &  25 &  72 & 178 & 4105 &   8 &  42 &  15 &  36 & 113 & 163 & 248 & 1146 \\
& $T_E$ &  12 &  47 &  20 &  55 &  73 & 212 & 399 & 528 &  11 &  49 &  20 &  20 & 148 & 214 & 238 & 4279 &   6 &  40 &  14 &  17 &  55 & 115 & 1363 & 1363 \\
& $T_M$ &  16 &  75 &  61 &  92 & 145 & 184 & 449 & 551 &  14 &  57 &  47 &  48 &  53 & 150 & 223 & 4288 &  12 &  63 &  62 &  79 &  92 & 129 & 333 & 1214 \\
& $T_H$ &  11 &  95 &  96 & 123 & 200 & 215 & 394 & 503 &  10 &  90 &  99 & 102 & 120 & 156 & 202 & 4329 &  11 &  86 & 107 & 122 & 130 & 147 & 199 & 775 \\
\midrule
\multirow{4}{*}{$D_R$}
& $T_T$ & 0.05 & 0.14 & 0.07 & 0.19 & 0.33 & 0.36 & 0.60 & 0.99 & 0.04 & 0.10 & 0.08 & 0.08 & 0.10 & 0.27 & 0.56 & 1.00 & 0.03 & 0.18 & 0.06 & 0.15 & 0.56 & 0.98 & 1.00 & 1.00 \\
& $T_E$ & 0.03 & 0.16 & 0.08 & 0.18 & 0.27 & 0.77 & 1.00 & 1.00 & 0.04 & 0.18 & 0.08 & 0.09 & 0.69 & 1.00 & 1.00 & 1.00 & 0.02 & 0.12 & 0.05 & 0.08 & 0.25 & 0.59 & 1.00 & 1.00 \\
& $T_M$ & 0.07 & 0.25 & 0.22 & 0.31 & 0.48 & 0.66 & 0.86 & 0.98 & 0.05 & 0.19 & 0.19 & 0.19 & 0.21 & 0.52 & 0.72 & 0.99 & 0.05 & 0.25 & 0.24 & 0.30 & 0.45 & 0.57 & 1.00 & 1.00 \\
& $T_H$ & 0.04 & 0.32 & 0.37 & 0.44 & 0.63 & 0.65 & 0.72 & 0.97 & 0.04 & 0.34 & 0.41 & 0.42 & 0.47 & 0.63 & 0.69 & 0.98 & 0.04 & 0.34 & 0.38 & 0.46 & 0.54 & 0.61 & 0.68 & 1.00 \\
\bottomrule
\end{tabular}
\end{adjustbox}
\end{table*}

\begin{table*}[htbp]
\caption{$M_B$, word Tokenization, 10\% of $T_x$}
\begin{adjustbox}{max width=\textwidth}
\begin{tabular}{@{}ll|rrrrrrrr|rrrrrrrr|rrrrrrrr@{}}
\toprule
\multicolumn{2}{c|}{Trained} & \multicolumn{8}{c|}{$V_A$} & \multicolumn{8}{c|}{$V_B$} & \multicolumn{8}{c}{$V_C$}
 \\ \multicolumn{2}{c|}{on} & min & avg & $q_{.50}$ & $q_{.75}$ & $q_{.90}$ & $q_{.95}$ & $q_{.99}$ & max & min & avg & $q_{.50}$ & $q_{.75}$ & $q_{.90}$ & $q_{.95}$ & $q_{.99}$ & max & min & avg & $q_{.50}$ & $q_{.75}$ & $q_{.90}$ & $q_{.95}$ & $q_{.99}$ & max
 \\ \midrule
\multirow{4}{*}{$D_A$}
& $T_T$ &  23 &  88 &  82 & 105 & 129 & 154 & 318 & 479 &  18 &  87 &  71 & 103 & 109 & 119 & 179 & 4282 &  24 & 110 & 107 & 139 & 165 & 179 & 203 & 1288 \\
& $T_E$ &  30 &  73 &  60 &  85 & 124 & 147 & 269 & 429 &  28 &  67 &  55 &  62 &  85 & 104 & 166 & 4334 &  19 &  96 &  91 & 117 & 142 & 155 & 217 & 1217 \\
& $T_M$ &  14 &  89 &  74 & 111 & 148 & 178 & 387 & 556 &  17 &  76 &  64 &  85 &  91 & 117 & 185 & 4316 &   9 &  93 &  87 & 116 & 152 & 172 & 278 & 1156 \\
& $T_H$ &  13 &  93 &  79 & 109 & 146 & 187 & 436 & 551 &  14 &  82 &  72 &  85 & 106 & 135 & 214 & 4336 &   6 &  88 &  82 & 107 & 142 & 171 & 274 & 1178 \\
\midrule
\multirow{4}{*}{$D_R$}
& $T_T$ & 0.08 & 0.32 & 0.31 & 0.41 & 0.49 & 0.53 & 0.63 & 0.80 & 0.08 & 0.33 & 0.32 & 0.43 & 0.46 & 0.47 & 0.56 & 0.98 & 0.08 & 0.47 & 0.48 & 0.54 & 0.58 & 0.61 & 0.66 & 0.94 \\
& $T_E$ & 0.12 & 0.26 & 0.23 & 0.29 & 0.40 & 0.45 & 0.58 & 0.74 & 0.09 & 0.25 & 0.24 & 0.27 & 0.35 & 0.36 & 0.51 & 0.99 & 0.10 & 0.42 & 0.40 & 0.49 & 0.61 & 0.70 & 0.87 & 0.93 \\
& $T_M$ & 0.06 & 0.31 & 0.29 & 0.39 & 0.48 & 0.54 & 0.81 & 0.97 & 0.07 & 0.28 & 0.28 & 0.35 & 0.38 & 0.40 & 0.55 & 0.98 & 0.06 & 0.39 & 0.39 & 0.48 & 0.57 & 0.63 & 0.73 & 0.93 \\
& $T_H$ & 0.07 & 0.33 & 0.31 & 0.40 & 0.49 & 0.57 & 0.82 & 0.93 & 0.07 & 0.30 & 0.30 & 0.36 & 0.44 & 0.47 & 0.64 & 0.99 & 0.04 & 0.36 & 0.35 & 0.44 & 0.54 & 0.61 & 0.77 & 0.99 \\
\bottomrule
\end{tabular}
\end{adjustbox}
\end{table*}

\begin{table*}[htbp]
\caption{$M_B$, word Tokenization, 50\% of $T_x$}
\begin{adjustbox}{max width=\textwidth}
\begin{tabular}{@{}ll|rrrrrrrr|rrrrrrrr|rrrrrrrr@{}}
\toprule
\multicolumn{2}{c|}{Trained} & \multicolumn{8}{c|}{$V_A$} & \multicolumn{8}{c|}{$V_B$} & \multicolumn{8}{c}{$V_C$}
 \\ \multicolumn{2}{c|}{on} & min & avg & $q_{.50}$ & $q_{.75}$ & $q_{.90}$ & $q_{.95}$ & $q_{.99}$ & max & min & avg & $q_{.50}$ & $q_{.75}$ & $q_{.90}$ & $q_{.95}$ & $q_{.99}$ & max & min & avg & $q_{.50}$ & $q_{.75}$ & $q_{.90}$ & $q_{.95}$ & $q_{.99}$ & max
 \\ \midrule
\multirow{4}{*}{$D_A$}
& $T_T$ &  34 & 125 & 108 & 145 & 195 & 241 & 346 & 545 &  17 & 110 &  96 & 119 & 154 & 169 & 256 & 4273 &  21 & 101 &  97 & 119 & 147 & 163 & 203 & 1146 \\
& $T_E$ &  20 & 100 &  95 & 116 & 138 & 158 & 317 & 510 &   8 &  97 &  91 & 109 & 119 & 131 & 185 & 4348 &  20 & 113 & 109 & 137 & 162 & 175 & 215 & 1135 \\
& $T_M$ &   6 &  89 &  77 & 120 & 175 & 214 & 431 & 547 &   6 &  78 &  66 &  81 & 108 & 158 & 229 & 4360 &   4 &  84 &  81 & 105 & 134 & 162 & 254 & 1246 \\
& $T_H$ &   5 &  98 &  92 & 134 & 180 & 215 & 409 & 531 &   2 &  86 &  78 &  99 & 123 & 161 & 240 & 4372 &   0 &  88 &  86 & 115 & 142 & 168 & 272 & 1256 \\
\midrule
\multirow{4}{*}{$D_R$}
& $T_T$ & 0.12 & 0.46 & 0.43 & 0.54 & 0.74 & 0.77 & 0.82 & 0.91 & 0.11 & 0.43 & 0.40 & 0.50 & 0.60 & 0.72 & 0.80 & 0.98 & 0.09 & 0.44 & 0.42 & 0.50 & 0.59 & 0.65 & 0.73 & 0.97 \\
& $T_E$ & 0.10 & 0.37 & 0.37 & 0.44 & 0.53 & 0.58 & 0.67 & 0.88 & 0.04 & 0.37 & 0.38 & 0.46 & 0.49 & 0.52 & 0.60 & 1.00 & 0.09 & 0.48 & 0.49 & 0.55 & 0.61 & 0.64 & 0.73 & 0.84 \\
& $T_M$ & 0.03 & 0.30 & 0.30 & 0.43 & 0.58 & 0.67 & 0.82 & 0.95 & 0.03 & 0.28 & 0.27 & 0.36 & 0.45 & 0.52 & 0.70 & 0.99 & 0.03 & 0.34 & 0.35 & 0.42 & 0.49 & 0.56 & 0.68 & 0.91 \\
& $T_H$ & 0.02 & 0.34 & 0.36 & 0.47 & 0.59 & 0.66 & 0.80 & 0.88 & 0.01 & 0.32 & 0.36 & 0.41 & 0.50 & 0.54 & 0.73 & 0.99 & 0.00 & 0.36 & 0.36 & 0.47 & 0.58 & 0.64 & 0.77 & 1.00 \\
\bottomrule
\end{tabular}
\end{adjustbox}
\end{table*}

\begin{table*}[htbp]
\caption{$M_B$, word Tokenization, 100\% of $T_x$}
\begin{adjustbox}{max width=\textwidth}
\begin{tabular}{@{}ll|rrrrrrrr|rrrrrrrr|rrrrrrrr@{}}
\toprule
\multicolumn{2}{c|}{Trained} & \multicolumn{8}{c|}{$V_A$} & \multicolumn{8}{c|}{$V_B$} & \multicolumn{8}{c}{$V_C$}
 \\ \multicolumn{2}{c|}{on} & min & avg & $q_{.50}$ & $q_{.75}$ & $q_{.90}$ & $q_{.95}$ & $q_{.99}$ & max & min & avg & $q_{.50}$ & $q_{.75}$ & $q_{.90}$ & $q_{.95}$ & $q_{.99}$ & max & min & avg & $q_{.50}$ & $q_{.75}$ & $q_{.90}$ & $q_{.95}$ & $q_{.99}$ & max
 \\ \midrule
\multirow{4}{*}{$D_A$}
& $T_T$ &  57 & 139 & 133 & 163 & 193 & 212 & 259 & 436 &  46 & 115 & 100 & 114 & 152 & 178 & 214 & 4272 &  13 &  78 &  71 & 101 & 132 & 153 & 238 & 857 \\
& $T_E$ &  38 & 129 & 117 & 155 & 210 & 257 & 438 & 526 &  32 & 113 & 109 & 122 & 142 & 184 & 267 & 4324 &  47 &  95 &  86 & 109 & 127 & 139 & 223 & 743 \\
& $T_M$ &   5 &  89 &  84 & 127 & 173 & 231 & 461 & 571 &   4 & 113 & 137 & 147 & 151 & 185 & 249 & 4186 &  11 &  94 &  88 & 115 & 146 & 178 & 293 & 1250 \\
& $T_H$ &  55 & 106 &  94 & 110 & 141 & 176 & 350 & 511 &  43 & 103 & 100 & 103 & 104 & 122 & 181 & 3935 &  42 & 125 & 116 & 149 & 181 & 200 & 322 & 1092 \\
\midrule
\multirow{4}{*}{$D_R$}
& $T_T$ & 0.25 & 0.52 & 0.54 & 0.58 & 0.62 & 0.66 & 0.75 & 1.00 & 0.27 & 0.45 & 0.42 & 0.45 & 0.56 & 0.61 & 0.79 & 1.00 & 0.06 & 0.31 & 0.30 & 0.40 & 0.48 & 0.52 & 0.72 & 1.00 \\
& $T_E$ & 0.16 & 0.46 & 0.47 & 0.55 & 0.65 & 0.79 & 0.90 & 0.99 & 0.17 & 0.43 & 0.45 & 0.51 & 0.53 & 0.60 & 0.85 & 0.98 & 0.09 & 0.40 & 0.38 & 0.46 & 0.54 & 0.66 & 0.76 & 0.94 \\
& $T_M$ & 0.02 & 0.30 & 0.32 & 0.44 & 0.56 & 0.69 & 0.90 & 0.96 & 0.02 & 0.43 & 0.54 & 0.62 & 0.62 & 0.62 & 0.74 & 0.95 & 0.06 & 0.37 & 0.37 & 0.44 & 0.51 & 0.59 & 0.73 & 1.00 \\
& $T_H$ & 0.22 & 0.39 & 0.39 & 0.43 & 0.47 & 0.54 & 0.70 & 0.89 & 0.18 & 0.40 & 0.42 & 0.43 & 0.44 & 0.44 & 0.56 & 0.90 & 0.22 & 0.50 & 0.51 & 0.56 & 0.64 & 0.67 & 0.75 & 0.91 \\
\bottomrule
\end{tabular}
\end{adjustbox}
\end{table*}

\begin{table*}[htbp]
\caption{$M_M$, char Tokenization, 10\% of $T_x$}
\begin{adjustbox}{max width=\textwidth}
\begin{tabular}{@{}ll|rrrrrrrr|rrrrrrrr|rrrrrrrr@{}}
\toprule
\multicolumn{2}{c|}{Trained} & \multicolumn{8}{c|}{$V_A$} & \multicolumn{8}{c|}{$V_B$} & \multicolumn{8}{c}{$V_C$}
 \\ \multicolumn{2}{c|}{on} & min & avg & $q_{.50}$ & $q_{.75}$ & $q_{.90}$ & $q_{.95}$ & $q_{.99}$ & max & min & avg & $q_{.50}$ & $q_{.75}$ & $q_{.90}$ & $q_{.95}$ & $q_{.99}$ & max & min & avg & $q_{.50}$ & $q_{.75}$ & $q_{.90}$ & $q_{.95}$ & $q_{.99}$ & max
 \\ \midrule
\multirow{4}{*}{$D_A$}
& $T_T$ &   2 &   7 &   5 &   6 &  10 &  17 &  52 &  70 &   2 &  11 &   4 &   4 &   6 &   8 &  16 & 4234 &   2 &   5 &   3 &   4 &   9 &  16 &  52 & 485 \\
& $T_E$ &   0 &  11 &   9 &  13 &  21 &  30 &  68 & 236 &   0 &  16 &   8 &  11 &  18 &  19 &  49 & 4262 &   0 &  10 &   3 &  13 &  31 &  50 &  62 & 813 \\
& $T_M$ &   0 &  30 &  31 &  46 &  54 &  62 & 168 & 293 &   0 &  28 &  25 &  26 &  27 &  33 &  77 & 4274 &   0 &  40 &  43 &  59 &  66 &  82 & 116 & 1056 \\
& $T_H$ &   2 &  38 &  37 &  52 &  67 &  78 & 135 & 325 &   1 &  32 &  28 &  31 &  37 &  48 & 109 & 4272 &   1 &  43 &  43 &  63 &  75 &  87 & 141 & 792 \\
\midrule
\multirow{4}{*}{$D_R$}
& $T_T$ & 0.01 & 0.02 & 0.02 & 0.02 & 0.03 & 0.06 & 0.18 & 0.25 & 0.00 & 0.02 & 0.02 & 0.02 & 0.03 & 0.03 & 0.05 & 0.97 & 0.00 & 0.02 & 0.01 & 0.02 & 0.05 & 0.08 & 0.18 & 0.56 \\
& $T_E$ & 0.00 & 0.04 & 0.03 & 0.05 & 0.07 & 0.09 & 0.20 & 0.40 & 0.00 & 0.04 & 0.03 & 0.05 & 0.07 & 0.08 & 0.15 & 0.97 & 0.00 & 0.05 & 0.01 & 0.05 & 0.15 & 0.28 & 0.36 & 0.60 \\
& $T_M$ & 0.00 & 0.10 & 0.11 & 0.15 & 0.18 & 0.22 & 0.32 & 0.50 & 0.00 & 0.08 & 0.11 & 0.11 & 0.11 & 0.12 & 0.23 & 0.97 & 0.00 & 0.16 & 0.18 & 0.22 & 0.29 & 0.35 & 0.38 & 0.77 \\
& $T_H$ & 0.01 & 0.13 & 0.14 & 0.18 & 0.23 & 0.26 & 0.35 & 0.58 & 0.00 & 0.10 & 0.12 & 0.13 & 0.15 & 0.17 & 0.33 & 0.97 & 0.00 & 0.18 & 0.18 & 0.23 & 0.30 & 0.37 & 0.43 & 0.82 \\
\bottomrule
\end{tabular}
\end{adjustbox}
\end{table*}

\begin{table*}[htbp]
\caption{$M_M$, char Tokenization, 50\% of $T_x$}
\begin{adjustbox}{max width=\textwidth}
\begin{tabular}{@{}ll|rrrrrrrr|rrrrrrrr|rrrrrrrr@{}}
\toprule
\multicolumn{2}{c|}{Trained} & \multicolumn{8}{c|}{$V_A$} & \multicolumn{8}{c|}{$V_B$} & \multicolumn{8}{c}{$V_C$}
 \\ \multicolumn{2}{c|}{on} & min & avg & $q_{.50}$ & $q_{.75}$ & $q_{.90}$ & $q_{.95}$ & $q_{.99}$ & max & min & avg & $q_{.50}$ & $q_{.75}$ & $q_{.90}$ & $q_{.95}$ & $q_{.99}$ & max & min & avg & $q_{.50}$ & $q_{.75}$ & $q_{.90}$ & $q_{.95}$ & $q_{.99}$ & max
 \\ \midrule
\multirow{4}{*}{$D_A$}
& $T_T$ &   2 &   6 &   5 &   6 &  10 &  19 &  41 &  61 &   2 &  12 &   4 &   5 &   6 &   9 &  20 & 4232 &   2 &   6 &   3 &   3 &  14 &  19 &  58 & 609 \\
& $T_E$ &   0 &  11 &   8 &  15 &  22 &  33 &  88 & 287 &   0 &  15 &   9 &  12 &  13 &  16 &  56 & 4244 &   0 &  13 &   4 &  16 &  38 &  57 &  73 & 956 \\
& $T_M$ &   1 &  37 &  35 &  49 &  64 &  84 & 180 & 312 &   0 &  27 &  24 &  26 &  27 &  37 &  91 & 4329 &   0 &  38 &  43 &  58 &  65 &  78 & 126 & 1083 \\
& $T_H$ &   1 &  37 &  36 &  51 &  69 &  90 & 168 & 393 &   1 &  33 &  26 &  29 &  47 &  48 &  98 & 4276 &   0 &  45 &  48 &  65 &  79 &  90 & 140 & 934 \\
\midrule
\multirow{4}{*}{$D_R$}
& $T_T$ & 0.01 & 0.02 & 0.02 & 0.02 & 0.03 & 0.06 & 0.09 & 0.22 & 0.00 & 0.02 & 0.02 & 0.02 & 0.02 & 0.03 & 0.06 & 0.96 & 0.01 & 0.03 & 0.01 & 0.02 & 0.06 & 0.10 & 0.26 & 0.58 \\
& $T_E$ & 0.00 & 0.04 & 0.03 & 0.05 & 0.07 & 0.10 & 0.23 & 0.50 & 0.00 & 0.03 & 0.04 & 0.05 & 0.05 & 0.06 & 0.17 & 0.97 & 0.00 & 0.06 & 0.02 & 0.07 & 0.17 & 0.27 & 0.35 & 0.70 \\
& $T_M$ & 0.00 & 0.13 & 0.13 & 0.17 & 0.22 & 0.27 & 0.39 & 0.55 & 0.00 & 0.08 & 0.10 & 0.11 & 0.11 & 0.12 & 0.28 & 0.98 & 0.00 & 0.15 & 0.17 & 0.22 & 0.28 & 0.32 & 0.37 & 0.81 \\
& $T_H$ & 0.00 & 0.13 & 0.13 & 0.18 & 0.23 & 0.29 & 0.41 & 0.68 & 0.00 & 0.10 & 0.11 & 0.12 & 0.19 & 0.20 & 0.30 & 0.97 & 0.00 & 0.18 & 0.20 & 0.25 & 0.32 & 0.37 & 0.41 & 0.82 \\
\bottomrule
\end{tabular}
\end{adjustbox}
\end{table*}

\begin{table*}[htbp]
\caption{$M_M$, char Tokenization, 100\% of $T_x$}
\begin{adjustbox}{max width=\textwidth}
\begin{tabular}{@{}ll|rrrrrrrr|rrrrrrrr|rrrrrrrr@{}}
\toprule
\multicolumn{2}{c|}{Trained} & \multicolumn{8}{c|}{$V_A$} & \multicolumn{8}{c|}{$V_B$} & \multicolumn{8}{c}{$V_C$}
 \\ \multicolumn{2}{c|}{on} & min & avg & $q_{.50}$ & $q_{.75}$ & $q_{.90}$ & $q_{.95}$ & $q_{.99}$ & max & min & avg & $q_{.50}$ & $q_{.75}$ & $q_{.90}$ & $q_{.95}$ & $q_{.99}$ & max & min & avg & $q_{.50}$ & $q_{.75}$ & $q_{.90}$ & $q_{.95}$ & $q_{.99}$ & max
 \\ \midrule
\multirow{4}{*}{$D_A$}
& $T_T$ &   2 &   6 &   5 &   6 &   9 &  19 &  27 &  52 &   2 &  11 &   4 &   4 &   5 &   7 &  16 & 3734 &   2 &   6 &   3 &   3 &  10 &  22 &  43 & 466 \\
& $T_E$ &   1 &  25 &  19 &  30 &  37 &  49 & 197 & 254 &   0 &  13 &   5 &   6 &   7 &  14 &  33 & 4243 &   0 &  10 &   3 &   7 &  28 &  59 &  78 & 514 \\
& $T_M$ &   2 &  29 &  32 &  48 &  54 &  64 &  97 & 115 &   0 &  29 &  28 &  28 &  28 &  33 &  82 & 4277 &   0 &  37 &  44 &  59 &  65 &  77 & 107 & 712 \\
& $T_H$ &   3 &  40 &  40 &  54 &  71 &  89 & 270 & 327 &   1 &  43 &  45 &  47 &  48 &  49 &  87 & 3350 &   1 &  41 &  47 &  63 &  68 &  81 & 125 & 677 \\
\midrule
\multirow{4}{*}{$D_R$}
& $T_T$ & 0.01 & 0.02 & 0.02 & 0.02 & 0.03 & 0.06 & 0.08 & 0.21 & 0.01 & 0.02 & 0.02 & 0.02 & 0.02 & 0.02 & 0.06 & 0.85 & 0.01 & 0.02 & 0.01 & 0.02 & 0.04 & 0.13 & 0.15 & 0.41 \\
& $T_E$ & 0.01 & 0.09 & 0.08 & 0.10 & 0.12 & 0.15 & 0.36 & 0.43 & 0.00 & 0.02 & 0.02 & 0.03 & 0.03 & 0.05 & 0.12 & 0.97 & 0.00 & 0.04 & 0.01 & 0.03 & 0.12 & 0.20 & 0.36 & 0.50 \\
& $T_M$ & 0.01 & 0.10 & 0.11 & 0.16 & 0.19 & 0.21 & 0.28 & 0.34 & 0.00 & 0.09 & 0.11 & 0.12 & 0.12 & 0.12 & 0.25 & 0.97 & 0.00 & 0.15 & 0.17 & 0.22 & 0.27 & 0.31 & 0.34 & 0.81 \\
& $T_H$ & 0.02 & 0.14 & 0.14 & 0.19 & 0.26 & 0.30 & 0.47 & 0.58 & 0.00 & 0.15 & 0.18 & 0.20 & 0.20 & 0.21 & 0.29 & 0.76 & 0.00 & 0.16 & 0.18 & 0.23 & 0.29 & 0.34 & 0.38 & 0.79 \\
\bottomrule
\end{tabular}
\end{adjustbox}
\end{table*}

\begin{table*}[htbp]
\caption{$M_M$, word Tokenization, 10\% of $T_x$}
\begin{adjustbox}{max width=\textwidth}
\begin{tabular}{@{}ll|rrrrrrrr|rrrrrrrr|rrrrrrrr@{}}
\toprule
\multicolumn{2}{c|}{Trained} & \multicolumn{8}{c|}{$V_A$} & \multicolumn{8}{c|}{$V_B$} & \multicolumn{8}{c}{$V_C$}
 \\ \multicolumn{2}{c|}{on} & min & avg & $q_{.50}$ & $q_{.75}$ & $q_{.90}$ & $q_{.95}$ & $q_{.99}$ & max & min & avg & $q_{.50}$ & $q_{.75}$ & $q_{.90}$ & $q_{.95}$ & $q_{.99}$ & max & min & avg & $q_{.50}$ & $q_{.75}$ & $q_{.90}$ & $q_{.95}$ & $q_{.99}$ & max
 \\ \midrule
\multirow{4}{*}{$D_A$}
& $T_T$ &   2 &  17 &   9 &  16 &  41 &  64 &  97 & 255 &   2 &  20 &   8 &  10 &  33 &  36 &  99 & 4374 &   2 &  22 &  11 &  34 &  55 &  70 &  97 & 458 \\
& $T_E$ &   1 &  41 &  34 &  51 &  78 & 101 & 191 & 305 &   0 &  28 &  19 &  28 &  34 &  43 &  92 & 4374 &   0 &  43 &  39 &  54 &  76 &  91 & 127 & 966 \\
& $T_M$ &   2 &  50 &  41 &  68 &  99 & 116 & 223 & 338 &   0 &  38 &  24 &  38 &  53 &  84 & 123 & 4374 &   0 &  56 &  53 &  76 & 101 & 114 & 153 & 943 \\
& $T_H$ &   6 &  51 &  43 &  63 &  87 & 106 & 200 & 327 &   2 &  42 &  34 &  48 &  59 &  80 & 116 & 4374 &   0 &  58 &  55 &  78 & 101 & 116 & 148 & 1008 \\
\midrule
\multirow{4}{*}{$D_R$}
& $T_T$ & 0.01 & 0.06 & 0.04 & 0.06 & 0.17 & 0.22 & 0.29 & 0.43 & 0.01 & 0.05 & 0.03 & 0.04 & 0.14 & 0.15 & 0.32 & 0.99 & 0.00 & 0.10 & 0.04 & 0.16 & 0.28 & 0.32 & 0.47 & 0.66 \\
& $T_E$ & 0.00 & 0.14 & 0.13 & 0.18 & 0.26 & 0.31 & 0.39 & 0.52 & 0.00 & 0.09 & 0.08 & 0.12 & 0.14 & 0.15 & 0.28 & 0.99 & 0.00 & 0.18 & 0.16 & 0.22 & 0.32 & 0.38 & 0.47 & 0.71 \\
& $T_M$ & 0.01 & 0.17 & 0.16 & 0.24 & 0.31 & 0.35 & 0.45 & 0.56 & 0.00 & 0.12 & 0.11 & 0.16 & 0.20 & 0.28 & 0.36 & 0.99 & 0.00 & 0.23 & 0.23 & 0.31 & 0.37 & 0.44 & 0.52 & 0.69 \\
& $T_H$ & 0.02 & 0.18 & 0.17 & 0.22 & 0.28 & 0.33 & 0.42 & 0.54 & 0.01 & 0.14 & 0.14 & 0.20 & 0.22 & 0.27 & 0.34 & 0.99 & 0.00 & 0.24 & 0.24 & 0.30 & 0.37 & 0.42 & 0.52 & 0.74 \\
\bottomrule
\end{tabular}
\end{adjustbox}
\end{table*}

\begin{table*}[htbp]
\caption{$M_M$, word Tokenization, 50\% of $T_x$}
\begin{adjustbox}{max width=\textwidth}
\begin{tabular}{@{}ll|rrrrrrrr|rrrrrrrr|rrrrrrrr@{}}
\toprule
\multicolumn{2}{c|}{Trained} & \multicolumn{8}{c|}{$V_A$} & \multicolumn{8}{c|}{$V_B$} & \multicolumn{8}{c}{$V_C$}
 \\ \multicolumn{2}{c|}{on} & min & avg & $q_{.50}$ & $q_{.75}$ & $q_{.90}$ & $q_{.95}$ & $q_{.99}$ & max & min & avg & $q_{.50}$ & $q_{.75}$ & $q_{.90}$ & $q_{.95}$ & $q_{.99}$ & max & min & avg & $q_{.50}$ & $q_{.75}$ & $q_{.90}$ & $q_{.95}$ & $q_{.99}$ & max
 \\ \midrule
\multirow{4}{*}{$D_A$}
& $T_T$ &   6 &  31 &  13 &  46 &  72 &  90 & 146 & 307 &   2 &  30 &  14 &  31 &  45 &  66 & 119 & 4374 &   2 &  25 &  15 &  38 &  59 &  72 & 102 & 289 \\
& $T_E$ &   6 &  40 &  32 &  46 &  74 &  91 & 221 & 288 &   0 &  36 &  27 &  32 &  45 &  70 & 111 & 4374 &   0 &  41 &  35 &  56 &  73 &  87 & 140 & 854 \\
& $T_M$ &   0 &  46 &  39 &  61 &  94 & 112 & 223 & 360 &   0 &  31 &  18 &  29 &  45 &  71 & 114 & 4381 &   0 &  53 &  51 &  70 &  95 & 109 & 142 & 811 \\
& $T_H$ &   4 &  40 &  35 &  50 &  73 &  89 & 220 & 306 &   0 &  31 &  21 &  29 &  34 &  56 & 104 & 4374 &   0 &  49 &  49 &  64 &  84 &  96 & 127 & 615 \\
\midrule
\multirow{4}{*}{$D_R$}
& $T_T$ & 0.02 & 0.11 & 0.05 & 0.19 & 0.26 & 0.29 & 0.37 & 0.52 & 0.01 & 0.10 & 0.06 & 0.14 & 0.20 & 0.24 & 0.37 & 0.99 & 0.00 & 0.11 & 0.07 & 0.18 & 0.27 & 0.33 & 0.48 & 0.67 \\
& $T_E$ & 0.02 & 0.14 & 0.13 & 0.17 & 0.24 & 0.28 & 0.41 & 0.49 & 0.00 & 0.12 & 0.11 & 0.13 & 0.17 & 0.24 & 0.33 & 0.99 & 0.00 & 0.17 & 0.15 & 0.23 & 0.31 & 0.41 & 0.49 & 0.68 \\
& $T_M$ & 0.00 & 0.16 & 0.14 & 0.22 & 0.30 & 0.34 & 0.44 & 0.60 & 0.00 & 0.10 & 0.08 & 0.12 & 0.17 & 0.24 & 0.35 & 1.00 & 0.00 & 0.22 & 0.21 & 0.28 & 0.36 & 0.40 & 0.49 & 0.70 \\
& $T_H$ & 0.02 & 0.14 & 0.14 & 0.18 & 0.24 & 0.28 & 0.41 & 0.51 & 0.00 & 0.10 & 0.09 & 0.12 & 0.14 & 0.19 & 0.31 & 0.99 & 0.00 & 0.20 & 0.20 & 0.26 & 0.32 & 0.37 & 0.46 & 0.68 \\
\bottomrule
\end{tabular}
\end{adjustbox}
\end{table*}

\begin{table*}[htbp]
\caption{$M_M$, word Tokenization, 100\% of $T_x$}
\begin{adjustbox}{max width=\textwidth}
\begin{tabular}{@{}ll|rrrrrrrr|rrrrrrrr|rrrrrrrr@{}}
\toprule
\multicolumn{2}{c|}{Trained} & \multicolumn{8}{c|}{$V_A$} & \multicolumn{8}{c|}{$V_B$} & \multicolumn{8}{c}{$V_C$}
 \\ \multicolumn{2}{c|}{on} & min & avg & $q_{.50}$ & $q_{.75}$ & $q_{.90}$ & $q_{.95}$ & $q_{.99}$ & max & min & avg & $q_{.50}$ & $q_{.75}$ & $q_{.90}$ & $q_{.95}$ & $q_{.99}$ & max & min & avg & $q_{.50}$ & $q_{.75}$ & $q_{.90}$ & $q_{.95}$ & $q_{.99}$ & max
 \\ \midrule
\multirow{4}{*}{$D_A$}
& $T_T$ &  36 &  80 &  70 &  92 & 118 & 132 & 280 & 332 &  27 &  66 &  53 &  54 &  92 & 116 & 150 & 4374 &   8 &  49 &  45 &  67 &  81 &  94 & 151 & 466 \\
& $T_E$ &   6 &  30 &  22 &  31 &  64 &  73 & 235 & 277 &   0 &  18 &  10 &  10 &  18 &  44 &  85 & 4374 &   0 &  32 &  25 &  46 &  62 &  77 & 143 & 591 \\
& $T_M$ &   6 &  31 &  27 &  42 &  51 &  66 & 212 & 257 &   0 &  23 &  18 &  18 &  21 &  39 &  90 & 4374 &   0 &  44 &  47 &  58 &  74 &  88 & 111 & 569 \\
& $T_H$ &  10 &  73 &  62 &  97 & 123 & 143 & 272 & 337 &   6 &  59 &  50 &  51 &  87 & 111 & 144 & 4345 &   4 &  78 &  75 & 105 & 124 & 136 & 172 & 938 \\
\midrule
\multirow{4}{*}{$D_R$}
& $T_T$ & 0.14 & 0.29 & 0.29 & 0.32 & 0.37 & 0.39 & 0.53 & 0.55 & 0.08 & 0.24 & 0.23 & 0.25 & 0.32 & 0.38 & 0.43 & 0.99 & 0.03 & 0.22 & 0.19 & 0.28 & 0.40 & 0.47 & 0.57 & 0.69 \\
& $T_E$ & 0.02 & 0.10 & 0.09 & 0.12 & 0.20 & 0.24 & 0.43 & 0.49 & 0.00 & 0.04 & 0.04 & 0.04 & 0.06 & 0.15 & 0.26 & 0.99 & 0.00 & 0.13 & 0.11 & 0.19 & 0.27 & 0.31 & 0.50 & 0.66 \\
& $T_M$ & 0.02 & 0.10 & 0.10 & 0.14 & 0.17 & 0.25 & 0.39 & 0.44 & 0.00 & 0.06 & 0.08 & 0.08 & 0.09 & 0.14 & 0.27 & 0.99 & 0.00 & 0.18 & 0.19 & 0.23 & 0.28 & 0.33 & 0.44 & 0.66 \\
& $T_H$ & 0.05 & 0.25 & 0.24 & 0.32 & 0.38 & 0.41 & 0.51 & 0.56 & 0.03 & 0.21 & 0.21 & 0.21 & 0.30 & 0.36 & 0.41 & 0.99 & 0.03 & 0.33 & 0.32 & 0.39 & 0.48 & 0.49 & 0.61 & 0.70 \\
\bottomrule
\end{tabular}
\end{adjustbox}
\end{table*}

\begin{table*}[htbp]
\caption{$M_T$, word Tokenization, Sequence Length of 256, and 10\% of $T_x$}
\begin{adjustbox}{max width=\textwidth}
\begin{tabular}{@{}ll|rrrrrrrr|rrrrrrrr|rrrrrrrr@{}}
\toprule
\multicolumn{2}{c|}{Trained} & \multicolumn{8}{c|}{$V_A$} & \multicolumn{8}{c|}{$V_B$} & \multicolumn{8}{c}{$V_C$}
 \\ \multicolumn{2}{c|}{on} & min & avg & $q_{.50}$ & $q_{.75}$ & $q_{.90}$ & $q_{.95}$ & $q_{.99}$ & max & min & avg & $q_{.50}$ & $q_{.75}$ & $q_{.90}$ & $q_{.95}$ & $q_{.99}$ & max & min & avg & $q_{.50}$ & $q_{.75}$ & $q_{.90}$ & $q_{.95}$ & $q_{.99}$ & max
 \\ \midrule
\multirow{4}{*}{$D_A$}
& $T_T$ &  11 &  57 &  20 &  56 & 203 & 217 & 252 & 426 &  11 &  63 &  17 &  39 & 206 & 229 & 246 & 4321 &  21 &  32 &  24 &  35 &  51 &  65 & 107 & 1162 \\
& $T_E$ &   0 &  27 &  12 &  28 &  61 & 129 & 229 & 440 &   0 &  24 &  11 &  15 &  26 &  50 & 183 & 4350 &   0 &  33 &  26 &  44 &  62 &  84 & 173 & 1353 \\
& $T_M$ &   0 &  32 &  19 &  37 &  75 & 148 & 189 & 371 &   0 &  39 &  18 &  23 & 137 & 137 & 160 & 4339 &   0 &  49 &  37 &  54 & 120 & 169 & 219 & 1321 \\
& $T_H$ &   0 &  39 &  32 &  45 &  76 & 120 & 227 & 445 &   0 &  32 &  21 &  22 &  41 &  82 & 155 & 4338 &   0 &  50 &  44 &  58 &  87 & 140 & 225 & 1296 \\
\midrule
\multirow{4}{*}{$D_R$}
& $T_T$ & 0.04 & 0.23 & 0.07 & 0.20 & 0.95 & 0.96 & 0.96 & 0.97 & 0.03 & 0.24 & 0.07 & 0.13 & 0.96 & 0.96 & 0.97 & 0.99 & 0.02 & 0.15 & 0.11 & 0.16 & 0.27 & 0.39 & 0.51 & 0.93 \\
& $T_E$ & 0.00 & 0.09 & 0.05 & 0.10 & 0.21 & 0.36 & 0.61 & 0.78 & 0.00 & 0.06 & 0.05 & 0.06 & 0.10 & 0.17 & 0.60 & 1.00 & 0.00 & 0.14 & 0.11 & 0.18 & 0.30 & 0.39 & 0.58 & 0.99 \\
& $T_M$ & 0.00 & 0.11 & 0.07 & 0.12 & 0.23 & 0.50 & 0.62 & 0.82 & 0.00 & 0.13 & 0.08 & 0.10 & 0.57 & 0.58 & 0.60 & 0.99 & 0.00 & 0.19 & 0.15 & 0.24 & 0.47 & 0.57 & 0.65 & 0.98 \\
& $T_H$ & 0.00 & 0.13 & 0.12 & 0.18 & 0.26 & 0.37 & 0.57 & 0.75 & 0.00 & 0.10 & 0.09 & 0.09 & 0.17 & 0.26 & 0.65 & 0.99 & 0.00 & 0.20 & 0.18 & 0.25 & 0.38 & 0.51 & 0.66 & 0.97 \\
\bottomrule
\end{tabular}
\end{adjustbox}
\end{table*}

\begin{table*}[htbp]
\caption{$M_T$, word Tokenization, Sequence Length of 256, and 50\% of $T_x$}
\begin{adjustbox}{max width=\textwidth}
\begin{tabular}{@{}ll|rrrrrrrr|rrrrrrrr|rrrrrrrr@{}}
\toprule
\multicolumn{2}{c|}{Trained} & \multicolumn{8}{c|}{$V_A$} & \multicolumn{8}{c|}{$V_B$} & \multicolumn{8}{c}{$V_C$}
 \\ \multicolumn{2}{c|}{on} & min & avg & $q_{.50}$ & $q_{.75}$ & $q_{.90}$ & $q_{.95}$ & $q_{.99}$ & max & min & avg & $q_{.50}$ & $q_{.75}$ & $q_{.90}$ & $q_{.95}$ & $q_{.99}$ & max & min & avg & $q_{.50}$ & $q_{.75}$ & $q_{.90}$ & $q_{.95}$ & $q_{.99}$ & max
 \\ \midrule
\multirow{4}{*}{$D_A$}
& $T_T$ &  11 &  71 &  32 & 111 & 198 & 220 & 281 & 593 &   8 &  61 &  16 &  80 & 129 & 207 & 270 & 4302 &  21 &  38 &  25 &  45 &  70 & 101 & 172 & 1157 \\
& $T_E$ &   0 &  24 &  11 &  30 &  50 &  96 & 221 & 480 &   0 &  25 &  10 &  15 &  27 &  70 & 173 & 4342 &   0 &  29 &  21 &  41 &  59 &  82 & 191 & 1275 \\
& $T_M$ &   0 &  34 &  23 &  39 &  76 & 155 & 228 & 414 &   0 &  31 &  18 &  21 &  39 & 131 & 188 & 4349 &   0 &  48 &  43 &  56 &  83 & 138 & 208 & 1214 \\
& $T_H$ &   0 &  28 &  23 &  37 &  52 &  82 & 191 & 382 &   0 &  29 &  19 &  21 &  35 &  74 & 144 & 4330 &   0 &  45 &  39 &  54 &  78 & 115 & 206 & 1282 \\
\midrule
\multirow{4}{*}{$D_R$}
& $T_T$ & 0.03 & 0.26 & 0.10 & 0.44 & 0.65 & 0.97 & 0.97 & 0.99 & 0.04 & 0.22 & 0.07 & 0.32 & 0.54 & 0.95 & 0.97 & 0.98 & 0.02 & 0.17 & 0.12 & 0.20 & 0.39 & 0.50 & 0.70 & 0.95 \\
& $T_E$ & 0.00 & 0.08 & 0.04 & 0.10 & 0.20 & 0.30 & 0.69 & 0.89 & 0.00 & 0.07 & 0.05 & 0.06 & 0.09 & 0.22 & 0.65 & 0.99 & 0.00 & 0.12 & 0.09 & 0.17 & 0.28 & 0.35 & 0.52 & 0.95 \\
& $T_M$ & 0.00 & 0.11 & 0.08 & 0.13 & 0.23 & 0.54 & 0.64 & 0.79 & 0.00 & 0.09 & 0.08 & 0.09 & 0.13 & 0.49 & 0.62 & 1.00 & 0.00 & 0.19 & 0.17 & 0.23 & 0.34 & 0.49 & 0.62 & 0.96 \\
& $T_H$ & 0.00 & 0.09 & 0.08 & 0.13 & 0.18 & 0.26 & 0.59 & 0.74 & 0.00 & 0.08 & 0.08 & 0.09 & 0.12 & 0.24 & 0.60 & 0.99 & 0.00 & 0.18 & 0.17 & 0.23 & 0.32 & 0.44 & 0.57 & 0.96 \\
\bottomrule
\end{tabular}
\end{adjustbox}
\end{table*}

\begin{table*}[htbp]
\caption{$M_T$, word Tokenization, Sequence Length of 256, and 100\% of $T_x$}
\begin{adjustbox}{max width=\textwidth}
\begin{tabular}{@{}ll|rrrrrrrr|rrrrrrrr|rrrrrrrr@{}}
\toprule
\multicolumn{2}{c|}{Trained} & \multicolumn{8}{c|}{$V_A$} & \multicolumn{8}{c|}{$V_B$} & \multicolumn{8}{c}{$V_C$}
 \\ \multicolumn{2}{c|}{on} & min & avg & $q_{.50}$ & $q_{.75}$ & $q_{.90}$ & $q_{.95}$ & $q_{.99}$ & max & min & avg & $q_{.50}$ & $q_{.75}$ & $q_{.90}$ & $q_{.95}$ & $q_{.99}$ & max & min & avg & $q_{.50}$ & $q_{.75}$ & $q_{.90}$ & $q_{.95}$ & $q_{.99}$ & max
 \\ \midrule
\multirow{4}{*}{$D_A$}
& $T_T$ &  14 &  24 &  15 &  22 &  43 &  60 & 173 & 297 &  12 &  25 &  16 &  16 &  20 &  36 &  87 & 4237 &  21 &  31 &  24 &  33 &  57 &  69 &  92 & 1137 \\
& $T_E$ &  13 &  68 &  49 &  80 & 112 & 176 & 283 & 468 &   9 &  66 &  39 &  46 & 138 & 183 & 253 & 4341 &   0 &  58 &  47 &  66 & 133 & 183 & 228 & 1243 \\
& $T_M$ &   0 &  47 &  26 &  51 & 152 & 162 & 249 & 368 &   0 &  48 &  18 &  27 & 139 & 140 & 173 & 4318 &   0 &  51 &  47 &  62 &  83 & 120 & 235 & 1202 \\
& $T_H$ &   0 &  28 &  15 &  37 &  76 & 115 & 172 & 419 &   0 &  31 &  18 &  19 &  58 &  93 & 139 & 4346 &   0 &  41 &  38 &  54 &  66 &  87 & 212 & 1193 \\
\midrule
\multirow{4}{*}{$D_R$}
& $T_T$ & 0.03 & 0.08 & 0.06 & 0.08 & 0.16 & 0.18 & 0.31 & 0.70 & 0.04 & 0.07 & 0.07 & 0.07 & 0.09 & 0.12 & 0.26 & 0.97 & 0.04 & 0.14 & 0.11 & 0.13 & 0.32 & 0.42 & 0.47 & 0.92 \\
& $T_E$ & 0.07 & 0.24 & 0.21 & 0.27 & 0.33 & 0.52 & 0.74 & 0.81 & 0.05 & 0.24 & 0.16 & 0.19 & 0.63 & 0.73 & 0.77 & 0.99 & 0.00 & 0.24 & 0.20 & 0.29 & 0.48 & 0.70 & 0.78 & 0.91 \\
& $T_M$ & 0.00 & 0.15 & 0.11 & 0.18 & 0.50 & 0.56 & 0.61 & 0.70 & 0.00 & 0.16 & 0.08 & 0.09 & 0.59 & 0.60 & 0.60 & 0.99 & 0.00 & 0.20 & 0.19 & 0.24 & 0.36 & 0.47 & 0.61 & 0.97 \\
& $T_H$ & 0.00 & 0.09 & 0.06 & 0.12 & 0.25 & 0.31 & 0.59 & 0.72 & 0.00 & 0.09 & 0.08 & 0.08 & 0.20 & 0.29 & 0.40 & 0.99 & 0.00 & 0.16 & 0.16 & 0.21 & 0.29 & 0.35 & 0.50 & 0.90 \\
\bottomrule
\end{tabular}
\end{adjustbox}
\end{table*}

\begin{table*}[htbp]
\caption{$M_T$, word Tokenization, Sequence Length of 756, and 10\% of $T_x$}
\begin{adjustbox}{max width=\textwidth}
\begin{tabular}{@{}ll|rrrrrrrr|rrrrrrrr|rrrrrrrr@{}}
\toprule
\multicolumn{2}{c|}{Trained} & \multicolumn{8}{c|}{$V_A$} & \multicolumn{8}{c|}{$V_B$} & \multicolumn{8}{c}{$V_C$}
 \\ \multicolumn{2}{c|}{on} & min & avg & $q_{.50}$ & $q_{.75}$ & $q_{.90}$ & $q_{.95}$ & $q_{.99}$ & max & min & avg & $q_{.50}$ & $q_{.75}$ & $q_{.90}$ & $q_{.95}$ & $q_{.99}$ & max & min & avg & $q_{.50}$ & $q_{.75}$ & $q_{.90}$ & $q_{.95}$ & $q_{.99}$ & max
 \\ \midrule
\multirow{4}{*}{$D_A$}
& $T_T$ &   8 & 161 & 184 & 215 & 257 & 287 & 387 & 589 &   5 & 159 & 175 & 218 & 227 & 241 & 313 & 4166 &   4 &  29 &  21 &  25 &  54 &  70 & 105 & 1171 \\
& $T_E$ &   0 &  21 &   8 &  14 &  38 & 147 & 196 & 349 &   0 &  20 &   7 &  10 &  14 &  34 & 178 & 4385 &   0 &  39 &  24 &  47 &  95 & 152 & 220 & 1262 \\
& $T_M$ &   0 &  31 &  19 &  37 &  81 & 148 & 181 & 385 &   0 &  38 &  18 &  19 & 137 & 137 & 144 & 4378 &   0 &  54 &  40 &  60 & 146 & 176 & 222 & 1190 \\
& $T_H$ &   0 &  30 &  19 &  38 &  59 &  84 & 216 & 420 &   0 &  38 &  19 &  23 & 136 & 140 & 144 & 4377 &   0 &  49 &  42 &  57 &  97 & 140 & 205 & 1288 \\
\midrule
\multirow{4}{*}{$D_R$}
& $T_T$ & 0.02 & 0.61 & 0.64 & 0.94 & 0.96 & 0.96 & 0.97 & 0.98 & 0.01 & 0.64 & 0.74 & 0.95 & 0.95 & 0.95 & 0.96 & 0.98 & 0.01 & 0.14 & 0.10 & 0.13 & 0.30 & 0.41 & 0.52 & 0.89 \\
& $T_E$ & 0.00 & 0.06 & 0.03 & 0.05 & 0.13 & 0.45 & 0.57 & 0.70 & 0.00 & 0.05 & 0.03 & 0.04 & 0.06 & 0.12 & 0.58 & 1.00 & 0.00 & 0.16 & 0.10 & 0.21 & 0.43 & 0.56 & 0.68 & 0.93 \\
& $T_M$ & 0.00 & 0.10 & 0.07 & 0.12 & 0.24 & 0.50 & 0.59 & 0.74 & 0.00 & 0.13 & 0.06 & 0.08 & 0.57 & 0.58 & 0.62 & 1.00 & 0.00 & 0.21 & 0.17 & 0.28 & 0.54 & 0.60 & 0.68 & 0.87 \\
& $T_H$ & 0.00 & 0.10 & 0.07 & 0.13 & 0.20 & 0.26 & 0.59 & 0.80 & 0.00 & 0.12 & 0.08 & 0.10 & 0.56 & 0.57 & 0.60 & 1.00 & 0.00 & 0.20 & 0.17 & 0.24 & 0.38 & 0.56 & 0.67 & 0.94 \\
\bottomrule
\end{tabular}
\end{adjustbox}
\end{table*}

\begin{table*}[htbp]
\caption{$M_T$, word Tokenization, Sequence Length of 756, and 50\% of $T_x$}
\begin{adjustbox}{max width=\textwidth}
\begin{tabular}{@{}ll|rrrrrrrr|rrrrrrrr|rrrrrrrr@{}}
\toprule
\multicolumn{2}{c|}{Trained} & \multicolumn{8}{c|}{$V_A$} & \multicolumn{8}{c|}{$V_B$} & \multicolumn{8}{c}{$V_C$}
 \\ \multicolumn{2}{c|}{on} & min & avg & $q_{.50}$ & $q_{.75}$ & $q_{.90}$ & $q_{.95}$ & $q_{.99}$ & max & min & avg & $q_{.50}$ & $q_{.75}$ & $q_{.90}$ & $q_{.95}$ & $q_{.99}$ & max & min & avg & $q_{.50}$ & $q_{.75}$ & $q_{.90}$ & $q_{.95}$ & $q_{.99}$ & max
 \\ \midrule
\multirow{4}{*}{$D_A$}
& $T_T$ &   7 &  90 &  83 & 129 & 175 & 211 & 342 & 559 &   7 &  88 &  52 & 131 & 180 & 185 & 245 & 4323 &   5 &  57 &  38 &  71 & 121 & 152 & 227 & 1158 \\
& $T_E$ &   0 &  15 &   7 &  15 &  37 &  54 & 172 & 403 &   0 &  18 &   7 &  11 &  18 &  36 & 144 & 4391 &   0 &  25 &  18 &  32 &  51 &  74 & 180 & 1263 \\
& $T_M$ &   0 &  23 &  18 &  33 &  45 &  65 & 170 & 360 &   0 &  24 &  18 &  18 &  21 &  40 & 159 & 4384 &   0 &  45 &  38 &  54 &  77 & 135 & 214 & 1249 \\
& $T_H$ &   0 &  29 &  23 &  39 &  56 &  98 & 198 & 404 &   0 &  26 &  18 &  22 &  24 &  46 & 121 & 4368 &   0 &  46 &  43 &  58 &  79 & 115 & 207 & 1289 \\
\midrule
\multirow{4}{*}{$D_R$}
& $T_T$ & 0.02 & 0.32 & 0.32 & 0.46 & 0.63 & 0.68 & 0.82 & 0.98 & 0.02 & 0.33 & 0.22 & 0.55 & 0.76 & 0.78 & 0.87 & 0.99 & 0.02 & 0.25 & 0.18 & 0.36 & 0.52 & 0.61 & 0.77 & 0.91 \\
& $T_E$ & 0.00 & 0.05 & 0.02 & 0.05 & 0.12 & 0.18 & 0.44 & 0.76 & 0.00 & 0.04 & 0.03 & 0.05 & 0.08 & 0.12 & 0.49 & 1.00 & 0.00 & 0.10 & 0.07 & 0.14 & 0.27 & 0.32 & 0.51 & 0.93 \\
& $T_M$ & 0.00 & 0.07 & 0.06 & 0.11 & 0.15 & 0.20 & 0.53 & 0.73 & 0.00 & 0.07 & 0.07 & 0.08 & 0.09 & 0.13 & 0.56 & 1.00 & 0.00 & 0.18 & 0.17 & 0.22 & 0.33 & 0.46 & 0.57 & 0.92 \\
& $T_H$ & 0.00 & 0.10 & 0.09 & 0.13 & 0.19 & 0.30 & 0.53 & 0.78 & 0.00 & 0.07 & 0.08 & 0.09 & 0.10 & 0.16 & 0.35 & 0.99 & 0.00 & 0.18 & 0.18 & 0.23 & 0.33 & 0.41 & 0.55 & 0.97 \\
\bottomrule
\end{tabular}
\end{adjustbox}
\end{table*}

\begin{table*}[htbp]
\caption{$M_T$, word Tokenization, Sequence Length of 756, and 100\% of $T_x$}
\begin{adjustbox}{max width=\textwidth}
\begin{tabular}{@{}ll|rrrrrrrr|rrrrrrrr|rrrrrrrr@{}}
\toprule
\multicolumn{2}{c|}{Trained} & \multicolumn{8}{c|}{$V_A$} & \multicolumn{8}{c|}{$V_B$} & \multicolumn{8}{c}{$V_C$}
 \\ \multicolumn{2}{c|}{on} & min & avg & $q_{.50}$ & $q_{.75}$ & $q_{.90}$ & $q_{.95}$ & $q_{.99}$ & max & min & avg & $q_{.50}$ & $q_{.75}$ & $q_{.90}$ & $q_{.95}$ & $q_{.99}$ & max & min & avg & $q_{.50}$ & $q_{.75}$ & $q_{.90}$ & $q_{.95}$ & $q_{.99}$ & max
 \\ \midrule
\multirow{4}{*}{$D_A$}
& $T_T$ & 140 & 251 & 239 & 281 & 314 & 341 & 448 & 520 &  75 & 234 & 225 & 226 & 273 & 305 & 344 & 4356 &   4 &  27 &  21 &  27 &  35 &  55 &  79 & 926 \\
& $T_E$ &   0 &  25 &  18 &  33 &  44 &  62 & 285 & 375 &   0 &  26 &  19 &  19 &  21 &  48 & 180 & 4174 &   0 &  28 &  22 &  37 &  51 &  72 & 183 & 1100 \\
& $T_M$ &   0 &  25 &  19 &  37 &  56 &  80 & 156 & 419 &   0 &  29 &  19 &  22 &  49 &  80 & 120 & 4310 &   0 &  41 &  39 &  52 &  64 &  85 & 209 & 962 \\
& $T_H$ &   0 &  21 &  21 &  36 &  42 &  44 & 123 & 210 &   0 &  23 &  18 &  21 &  21 &  26 &  75 & 4323 &   0 &  40 &  38 &  54 &  68 &  80 & 207 & 1118 \\
\midrule
\multirow{4}{*}{$D_R$}
& $T_T$ & 0.53 & 0.94 & 0.94 & 0.95 & 0.96 & 0.96 & 0.96 & 0.97 & 0.74 & 0.94 & 0.95 & 0.95 & 0.95 & 0.96 & 0.96 & 0.99 & 0.01 & 0.13 & 0.10 & 0.13 & 0.23 & 0.33 & 0.41 & 0.68 \\
& $T_E$ & 0.00 & 0.08 & 0.06 & 0.11 & 0.15 & 0.21 & 0.58 & 0.73 & 0.00 & 0.07 & 0.08 & 0.08 & 0.09 & 0.17 & 0.63 & 0.96 & 0.00 & 0.12 & 0.09 & 0.16 & 0.25 & 0.28 & 0.50 & 0.82 \\
& $T_M$ & 0.00 & 0.08 & 0.08 & 0.12 & 0.20 & 0.25 & 0.34 & 0.70 & 0.00 & 0.08 & 0.08 & 0.09 & 0.18 & 0.26 & 0.34 & 0.98 & 0.00 & 0.16 & 0.16 & 0.21 & 0.28 & 0.36 & 0.50 & 0.75 \\
& $T_H$ & 0.00 & 0.07 & 0.07 & 0.12 & 0.14 & 0.16 & 0.24 & 0.35 & 0.00 & 0.06 & 0.08 & 0.09 & 0.09 & 0.10 & 0.24 & 0.99 & 0.00 & 0.16 & 0.17 & 0.21 & 0.28 & 0.36 & 0.48 & 0.82 \\
\bottomrule
\end{tabular}
\end{adjustbox}
\end{table*}

\begin{table*}[htbp]
\caption{$M_T$, word Tokenization, Sequence Length of 4188, and 10\% of $T_x$}
\begin{adjustbox}{max width=\textwidth}
\begin{tabular}{@{}ll|rrrrrrrr|rrrrrrrr|rrrrrrrr@{}}
\toprule
\multicolumn{2}{c|}{Trained} & \multicolumn{8}{c|}{$V_A$} & \multicolumn{8}{c|}{$V_B$} & \multicolumn{8}{c}{$V_C$}
 \\ \multicolumn{2}{c|}{on} & min & avg & $q_{.50}$ & $q_{.75}$ & $q_{.90}$ & $q_{.95}$ & $q_{.99}$ & max & min & avg & $q_{.50}$ & $q_{.75}$ & $q_{.90}$ & $q_{.95}$ & $q_{.99}$ & max & min & avg & $q_{.50}$ & $q_{.75}$ & $q_{.90}$ & $q_{.95}$ & $q_{.99}$ & max
 \\ \midrule
\multirow{4}{*}{$D_A$}
& $T_T$ &   9 & 119 &  35 & 219 & 281 & 309 & 391 & 594 &   8 & 107 &  16 & 211 & 224 & 252 & 324 & 4306 &  16 &  32 &  22 &  33 &  59 &  70 & 110 & 1329 \\
& $T_E$ &   0 &  20 &   9 &  17 &  38 & 123 & 207 & 388 &   0 &  20 &   8 &  10 &  14 &  35 & 172 & 4352 &   0 &  32 &  24 &  40 &  61 &  90 & 189 & 1345 \\
& $T_M$ &   0 &  23 &  14 &  31 &  42 &  71 & 182 & 366 &   0 &  27 &  18 &  18 &  22 & 119 & 148 & 4349 &   0 &  45 &  38 &  54 &  79 & 125 & 210 & 1335 \\
& $T_H$ &   0 &  30 &  21 &  39 &  63 & 100 & 232 & 449 &   0 &  26 &  17 &  21 &  25 &  56 & 125 & 4347 &   0 &  41 &  34 &  53 &  69 &  87 & 213 & 1275 \\
\midrule
\multirow{4}{*}{$D_R$}
& $T_T$ & 0.02 & 0.44 & 0.12 & 0.94 & 0.96 & 0.96 & 0.98 & 0.99 & 0.03 & 0.41 & 0.07 & 0.94 & 0.95 & 0.95 & 0.97 & 0.99 & 0.02 & 0.15 & 0.11 & 0.15 & 0.28 & 0.40 & 0.51 & 0.98 \\
& $T_E$ & 0.00 & 0.06 & 0.03 & 0.06 & 0.13 & 0.31 & 0.55 & 0.74 & 0.00 & 0.05 & 0.03 & 0.04 & 0.05 & 0.12 & 0.57 & 1.00 & 0.00 & 0.13 & 0.10 & 0.17 & 0.29 & 0.39 & 0.58 & 0.99 \\
& $T_M$ & 0.00 & 0.08 & 0.05 & 0.11 & 0.14 & 0.22 & 0.57 & 0.74 & 0.00 & 0.08 & 0.08 & 0.08 & 0.09 & 0.39 & 0.60 & 1.00 & 0.00 & 0.18 & 0.17 & 0.22 & 0.35 & 0.48 & 0.64 & 0.98 \\
& $T_H$ & 0.00 & 0.10 & 0.07 & 0.13 & 0.21 & 0.30 & 0.50 & 0.77 & 0.00 & 0.07 & 0.07 & 0.09 & 0.10 & 0.19 & 0.46 & 1.00 & 0.00 & 0.16 & 0.15 & 0.21 & 0.30 & 0.38 & 0.56 & 0.94 \\
\bottomrule
\end{tabular}
\end{adjustbox}
\end{table*}

\begin{table*}[htbp]
\caption{$M_T$, word Tokenization, Sequence Length of 4188, and 50\% of $T_x$}
\begin{adjustbox}{max width=\textwidth}
\begin{tabular}{@{}ll|rrrrrrrr|rrrrrrrr|rrrrrrrr@{}}
\toprule
\multicolumn{2}{c|}{Trained} & \multicolumn{8}{c|}{$V_A$} & \multicolumn{8}{c|}{$V_B$} & \multicolumn{8}{c}{$V_C$}
 \\ \multicolumn{2}{c|}{on} & min & avg & $q_{.50}$ & $q_{.75}$ & $q_{.90}$ & $q_{.95}$ & $q_{.99}$ & max & min & avg & $q_{.50}$ & $q_{.75}$ & $q_{.90}$ & $q_{.95}$ & $q_{.99}$ & max & min & avg & $q_{.50}$ & $q_{.75}$ & $q_{.90}$ & $q_{.95}$ & $q_{.99}$ & max
 \\ \midrule
\multirow{4}{*}{$D_A$}
& $T_T$ &  15 & 126 & 105 & 188 & 261 & 301 & 395 & 600 &  14 & 118 &  89 & 133 & 237 & 238 & 320 & 4286 &   4 &  85 &  77 & 126 & 170 & 197 & 281 & 1281 \\
& $T_E$ &   0 &  17 &   7 &  16 &  36 &  60 & 176 & 454 &   0 &  18 &   7 &  10 &  18 &  39 & 144 & 4390 &   0 &  25 &  17 &  33 &  54 &  84 & 175 & 1270 \\
& $T_M$ &   0 &  26 &  22 &  37 &  51 &  68 & 155 & 377 &   0 &  25 &  18 &  21 &  23 &  43 & 146 & 4358 &   0 &  45 &  41 &  54 &  77 & 114 & 207 & 1332 \\
& $T_H$ &   0 &  25 &  22 &  37 &  47 &  68 & 165 & 357 &   0 &  27 &  18 &  22 &  28 &  48 & 140 & 4349 &   0 &  42 &  38 &  54 &  73 &  91 & 207 & 1306 \\
\midrule
\multirow{4}{*}{$D_R$}
& $T_T$ & 0.04 & 0.46 & 0.42 & 0.62 & 1.00 & 1.00 & 1.00 & 1.00 & 0.04 & 0.46 & 0.39 & 0.55 & 1.00 & 1.00 & 1.00 & 1.00 & 0.01 & 0.36 & 0.40 & 0.57 & 0.66 & 0.72 & 0.78 & 0.96 \\
& $T_E$ & 0.00 & 0.05 & 0.03 & 0.06 & 0.12 & 0.19 & 0.47 & 0.77 & 0.00 & 0.04 & 0.03 & 0.04 & 0.06 & 0.13 & 0.56 & 1.00 & 0.00 & 0.11 & 0.07 & 0.15 & 0.28 & 0.38 & 0.58 & 0.93 \\
& $T_M$ & 0.00 & 0.09 & 0.08 & 0.13 & 0.17 & 0.21 & 0.59 & 0.70 & 0.00 & 0.07 & 0.08 & 0.09 & 0.10 & 0.15 & 0.62 & 1.00 & 0.00 & 0.18 & 0.17 & 0.22 & 0.33 & 0.47 & 0.61 & 0.98 \\
& $T_H$ & 0.00 & 0.08 & 0.08 & 0.12 & 0.16 & 0.21 & 0.37 & 0.68 & 0.00 & 0.08 & 0.08 & 0.09 & 0.10 & 0.16 & 0.39 & 0.99 & 0.00 & 0.17 & 0.16 & 0.22 & 0.32 & 0.41 & 0.52 & 0.96 \\
\bottomrule
\end{tabular}%
\end{adjustbox}
\end{table*}

\begin{table*}[htbp]
\caption{$M_L$, word Tokenization, 100\% of $T_T$, evaluated on $V*$}
\begin{adjustbox}{max width=\textwidth}
%
\end{adjustbox}
\end{table*}

\begin{table*}[htbp]
\caption{$M_B$, word Tokenization, 100\% of $T_T$, evaluated on $V*$}
\begin{adjustbox}{max width=\textwidth}
%
\end{adjustbox}
\end{table*}

\begin{table*}[htbp]
\caption{$M_T$, word Tokenization, Sequence Length of 256, and 100\% of $T_T$, evaluated on $V*$}
\begin{adjustbox}{max width=\textwidth}
%
\end{adjustbox}
\end{table*}

\end{document}